\documentclass[12pt,oneside,a4paper,onecolumn]{elsarticle}
\usepackage{times}
\usepackage{fullpage}

\usepackage{xcolor}
\usepackage{hyperref}
\usepackage{amsmath}
\usepackage{amssymb}
\usepackage{graphicx}
\usepackage{multicol}
\usepackage{caption}
\usepackage{algorithm}
\usepackage{algorithmic}
\usepackage{cases}
\usepackage{subcaption}
\usepackage{tikz}
\usetikzlibrary{angles,quotes}
\usepackage{slashed}
\usetikzlibrary{patterns}
\usetikzlibrary{shapes.arrows}
\usepackage{soul}
\usepackage{pgfplots}
\usepackage{placeins}
\pgfplotsset{compat=newest}
\usepackage{mathtools} 
\usepackage{diagbox}
\usepackage{fixltx2e}
\usetikzlibrary{arrows.meta}

\newtheorem{rem}{Remark}

\usepackage{adjustbox}

\journal{Journal of \LaTeX\ Templates}

\newcommand{\INDSTATE}[1][1]{\STATE\hspace{#1\algorithmicindent}}

\begin{document}

\begin{frontmatter}

\title{A 3D phase-field based Eulerian variational framework for multiphase fluid-structure interaction with contact dynamics}

\author[UBC]{Xiaoyu Mao}

\author[UBC]{Rajeev Jaiman\corref{cor1}}
\ead{rjaiman@mech.ubc.ca}
\cortext[cor1]{Corresponding author}
\address[UBC]{Department of Mechanical Engineering, University of British Columbia, Vancouver, Canada}




\begin{abstract}	

Using a fixed Eulerian mesh, interface-capturing approaches such as volume-of-fluid, level-set and phase-field methods have been successfully utilized for a broad range of moving boundary problems involving multiphase fluids and single-phase fluid-structure interaction. Nevertheless, multiphase fluids interacting with multiple solids are rarely explored, especially for large-scale finite element simulations with contact dynamics. 
In this work, we introduce a novel parallelized three-dimensional fully Eulerian variational framework for simulating multiphase fluids interacting with multiple deformable solids subjected to solid-to-solid contact. In the framework, each solid or fluid phase is identified by a standalone phase indicator. Moreover the phase indicators are initialized by the grid cell method, which restricts the calculation to several grid cells instead of the entire domain and provides efficiency for modeling evolving interfaces on a fixed mesh. 
A diffuse interface description is employed for a smooth interpolation of the physical properties across the phases, yielding unified mass and momentum conservation equations for the coupled dynamical interactions. For each solid object, temporal integration is carried out to track the strain evolution in an Eulerian frame of reference. The coupled differential equations are solved in a partitioned manner and integrated via nonlinear iterations. 
We first verify the framework against reference numerical data in a two-dimensional case of a rotational disk in a lid-driven cavity flow. The case is generalized to a rotational sphere in a lid-driven cavity flow to showcase large deformation and rotational motion of solids and examine the convergence in three dimensions. We then simulate the falling of an immersed solid sphere on an elastic block under gravitational force to demonstrate the translational motion and the solid-to-solid contact in a fluid environment. Finally, we demonstrate the capability of our proposed framework for a ship-ice interaction problem involving multiphase fluids with an air-water interface and contact between a floating ship and ice floes.
\end{abstract}

\begin{keyword} 
	Fully Eulerian FSI,
	Phase-field method,
	Multiphase fluids,
	Solid-to-solid contact,
	Partitioned iterative coupling
\end{keyword}

\end{frontmatter}

\section{Introduction}
Multiphase fluid-structure interaction is a bi-directional coupling of multiphase fluids with moving/deforming structures, where the structures can contact each other. It is frequently encountered in many natural phenomena and engineering applications. Practical applications include the formation and clearance of blood clots \cite{montgomery2023clotfoam,mahdy2019clearing}, soft robotics \cite{nojiri2019development} in multiphase flow, floating offshore structures, subsea pipelines and of particular interest of the current work, the ship-ice interaction in the Arctic region \cite{xue2020review}. These problems usually involve large translational and rotational motion of solids, the resulting fluid-structure interaction (FSI), evolving interfaces of multiphase fluids, and contact between solids.  In such scenarios, robust and accurate modeling  of fluid-solid and fluid-fluid interfaces with solid-to-solid contact interactions  is very challenging for the state-of-the-art numerical methods.  The simulation of such cases requires a general multiphase and multiphysics numerical framework to treat evolving interfaces consistently and overcome intrinsic differences between the kinematics and dynamics of fluids and solids.

Various techniques treating the fluid-solid interface have been developed in the past several decades. As one of the most accurate approaches, the arbitrary Lagrangian-Eulerian method \cite{donea2017arbitrary} has been wildly used and demonstrated its success in solving FSI problems with an oscillatory motion of solids. However, it may fail for solids subjected to large translational and rotational motion, which can cause significant mesh distortion during the evolution of the fluid-solid interface. Topological changes in the contact between solids are also quite challenging to handle. To resolve the mesh distortion, one can use independent meshes for fluids and solids and couple the solutions with interpolation or Lagrangian multiplier, as conducted in the immersed boundary methods \cite{peskin2002}, overset grid methods \cite{benek1983flexible,tang2003overset} and fictitious domain methods \cite{glowinski1994fictitious,baaijens2001fictitious}. In these methods, the fluid-solid coupling in non-matching grids can violate conservation laws and boundary conditions and thus must be judiciously constructed. Another approach to resolve the inconsistency between fluids and solids is to unify the kinematics and dynamics in an Eulerian frame of reference \cite{dunne2006eulerian,wick2013fully}. Through an interface-capturing approach on a fixed mesh, the fully Eulerian approach allows for arbitrary motion and topological changes. The velocity and stress continuity at fluid-fluid and fluid-solid interfaces are naturally satisfied via unified velocity field and momentum equations. These traits bring convenience and robustness for numerical simulations of multiphase FSI problems with reasonable accuracy, thus being the focus of the current work.

\begin{figure}
	\centering
	\includegraphics[scale=0.4,trim=0 0 0 0,clip]{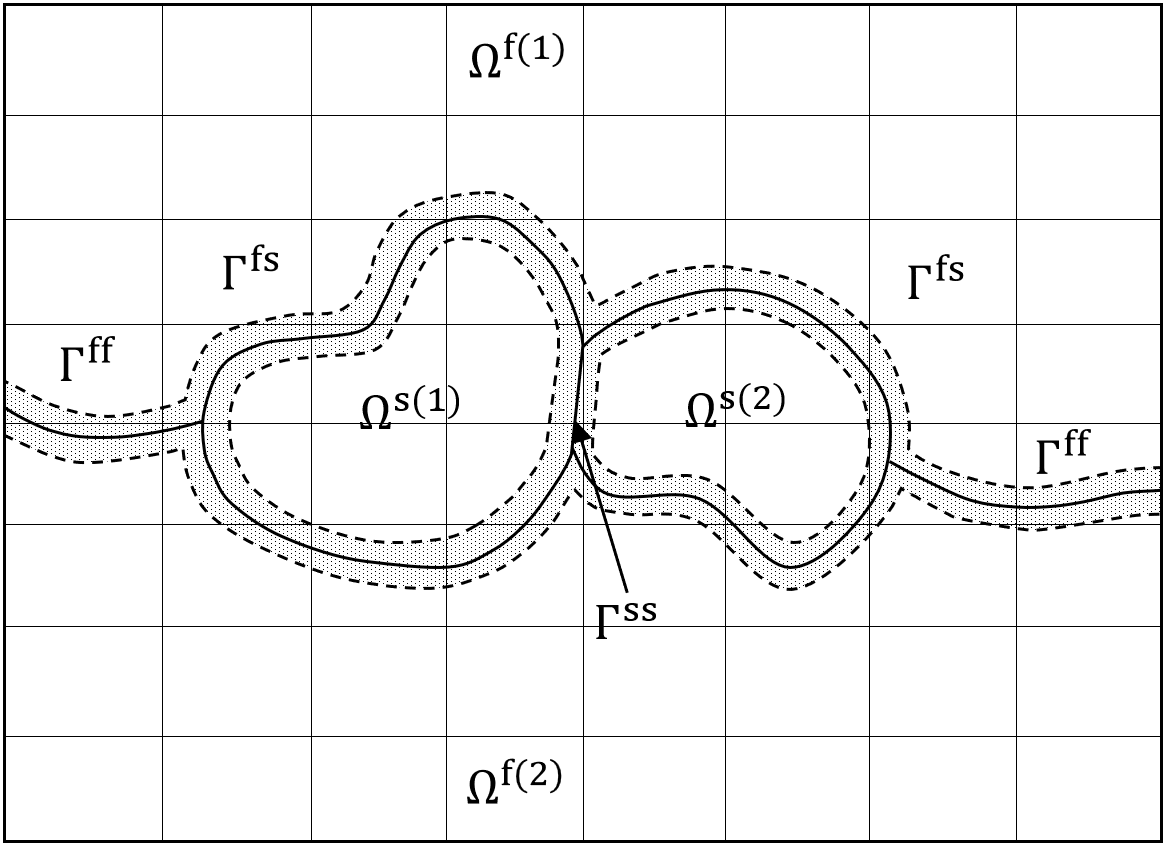}
	\caption{An abstract representation of fluid-fluid and fluid-solid interfaces in multiphase fluid-solid interaction problems. The diffuse interface in
		the Eulerian mesh is shown as the shaded region.}
	\label{absrep}
\end{figure}

While alleviating the issues in interface motion and dynamics, two major challenges arise from the fully Eulerian description namely (i) accuracy of interface representation and evolution, and (ii) evaluation of displacement and strain of solids in an Eulerian frame of reference. For the first challenge, we employ our recently proposed interface-preserving phase-field method \cite{mao2021variational}. The phase-field method involves a diffuse interface description, where the interface is modeled as a transitional region with finite thickness, across which the physical properties vary smoothly. The transition pattern is adaptively regularized to maintain a consistent and smooth interface transition in the diffuse interface description. For the first time, we apply this description to all the interfaces involved in multiphase FSI problems, namely the fluid-fluid interface $\Gamma^\mathrm{ff}$, the fluid-solid interface $\Gamma^\mathrm{fs}$ and the solid-solid interface $\Gamma^\mathrm{ss}$ between the fluid domains $\Omega^\mathrm{f}$ and the solid domains $\Omega^\mathrm{s}$ as illustrated in Fig.~\ref{absrep}. With this approach, the requirement of resolving the interface to satisfy the velocity and traction continuity conditions with the standard finite element method can be circumvented. For the second challenge, one can consider the initial point set method \cite{dunne2006eulerian}, the reference map technique \cite{kamrin2012reference}, and the evolution of left Cauchy-Green tensor \cite{sun2014full,mokbel2018phase}. To maintain accuracy during the evaluation from the solid displacement to strain, we directly evolve the left Cauchy-Green tensor in the current work.

The present work is an extension of the work carried out in our earlier work \cite{mao2023interface}, wherein a two-dimensional (2D) variational framework for a fully Eulerian single-phase FSI is established. In the current work, the framework is extended to three-dimensional (3D) multiphase fluids interacting with multiple solid objects with solid-to-solid contact dynamics. The framework employs a combined formulation for multiphase flow and fluid-structure interaction using our interface-preserving method for unstructured meshes. We consider a 3D finite element discretization in a general computational domain with given boundary conditions. The interfaces are identified by multiple order parameters indicating the local composition while evolving by the interface-preserving phase-field equations. The resulting order parameters are used to unify the mass and momentum conservation equations via phase-dependent interpolation. In the regions indicated as solid domain, the left Cauchy-Green tensor is tracked to complement the solid strain, hence completing the Eulerian multiphase FSI framework. The equations are solved via the Newton-Raphson method and integrated in a partitioned iterative manner. Solving these partial differential equations for the coupled multiphase FSI system in a 3D domain can be computationally very expensive. To make the simulations practical and robust, we employ a stabilized Petrov-Galerkin discretization and a fully implicit time marching scheme to enable a coarser level of spatial and temporal resolutions. The framework is realized through multi-processor parallelized implementation to further accelerate the simulations. 

Through the aforementioned computational efforts, we have successfully built a fully integrated, stable and parallelized 3D fully Eulerian finite element solver for general-purpose multiphase FSI problems. We showcase the capability and validity of the framework for the test cases of increasing complexity, namely a rotating disk and a sphere in a lid-driven cavity flow, falling and bouncing back of a solid sphere on an elastic block. We highlight the accuracy and convergence of the framework to capture the fluid-solid interface with solid–to-solid contact interactions. In practical multiphase FSI problems, diffuse interfaces for solid objects can have complex geometries that cannot be generated through simple analytical expressions. In these cases, the initialization of the order parameter field needs to be achieved in a general and robust manner during the pre-processing stage. To accomplish this goal, we suggest a phase-field interface processing algorithm by integrating the grid cell method and ray-casting algorithm. A novel non-duplicated plane partitioning is developed for the discrete implementation of the ray-casting algorithm. Finally, we demonstrate the geometry processing utilities with our proposed fully Eulerian finite element framework for a simplified situation of ship-ice interaction.

The organization of this paper is as follows: Section \ref{formulation} describes the formulations of the unified Eulerian variational framework for multiphase FSI. Section \ref{odp_Init} presents the methodologies to initialize the order parameter field for complex geometries. We first examine the initialization of the order parameters in Section \ref{exp_Init}. After that, the multiphase FSI framework is verified and demonstrated in cases with increasing complexity in Section \ref{test}. The ship-ice interaction is discussed and demonstrated in Section \ref{Sec:ship}. Finally, we conclude the work in Section \ref{conclusion}.

\section{Fully Eulerian formulation for multiphase FSI} \label{formulation} 
In this section, we present the continuous and semi-discrete formulations for the fully Eulerian multiphase FSI framework. We start with the unified mass and momentum conservation equations for the continuum. Then we describe the interface equations and the constitutive relation for the fluid phases. We next introduce the interface equations, the constitutive relation, and the kinematic equations for the solid phases. Specifically, we provide a detailed description of the linearization of the solid stress. Finally, we discuss some implementation details in the framework.

\subsection{Conservation laws for the multiphase FSI system}
We present the partial differential equations for the unified mass and momentum conservation at the continuous level and then review the semi-discrete variational formulations using a stabilized finite element discretization.
\subsubsection{Strong differential form}

Consider a physical domain $\Omega\times]0,T[$ with spatial coordinates $\boldsymbol{x}$ and temporal coordinate $t$. The domain comprises subdomains $\Omega^i_t,i=1,2,\cdots,n$, where $i$ and $n$ are the index and the total number of the subdomains. Each subdomain is considered to be an incompressible Newtonian fluid or Neo-Hookean solid of particular physical properties. The subdomains are distinguished by a fraction function $\alpha^i(\phi^i(\boldsymbol{x},t))$, where $\phi^i(\boldsymbol{x},t)$ being the order parameter in the phase-field method. Both $\alpha^i(\phi^i)$ and $\phi^i$ are used to indicate a binary system that the current location $(\boldsymbol{x},t)$ \emph{is} or \emph{is not} phase $i$, which leads to $\alpha^i(\phi^i(\boldsymbol{x},t))=1,\boldsymbol{x}\in \Omega^i_t$ and $\alpha^i(\phi^i(\boldsymbol{x},t))=0, \boldsymbol{x}\in \Omega\backslash\Omega^i_t$. In the current work, $\alpha^i(\phi^i)$=$\frac{1+\phi^i}{2}$ is used.   Through phase-dependent interpolation, the density, viscosity, stress, and the body force of the continua can be calculated as: $\rho=\sum\limits_{i=1}^n\alpha^i(\phi^i)\rho^i$,$\nu =\sum\limits_{i=1}^n\alpha^i(\phi^i)\nu^i$, $\boldsymbol{b}=\sum\limits_{i=1}^n\alpha^i(\phi^i)\boldsymbol{b}^i$ and $\boldsymbol{\sigma}=\sum\limits_{i=1}^n\alpha^i(\phi^i)\boldsymbol{\sigma}^i$. Along with the incompressible constraint, the momentum and mass conservation can be written as:
\begin{align}
	\rho(\partial_t \boldsymbol{v}+\boldsymbol{v}\cdot\nabla\boldsymbol{v})&= \nabla\cdot\boldsymbol{\sigma}+\boldsymbol{b},\\
	\nabla\cdot \boldsymbol{v}&=0,
\end{align}
where $\boldsymbol{v}$ is a unified velocity field for the continua.

\subsubsection{Semi-discrete variational form}

We employ the generalized-$\alpha$ technique \cite{jansen2000generalized} for the temporal discretization which enables a user-controlled high frequency damping desirable for coarse discretizations in space and time. For a first-order system of variable $\varphi$, the generalized-$\alpha$ method is given by: 
\begin{align}
	\partial_t \varphi^{n+\alpha_m}&=f(\varphi^{n+\alpha}),\\
	\varphi^{n+1}&=\varphi^{n}+\Delta t \partial_t \varphi^n +\Delta t\varsigma (\partial_t \varphi^{n+1}-\partial_t \varphi^n),\\
	\partial_t \varphi^{n+\alpha_m} &= \partial_t \varphi^{n}+\alpha_m (\partial_t \varphi^{n+1}-\partial_t \varphi^n),\\
	\varphi^{n+\alpha}&=\varphi^n+\alpha(\varphi^{n+1}-\varphi^{n}),
\end{align} 
where $\alpha$, $\alpha_\mathrm{m}$ and $\varsigma$ are the generalized-$\alpha$ parameters depending on the user-defined spectral radius $\rho_{\infty}$:
\begin{align}
	\alpha=\frac{1}{1+\rho_{\infty}},\ \alpha_m=\frac{1}{2}\left(\frac{3-\rho_{\infty}}{1+\rho_{\infty}}\right),\ \varsigma=\frac{1}{2}+\alpha_m-\alpha.
\end{align} 
The temporal evolution works in a predictor-multicorrector manner. In every nonlinear iteration, we first predict the solution at $n+1$, interpolate the solution to $n+\alpha$ and solve the first-order system. After that, we correct the solution at $n+1$ according to the solution at $n+\alpha$. In the current work, the spectral radius is selected as $\rho_{\infty}=0$. This applies to the temporal discretization of all equations hereafter.

Now we introduce the spatial discretization using the stabilized finite element method. Suppose $\mathcal{S}^\mathrm{h}_{\boldsymbol{v}}$ and $\mathcal{S}^\mathrm{h}_{p}$  denote the space of trial solutions such that:
\begin{align}
	\mathcal{S}^\mathrm{h}_{\boldsymbol{v}} &= \big\{ \boldsymbol{v}_\mathrm{h}\ |\ \boldsymbol{v}_\mathrm{h} \in \left(H^1(\Omega)\right)^{d}, \boldsymbol{v}_\mathrm{h} = \boldsymbol{v}_{D}\ \mathrm{on}\ \Gamma_{D} \big\},\\
	\mathcal{S}^\mathrm{h}_{p} &= \big\{ p_\mathrm{h}\ |\ p_\mathrm{h} \in L^2(\Omega) \big\},
\end{align}
where $\left(H^1(\Omega)\right)^{d}$ denotes the space of square-integrable $\mathbb{R}^{d}$-valued functions with square-integrable derivatives on $\Omega$, $L^2(\Omega)$ is the space of the scalar-valued functions that are square-integrable on $\Omega$ and $\Gamma_D$ represents the Dirichlet boundaries with the value of $\boldsymbol{v}_D$. Similarly, we define $\mathcal{V}^\mathrm{h}_{\boldsymbol{\psi}}$ and $\mathcal{V}^\mathrm{h}_{q}$ as the space of test functions such that:
\begin{align}
	\mathcal{V}^\mathrm{h}_{\boldsymbol{\psi}} &= \big\{ \boldsymbol{\psi}_\mathrm{h}\ |\ \boldsymbol{\psi}_\mathrm{h} \in \left(H^1(\Omega)\right)^{d}, \boldsymbol{\psi}_\mathrm{h} = \boldsymbol{0}\ \mathrm{on}\ \Gamma_{D} \big\},\\
	\mathcal{V}^\mathrm{h}_{q} &= \big\{ q_\mathrm{h}\ |\ q_\mathrm{h} \in L^2(\Omega) \big\}.
\end{align}
The variational statement of the momentum and mass conservation equations can be written as: \\
find $\left[\boldsymbol{v}_\mathrm{h}(t^\mathrm{n+\alpha}),p_\mathrm{h}(t^\mathrm{n+1})\right]\in \mathcal{S}^\mathrm{h}_{\boldsymbol{v}} \times \mathcal{S}^\mathrm{h}_{p}$ such that $\forall \left[\boldsymbol{\psi}_\mathrm{h},q_\mathrm{h}\right]\in \mathcal{V}^\mathrm{h}_{\boldsymbol{\psi}} \times \mathcal{V}^\mathrm{h}_{q}$,
\begin{align}
	&\int_{\Omega} \rho(\boldsymbol{\phi}_\mathrm{h}^{n+\alpha}) ( \partial_t{\boldsymbol{v}}^{n+\alpha_m}_\mathrm
	{h} + {\boldsymbol{v}}_{\mathrm{h}} \cdot\nabla{\boldsymbol{v}}_{\mathrm{h}})\cdot\boldsymbol{\psi}_{\mathrm{h}} \mathrm{d\Omega} +
	\int_{\Omega} {\boldsymbol{\sigma}}_{\mathrm{h}}(\boldsymbol{\phi}_\mathrm{h}^{n+\alpha})
	: \nabla\boldsymbol{\psi}_{\mathrm{h}} \mathrm{d\Omega} \nonumber \\
	+ &\displaystyle\sum_\mathrm{e=1}^\mathrm{n_{el}}\int_{\Omega
		^{\mathrm{e}}} \frac{\tau_\mathrm{m}}{\rho(\boldsymbol{\phi}_\mathrm{h}^{n+\alpha})} (\rho(\boldsymbol{\phi}_\mathrm{h}^{n+\alpha}
	){\boldsymbol{v}}_{\mathrm{h}}\cdot\nabla
	\boldsymbol{\psi}_{\mathrm{h}}+ \nabla q_{\mathrm{h}} )\cdot
	\boldsymbol{\mathcal{R}}_\mathrm{m} \mathrm
	{d\Omega^e} \nonumber \\
	+ &\int_{\Omega}q_{\mathrm{h}}(\nabla\cdot{\boldsymbol{v}}_{\mathrm{h}}) \mathrm{d\Omega} + \displaystyle\sum_\mathrm{e=1}^\mathrm{n_{el}}\int
	_{\Omega^{\mathrm{e}}} \nabla\cdot\boldsymbol{\psi}_{\mathrm{h}}\tau
	_\mathrm{c}\rho(\boldsymbol{\phi}_\mathrm{h}^{n+\alpha}) \mathcal{R}_\mathrm
	{c} \mathrm{d\Omega^e}\nonumber\\
	= & \int_{\Omega}\boldsymbol{b}(\boldsymbol{\phi}^{n+\alpha}_\mathrm{h})\cdot\boldsymbol{\psi}_{\mathrm{h}}\mathrm{d\Omega}, \label{PG_NS}
\end{align}
where $\boldsymbol{\phi}$ is a vector of all the order parameters used for the phase-dependent interpolation,  $\boldsymbol{A}:\boldsymbol{B}=A_{i,j}B_{i,j}$,  $\boldsymbol{\mathcal{R}}_\mathrm{m}$ and $\mathcal{R}_\mathrm{c}$  denote the element-wise residuals of the momentum and mass conservation equations respectively, and $n_\mathrm{el}$ is the number of elements. 
In Eq.~(\ref{PG_NS}), the terms in the first line represent the Galerkin projection of the momentum conservation equation in the test function space $\boldsymbol{\psi}_\mathrm{h}$ and the second line comprises of the Petrov-Galerkin stabilization term for the momentum conservation equation. The third line denotes the Galerkin projection and stabilization terms for the mass conservation equation. The terms in the last line are the Galerkin projection of the body force.  The stabilization parameters $\tau_\mathrm{m}$ and $\tau_\mathrm{c}$  in Eq.~(\ref{PG_NS}) are given by \cite{shakib1991new, brooks1982streamline}:
\begin{align}
	\tau_\mathrm{m} &= \left( \bigg( \frac{2}{\Delta t}\bigg)^2 + \boldsymbol{v}_\mathrm{h}\cdot\boldsymbol{G}\boldsymbol{v}_\mathrm{h} + C_I \bigg( \frac{\mu(\boldsymbol{\phi})}{\rho(\boldsymbol{\phi})}\bigg)^2 \boldsymbol{G}:\boldsymbol{G} \right)^{-1/2},\qquad \tau_\mathrm{c} = \frac{1}{\mathrm{tr}(\boldsymbol{G})\tau_\mathrm{m}},
\end{align}
where $C_I$ is a constant derived from the element-wise inverse estimates \cite{harari1992c}, $\boldsymbol{G}$ is the element contravariant metric tensor and $\mathrm{tr}(\boldsymbol{G})$ is the trace of the contravariant metric tensor. This stabilization in the variational form circumvents the Babu$\mathrm{\check{s}}$ka-Brezzi condition which needs to be satisfied by any standard mixed Galerkin method \cite{Johnson}.  

\subsection{Equations for the fluid phases} \label{subsec:feqs}
The fluids can be solved naturally in an Eulerian frame of reference. Therefore, we only need a phase-field equation to track the interface of each fluid phase. To maintain a consistent transition of physical properties across the diffuse interface, we employ the interface-preserving phase-field method developed in \cite{mao2021variational}. We first present the Allen-Cahn phase-field equation and the time-dependent mobility model. We next turn our attention to the stabilized semi-implicit form of our formulation.
\subsubsection{Strong differential form}
For each phase of a Newtonian fluid, we employ the interface-preserving phase-field method\cite{mao2021variational}, which is given as:
\begin{align}
	\frac{\partial\phi^i}{\partial t}+\boldsymbol{v}\cdot \nabla \phi^i &= - \gamma^i(t)\left(F'(\phi^i)-\varepsilon^2\nabla^2\phi^i-\beta(t)\sqrt{F(\phi^i)}\right), \label{Allen-Cahn}\\
	\gamma^i(t) &= \frac{1}{\eta}\mathcal{F}\left(\left|\frac{(\nabla \phi^i)^T \cdot \nabla\boldsymbol{v}\cdot \nabla \phi^i}{|\nabla\phi^i|^2}\right|\right),\label{continuous mobility}
\end{align} 
where $\varepsilon$ is the interface thickness parameter, $F(\phi^i)=\left(0.25\left(\left(\phi^i\right)^2-1\right)^2\right)$ is the bulk energy, $\beta(t)= \frac{\int_\Omega F'(\phi^i)d\Omega}{\int_\Omega 
	\sqrt{F(\phi^i)}d\Omega}$ is the Lagrange multiplier for mass conservation \cite{brassel2011modified}. The RHS of Eq.~\ref{Allen-Cahn} is the gradient flow of the free energy functional converging to a hyperbolic tangent profile, the magnitude of which is proportional to the mobility coefficient $\gamma$. The time-dependent mobility coefficient $\gamma^i(t)$ adaptively regularizes the hyperbolic tangent profile to address the convective distortion, which is the intensity of compression or extension on the diffuse interface due to the convection effect. The convective distortion is quantified by the projection of the velocity gradient in the normal direction of the diffuse interface region.
We consider $\mathcal{F}(\varphi(\boldsymbol{x},t))=\sqrt{\frac{\int (\varphi(\boldsymbol{x},t))^2 d\Omega}{\int 1 d\Omega}}$,  where $\ -\delta <\phi<\delta$ is the RMS value of the function field $\varphi(\boldsymbol{x},t)$ in the interface region, where the boundaries of the region are selected as $\delta = 0.9$. The parameter $\eta$ is a user-defined value to control the regularization and its selection is discussed in \cite{mao2021variational}.

\subsubsection{Discrete form of the time-dependent mobility model}
 We present the discrete form of the time-dependent mobility model in this subsection. The strong form of the time-dependent mobility model is given by Eq.~(\ref{continuous mobility}):
\begin{equation}
	\gamma(t)=\frac{1}{\eta}\mathcal{F}\left(\left|\frac{(\nabla\phi)^T\cdot\nabla\boldsymbol{v}\cdot\nabla\phi}{|\nabla\phi|^2}\right|\right), \label{discrete frame independnet}
\end{equation}
where  $\mathcal{F}(\varphi(\boldsymbol{x},t))=\sqrt{\frac{\int(\varphi(\boldsymbol{x},t))^2d\Omega}{\int 1 d\Omega}},\ \boldsymbol{x}\in\Gamma^{\phi}_{DI}(t)$ and $\Gamma^{\phi}_{DI}(t)=\{(\boldsymbol{x},t)||\phi(\boldsymbol{x},t)|\leqslant0.9\}$.
In Eq.~(\ref{discrete frame independnet}), $\mathcal{F}\left(\left|\frac{(\nabla\phi)^T\cdot\nabla\boldsymbol{v}\cdot\nabla\phi}{|\nabla\phi|^2}\right|\right)$ can be approximated as the RMS of $\left|(\nabla\phi)^T\cdot\nabla\boldsymbol{v}\cdot\nabla\phi/|\nabla\phi|^2 \right|$ at all the nodes located inside the diffuse interface region $\Gamma^{\phi}_{DI}(t)$.

The nodal value of $\left|(\nabla\phi)^T\cdot\nabla\boldsymbol{v}\cdot\nabla\phi/|\nabla\phi|^2 \right|$ is calculated as follows. The nodal value of $\boldsymbol{v}_\mathrm{h}(t^{\mathrm{n}+\alpha})$, $\phi_\mathrm{h}(t^{\mathrm{n}+\alpha})$ is used to interpolate the $\nabla\boldsymbol{v}_\mathrm{h}(t^{\mathrm{n}+\alpha})$ and $\nabla\phi_\mathrm{h}(t^{\mathrm{n}+\alpha})$ at the quadrature points, and $L^2$-projection is used to project the value on the quadrature points back to the nodes inside the diffuse interface region \cite{jaiman2016partitioned}. If the node lies outside the diffuse interface region, the value is assigned to be zero. For example, for the node $p$, we have:
\begin{align}\label{tdm-num}
	\left|\left(\frac{(\nabla\phi_\mathrm{h})^T\cdot\nabla\boldsymbol{v}_\mathrm{h}\cdot\nabla\phi_\mathrm{h}}{|\nabla\phi_\mathrm{h}|^2}\right)_{p}\right|=
	\begin{cases}
		\left|\dfrac{\sum_e\int_{\Omega^e}N_p( (\nabla\phi_\mathrm{h})^T\cdot\nabla\boldsymbol{v}_\mathrm{h}\cdot\nabla\phi_\mathrm{h})/|\nabla\phi_\mathrm{h}|^2 d\Omega^e}{\sum_e\int_{\Omega_e}N_p d\Omega^e}\right|
		& \mathrm{if}\ |\phi_p|\leqslant0.9 ,\\
		\hfil 0                    
		& \mathrm{if}\ |\phi_p|>0.9,
	\end{cases}
\end{align}
where $N_p$ represents the shape function at node $p$. The discrete form of the time-dependent mobility model is given by:
\begin{equation}
	\gamma(t^{\mathrm{n}+\alpha})=\frac{1}{\eta}\left(\sqrt{\frac{1}{n_{DI}}\sum\limits_{p=1}^{n_{DI}} \left|\left(\frac{(\nabla\phi_\mathrm{h})^T\cdot\nabla\boldsymbol{v}_\mathrm{h}\cdot\nabla\phi_\mathrm{h}}{|\nabla\phi_\mathrm{h}|^2}\right)_{p}\right|^2}\right),
\end{equation}
where $n_{DI}$ denotes the number of the nodes lying inside the diffuse interface region.

The $L^2$-projection in Eq.~(\ref{tdm-num}) needs extra attention in a multi-processor parallelized environment. In the current implementation, we first calculate the numerator and denominator at the RHS of Eq.~\ref{tdm-num} on each partitioned mesh block. For boundary nodes between the mesh blocks, message passing is performed to take the contribution from adjacent mesh blocks. After the numerator and the denominator are exchanged synchronously across the mesh, the division is calculated as the last step. A detailed discussion regarding the parallelized implementation of the $L^2$-projection can be found in \ref{backward projection}.

\subsubsection{Semi-discrete variational form}\label{dis-pf}

	Suppose $\mathcal{S}^\mathrm{h}_{\phi^i}$ and $\mathcal{V}^\mathrm{h}_{\psi}$ denote the space of trial solution and test functions such that:
	\begin{align}
		\mathcal{S}^\mathrm{h}_{\phi^i} &= \big\{ \phi^i_\mathrm{h}\ |\ \phi^i_\mathrm{h} \in H^1(\Omega), \phi^i_\mathrm{h} = \phi^i_{D}\ \mathrm{on}\ \Gamma^\mathrm{\phi^i}_{D} \big\},\\
		\mathcal{V}^\mathrm{h}_{\psi} &= \big\{ \psi_\mathrm{h}\ |\ \psi_\mathrm{h} \in H^1(\Omega), \psi_\mathrm{h} = 0\ \mathrm{on}\ \Gamma^\mathrm{\phi^i}_{D} \big\},
	\end{align}
	where $\Gamma^{\phi^i}_{D}$ denotes the Dirichlet boundaries for the order parameter with boundary value of $\phi^i_D$.
	The variational statement of the Allen-Cahn equations can be written as: find $\phi^i_\mathrm{h}(t^\mathrm{n+\alpha})\in\mathcal{S}^\mathrm{h}_{\phi^i}$ such that $\forall  \psi_\mathrm{h}\in \mathcal{V}^\mathrm{h}_{\psi}$ for the Allen-Cahn equation:
	\begin{align}
		\label{PPV_AC}
		\int_{\Omega}\Big( \psi_{\mathrm{h}}\partial_t{\phi}_{\mathrm{h}}^{i,n+\alpha_m} +
		\psi_{\mathrm{h}}\big(\boldsymbol{v}_\mathrm{h}^{n+\alpha}\cdot\nabla\phi^i_{\mathrm{h}}\big) +& \gamma^i(t^{n+\alpha}) \big(\nabla
		\psi_{\mathrm{h}}\cdot(\hat{k}\nabla\phi^i_{\mathrm{h}} ) + \psi_{\mathrm{h}}\hat{s}\phi^i
		_{\mathrm{h}} - \psi_{\mathrm{h}}\hat{f}\big) \Big) \mathrm{d}\Omega\nonumber\\
		+& \displaystyle\sum_\mathrm{e=1}^\mathrm{n_{el}}\int_{\Omega
			^{\mathrm{e}}} \left(\boldsymbol{v}_\mathrm{h}^{n+\alpha}\cdot\nabla \psi_{\mathrm{h}}
		\right)\tau_{\phi^i} \mathcal{R}_\phi^i  \mathrm{d}\Omega^{\mathrm{e}}=0,
	\end{align}
	where $\boldsymbol{v}$, $\hat{k}$, $\hat{s}$ and $\hat{f}$ are the flow velocity, the modified diffusion coefficient, modified reaction coefficient and modified source respectively which are defined in \cite{joshi2018positivity}, $\mathcal{R}_\phi^i$ denotes the element-wise residuals for the Allen-Cahn equation of the $i$-th fluid phase and the time-dependent mobility is denoted as $\gamma^i(t^{n+\alpha})$. In Eq.~(\ref{PPV_AC}), the first line is the Galerkin projection of the transient, convection, diffusion, reaction and source terms, the second line represents the Petrov-Galerkin stabilization terms where
	the stabilization parameters $\tau_\phi^i$ are given by \cite{shakib1991new, brooks1982streamline}:
	
	\begin{align}
		\tau_{\phi}^i &= \left( \bigg(\frac{2}{\Delta t} \bigg)^2 + \boldsymbol{v}_\mathrm{h}\cdot\boldsymbol{G}\boldsymbol{v}_\mathrm{h} + 9\hat{k}^2 \boldsymbol{G}:\boldsymbol{G} + \hat{s}^2 \right)^{-1/2}.
	\end{align}
For the current FSI model, the stress of the incompressible Newtonian fluid used  is given by:
\begin{equation}
	\boldsymbol{\sigma}^{i}=-p\boldsymbol{I}+\mu^i \left( \nabla \boldsymbol{v} + (\nabla \boldsymbol{v})^T\right),
\end{equation} 
where $p$ is the hydrostatic pressure and $\mu^i$ is the dynamic viscosity for the $i$-th fluid.

\subsection{Equations for the solid phases}\label{subsec:seqs}
For each phase of the Neo-Hookean solid, we employ the interface-preserving phase-field method to capture the interface on a fixed mesh. The equations and discretizations are the same as introduced in subsection \ref{subsec:feqs}. In addition, the strain of the solid needs to be tracked. In the current work, it is accomplished by evolving the left Cauchy-Green tensor. The governing equations are presented first, following the semi-discrete variational form.
While the contact between solids can be solved with the current framework via the phase-field method on a fixed mesh, its mechanism needs to be further clarified. In the collision of two solid objects, the contact solids are subjected to no-penetration and force balance conditions at the contact point or surface. In the current work, we only consider the elastic restoring force from the deformation of New-Hookean models with no-slip and no-penetration conditions at the contact point/surface. We want to demonstrate that by tracking each solid object with an independent order parameter and a left Cauchy-Green tensor, the force balance is naturally handled by the stress continuity, while the no-penetration condition and no-slip condition can be naturally satisfied with velocity continuity in a unified velocity field. 

\subsubsection{Strong differential form}
The solid strain is tracked by the evolution of the left Cauchy-Green tensor. However, the evolved components of the tensor persist in the computational field after the solid left the region. When the solid crosses the region again, the remaining values become unphysical initialization, which may result in numerical instability and wrong solutions. To resolve this issue, we try to recover $\boldsymbol{B}=\boldsymbol{I}$ outside of the solid domain at every time step employing an approach inspired by \cite{mokbel2018phase}:
\begin{align}
	\alpha^i(\phi^i)\left(\frac{\partial \boldsymbol{B}^i}{\partial t}+(\boldsymbol{v}\cdot\nabla)\boldsymbol{B^i}-\nabla\boldsymbol{v}\boldsymbol{B^i}-\boldsymbol{B}(\nabla\boldsymbol{v})^T\right)+(1-\alpha^i(\phi^i))(\boldsymbol{B}^i-\boldsymbol{I})=0.
\end{align}
With the left Cauchy-Green tensor readily available, the solid stress is calculated as $\boldsymbol{\sigma}^i=-p\boldsymbol{I}+\mu^i_{L}(\boldsymbol{B}^i-\boldsymbol{I})+\mu^i(\nabla\boldsymbol{v}+(\nabla\boldsymbol{v})^T)$, where $\mu^i_{L}$ and $\mu^i$ are the shear modulus and the viscosity for the $i$-th solid, respectively. 

\subsubsection{Semi-discrete variational form}
For the left Cauchy-Green tensor, we define the space of trial solution $\mathcal{S}_{\boldsymbol{B}^i}^{\mathrm{h}}$ and test functions $\mathcal{V}_{\boldsymbol{\psi}}^{\mathrm{h}}$ as:
\begin{align}
	\mathcal{S}^\mathrm{h}_{\boldsymbol{B}^i} &= \big\{ \boldsymbol{B}^i_\mathrm{h}\ |\ \boldsymbol{B}^i_\mathrm{h} \in \left(H^1(\Omega)\right)^{d}\big\},\\
	\mathcal{V}^\mathrm{h}_{\boldsymbol{\psi}} &= \big\{ \boldsymbol{\psi}_\mathrm{h}\ |\ \boldsymbol{\psi}_\mathrm{h} \in \left(H^1(\Omega)\right)^d \big\}.
\end{align}
The variational statement of the Cauchy-Green tensor can be written as:
find $\boldsymbol{B}^i_{\mathrm{h}}(t^{\mathrm{n+\alpha}})\in\mathcal{S}^{\mathrm{h}}_{\boldsymbol{B}^i}$ such that $\forall \boldsymbol{\psi}_{\mathrm{h}}\in\mathcal{V}^{\mathrm{h}}_{\boldsymbol{\psi}}$
\begin{align}\label{semiCGT}
	&\alpha(\phi^{n+\alpha})\int_{\Omega}\Big(\partial_t\boldsymbol{B}_{\mathrm{h}}^{i,n+\alpha_m}+(\boldsymbol{v}^{n+\alpha}_{\mathrm{h}}\cdot\nabla)\boldsymbol{B}^i_{\mathrm{h}}-\nabla\boldsymbol{v}^{n+\alpha}_\mathrm{h}\boldsymbol{B}^i_\mathrm{h}-\boldsymbol{B}^i_\mathrm{h}(\nabla\boldsymbol{v}^{n+\alpha}_\mathrm{h})^T\Big):\boldsymbol{\psi}_{\mathrm{h}}\mathrm{d\Omega}\nonumber\\
	+&\big(1-\alpha(\phi^{n+\alpha})\big)\int_{\Omega}\Big(\boldsymbol{B}^i_{\mathrm{h}}-\boldsymbol{I}\Big):\boldsymbol{\psi}_{\mathrm{h}}\mathrm{d\Omega}
	+\sum_{\mathrm{e=1}}^{n_{\mathrm{el}}}\int_{\Omega^{\mathrm{e}}}\tau_{\boldsymbol{B}^i}\left(\left(\boldsymbol{v}_\mathrm{h}^{n+\alpha}\cdot\nabla\right)\boldsymbol{\psi}_{\mathrm{h}}\right):\boldsymbol{\mathcal{R}}_{\boldsymbol{B}^i} \mathrm{d}\Omega^{\mathrm{e}}={0},
\end{align}
where $\boldsymbol{\mathcal{R}}_{\boldsymbol{B}^i}$ denotes the element-wise residuals of the evolution equation for the left Cauchy-Green tensor. In Eq.~(\ref{semiCGT}), the first line is the Galerkin projection, the second line represents the strain recovery term and the Petrov-Galerkin stabilization term. The stabilization parameter is selected as $\tau_{\boldsymbol{B}^i}=\left((2/\Delta t)^2+\boldsymbol{v}^{n+\alpha}_{\mathrm{h}}\cdot\boldsymbol{G}\boldsymbol{v}_{\mathrm{h}}^{n+\alpha}\right)$.

\subsection{Linearization of the solid stress}
We employ Newton-Raphson iterations to find the solution of the momentum conservation equation in our fully implicit implementation. Therefore, the residuals and directional derivatives associated with the solid stress are needed. While the discretizations of the hydrostatic pressure and viscous terms are quite standard, the shear components for the $i$-th solid $\mu_L^{i}(\boldsymbol{B}^{i,n+\alpha}-\boldsymbol{I})$, as a function of the left Cauchy-Green tensor, need to be handled carefully. In the calculation of the residuals associated with the left Cauchy-Green tensor, $\boldsymbol{B}_{h}^{i,n+\alpha}$ is directly used. The directional derivatives are calculated following \cite{sun2014full}, where $\boldsymbol{B}_h^{i,n+\alpha}$ is written as a function of the velocity utilizing the evolution equation of the left Cauchy-Green tensor. Hence, the Jacobian terms can be derived naturally. Furthermore, the momentum conservation equation and the evolution equation of the left Cauchy-Green tensor are decoupled with this treatment. Using the generalized-$\alpha$ discretization, $\boldsymbol{B}^i$ (the superscript $i$ is omitted in the following derivations) at $t^{\mathrm{n+\alpha}}$ can be derived as: 
\begin{align}\label{Bext}
	\boldsymbol{B}^{n+\alpha}&=\boldsymbol{B}^{n}+\alpha (\boldsymbol{B}^{n+1}-\boldsymbol{B}^{n}),\nonumber\\
	&=\boldsymbol{B}^{n}+\alpha \Delta t(\partial_t \boldsymbol{B}^n+\varsigma (\partial_t\boldsymbol{B}^{n+1}-\partial_t\boldsymbol{B}^n)),\nonumber\\
	&=\boldsymbol{B}^{n}+\alpha \Delta t(\partial_t\boldsymbol{B}^n+\frac{\varsigma}{\alpha_m} (\partial_t\boldsymbol{B}^{n+\alpha_m}-\partial_t\boldsymbol{B}^n)),\nonumber\\
	&=\boldsymbol{B}^{n}+\alpha \Delta t\left(\left(1-\frac{\varsigma}{\alpha_m}\right)\partial_t\boldsymbol{B}^n+\frac{\varsigma}{\alpha_m} \partial_t\boldsymbol{B}^{n+\alpha_m}\right).
\end{align}
Note that Eq.~(\ref{Bext}) is merely used for the derivation of the Jacobian matrix rather than the calculation of the residuals. To eliminate $\partial_t\boldsymbol{B}^{n+\alpha_m}$, we consider the evolution equation of the left Cauchy-Green tensor:
\begin{align}
	\partial_t\boldsymbol{B}^{n+\alpha_m}=-(\boldsymbol{v}^{n+\alpha}\cdot\nabla)\boldsymbol{B}^{n+\alpha}+\nabla\boldsymbol{v}^{n+\alpha}\boldsymbol{B}^{n+\alpha}+\boldsymbol{B}^{n+\alpha}(\nabla\boldsymbol{v}^{n+\alpha})^T.
\end{align}
The Jacobian terms can be calculated as:
\begin{align}
	\frac{\delta\boldsymbol{B}^{n+\alpha}}{\delta \boldsymbol{v}^{n+\alpha}}=\frac{\alpha\varsigma\Delta t}{\alpha_m}\left(-(\boldsymbol{N}\cdot\nabla)\boldsymbol{B}^{n+\alpha}+\nabla\boldsymbol{N}\boldsymbol{B}^{n+\alpha}+\boldsymbol{B}^{n+\alpha}\left(\nabla\boldsymbol{N}\right)^T\right),
\end{align} 
where $\boldsymbol{N}$ is a vector composed of the shape functions. More details can be found in \ref{solid stress}.

\subsection{Implementation details}
In this subsection, we present the implementation details of our variational framework. The fully Eulerian FSI framework is decoupled and solved in a  partitioned-block iterative manner which leads to flexibility and ease in its implementation to existing variational solvers. The root finding process of each block employs the Newton-Raphson method, which can be expressed in terms of the solution increments of the velocity and the pressure, the order parameters, and the left Cauchy-Green tensors ($\Delta \boldsymbol{u},\Delta p, \Delta \phi^i$ and $\Delta \boldsymbol{B}^i$ respectively).  
We start with the increment of the velocity and pressure: 
\begin{align} \label{LS_NS}
	\begin{bmatrix}
		\boldsymbol{K}_{\Omega}& & \boldsymbol{G}_{\Omega}\\ \noalign{\vspace{4pt}}
		-\boldsymbol{G}^T_{\Omega}& &\boldsymbol{C}_{\Omega}
	\end{bmatrix}
	\begin{Bmatrix}
		\Delta\boldsymbol{v}\\ \noalign{\vspace{4pt}}
		\Delta p
	\end{Bmatrix}
	&=
	-\begin{Bmatrix}
		\overline{\boldsymbol{\mathcal{R}}}_\mathrm{m} \\ \noalign{\vspace{4pt}}
		\overline{\mathcal{R}}_\mathrm{c}
	\end{Bmatrix},
\end{align}
where $\boldsymbol{K}_{\Omega}$ is the stiffness matrix of the momentum conservation equation consisting of transient, convection, viscous and Petrov-Galerkin stabilization terms, $\boldsymbol{G}_{\Omega}$ is the gradient operator, $\boldsymbol{G}^T_{\Omega}$ is the divergence operator for the mass conservation equation and $\boldsymbol{C}_{\Omega}$ is the stabilization term for the cross-coupling of pressure terms. $\overline{\boldsymbol{\mathcal{R}}}_\mathrm{m}$ and $\overline{\mathcal{R}}_\mathrm{c}$ represent the weighted residuals of the variational forms of the momentum and mass conservation equations. 

For fluid and solid interfaces, the order parameters are solved as:
\begin{align}
	\begin{bmatrix}
		\boldsymbol{K}_{\phi^i}
	\end{bmatrix}
	\begin{Bmatrix}
		\Delta\phi^i
	\end{Bmatrix}
	&=
	-\begin{Bmatrix}
		\overline{\mathcal{R}}(\phi^i)
	\end{Bmatrix} ,\label{LS_AC}\\
\end{align}
where $\boldsymbol{K}_{\phi^i}$ and  $\overline{\mathcal{R}(\phi^i)}$ are the stiffness matrix and the weighted residual of the $i$-th Allen-Cahn equation.
Lastly, we evolve of the left Cauchy-Green tensors:
\begin{align}
	\begin{bmatrix}
		\boldsymbol{K}_{\boldsymbol{B}^i}
	\end{bmatrix}
	\begin{Bmatrix}
		\Delta\boldsymbol{\boldsymbol{B}^i}
	\end{Bmatrix}
	&=
	-\begin{Bmatrix}
		\overline{\boldsymbol{\mathcal{R}}}(\boldsymbol{\boldsymbol{B}^i})
	\end{Bmatrix} ,\label{LS_MC}
\end{align}
where  $\boldsymbol{K}_{\boldsymbol{B}^i}$  and $	\overline{\boldsymbol{\mathcal{R}}}(\boldsymbol{B}^i)$ are the stiffness matrix and the weighted residual of the $i$-th left Cauchy-Green tensor. This finishes one nonlinear iteration.
The nonlinear iteration stops when the ratio between the $L^2$ norm of the increment and the current solution is less than $5\times 10^{-4}$ for all the blocks, or the nonlinear iteration number exceeds the upper limit, which is set as 20 in the current work. 

\begin{rem}
	For continua with $n_f$ number of fluid phases and $n_s$ number of solid objects, the current PDE system comprises a unified mass and momentum conservation equation, $n_f$ set of equations in subsection \ref{subsec:feqs} and $n_s$ set of equations in subsection \ref{subsec:seqs}.
\end{rem}
\begin{rem}
	The left Cauchy-Green tensor is solved as a symmetry tensor with six degrees of freedom. The equation is linearized via the Newton-Raphson iteration. In forming the block matrices, the solution vector is organized as $\left[\Delta B_{11},\Delta B_{22},\Delta B_{33},\Delta B_{12},\Delta B_{23},\Delta B_{13} \right]^T$. The subsequent linear system is solved with the GMRES method.
\end{rem}

\section{Phase-field interface processing} \label{odp_Init}
In the phase-field diffuse interface description, the local composition is determined by a phase indicator i.e. order parameter. The phase transition across the interface is dictated by a smooth yet highly localized variation of the order parameter. The variation is identical to give a consistent interface transition. This is achieved by a hyperbolic tangent composition of the signed distance function given by $\phi=\tanh(d/(\sqrt{2}\varepsilon))$, where $d$ denotes the signed distance function to the interface. The distance function provides a uniform variation in the normal direction of the interface; the sign indicates the phase or not; the hyperbolic tangent composition localizes the variation close to the interface, and the scaling factor $\varepsilon$ controls the thickness of the diffuse interface region.

When the geometry of a solid object is complex, such a field cannot be generated through simple analytical expressions. In numerical simulations, the domain and the interface are usually given in a discrete form: the computational domain is discretized by fixed Eulerian mesh, while the interface usually takes the form of a 2D interface mesh from computer-aided design software. In other words, the initialization of the order parameter field in a 3D Eulerian mesh according to a 2D interface mesh is required.

To initialize the field, we need the signed distance function $d$ to the target interface. This can be further decomposed into two sub-problems: (i) calculate the distance function to the interface and (ii) decide the sign of the distance according to the relative position of the current location to the interface. We introduce the grid cell method \cite{haines1994point} to restrict the calculation from the entire domain to several grid cells. A ray-casting algorithm is utilized to determine the relative position of an Eulerian mesh node to the surface mesh, and inside or outside the surface mesh indicating the phase or not.

\subsection{Calculation of the distance function}
We first introduce the calculation of the distance function. As discussed previously, the variation of the order parameter is highly localized. For example, for points which are $3\varepsilon$ away from the interface, $\phi(d=3\varepsilon)=0.9717$, while $\phi(d=\infty)=1$. At the same time, the phase-field method is inherently energy stable due to the free energy minimization. Therefore, we can safely truncate the calculation of the distance at points that are $3\varepsilon$ away from the interface without the concern of inaccuracy or instability, and assign values of $|\phi|=1$, such that the distance calculation is restricted to positions which are close to the interface.

The algorithm starts by decomposing the domain into a series of grid cells. After that, we identify the grid cells that contain the interface elements and associate these interface elements with the grid cell. Lastly, the calculation of the distance is restricted to the nodes in the grid cell and the associated interface element nodes. In dividing the grid cells, we employ the approach of \cite{haines1994point}. Denote the location of the Eulerian mesh node $p$ as $(x_p,y_p,z_p)$ where $x,y,z$ are the coordinates in the $X, Y$ and $Z$ directions, and the subscript $p=1,2,\cdots, n$ denotes the index of the node with $n$ being the total number of nodes. We can define a bounding box for the Eulerian mesh $[x_{\mathrm{min}},x_{\mathrm{max}}]\times[y_{\mathrm{min}},y_{\mathrm{max}}]\times[z_{\mathrm{min}},z_{\mathrm{max}}]$, where $x_{\mathrm{min}}=\min\limits_{p=1,2,\cdots,n}\{x_p\}$ and the rest values are defined similarly. If $n_x,n_y,n_z$ grid cells are specified in each of the directions, the size of the grid cells will be $h_x=(x_{\mathrm{max}}-x_{\mathrm{min}})/n_x$. To determine the grid cell index $(i,j,k)$ of a point $(x_p,y_p,z_p)$ where $i,j,k$ are grid cell indices in the $X, Y$ and $Z$ directions,  we can use $i=\lfloor x_p/h_x\rfloor,j=\lfloor y_p/h_y\rfloor,k=\lfloor z_p/h_z\rfloor$, where $\lfloor\cdot\rfloor$ denotes the floor operator which takes the integer part of a positive real number.

Once the division is established, we need to build up the relevance between the grid cells and the surface elements. For a given surface mesh, we first decompose the surface elements as triangle elements by rewriting the connectivity. This process helps to unify the calculation for all the surface elements. For each triangle element, we create a bounding box around it and assign the element to the grid cells that intersect the bounding box. For example, for a triangular element with vertices $(x_v,y_v,z_v),v=1,2,3$, we define the bounding box as $[x_{\mathrm{min}},x_{\mathrm{max}}]\times[y_{\mathrm{min}},y_{\mathrm{max}}]\times[z_{\mathrm{min}},z_{\mathrm{max}}]$, where the minimum and the maximum are picked among the three vertices. We further expand the bounding box by $3\varepsilon$, which ends up with $[x_{\mathrm{min}}-3\varepsilon,x_{\mathrm{max}}+3\varepsilon]\times[y_{\mathrm{min}}-3\varepsilon,y_{\mathrm{max}}+3\varepsilon]\times[z_{\mathrm{min}}-3\varepsilon,z_{\mathrm{max}}+3\varepsilon]$ so that all the nodes within $3\varepsilon$ distance from the surface element can be involved. The index of the corresponding grid cells can be denoted as $[i_{\mathrm{min}},i_{\mathrm{max}}]\times[j_{\mathrm{min}},j_{\mathrm{max}}]\times[k_{\mathrm{min}},k_{\mathrm{max}}]$, where $i_{\mathrm{min}}=\lfloor (x_{\mathrm{min}}-3\varepsilon)/h_x\rfloor$ and the rests are defined similarly. The surface elements are then assigned to these grid cells.

Once the grid cells and the associated nodes and surface elements are identified, we can restrict the scope of the calculation to each grid cell. For a given node belonging to the grid cell, the distance to the interface is taken as the shortest $L_2$ distance to the vertices of the surface elements assigned to the current grid cell. After the distance function is acquired, we compound the distance function by hyperbolic tangent function $|\phi|=\tanh(|d|/\sqrt{2}\varepsilon)$ to calculate the value of the order parameter field. For grid cells that do not contain any surface elements, we assign $|\phi|=1$ for all Eulerian mesh nodes. All the aforementioned steps are summarized in Algorithm~\ref{distance}.

\begin{algorithm}
	\caption{Distance calculation for fully Eulerian finite element framework}
	\label{distance}
	\begin{algorithmic}
		\STATE Allocate Eulerian mesh nodes and surface elements to the grid cells
		\STATE Loop over interface grid cells
		\INDSTATE Loop over Eulerian mesh nodes in the grid cell
		\INDSTATE[2] Loop over vertices of the surface elements
		\INDSTATE[3] Find the minimum distance $|d|$
		\INDSTATE Assign $|\phi|=\tanh(|d|/\sqrt{2}\varepsilon)$ \\
		\STATE Loop over grid cells without surface elements\\
		\INDSTATE Assign $|\phi| =1$ for all Eulerian mesh nodes\\
	\end{algorithmic}
\end{algorithm}

\subsection{Determination of the sign on a node}
Following the calculation of the magnitude for the order parameter value, we need to determine the sign on each node according to its relative position to the surface mesh. To be more specific, whether the point is inside or outside the given 2D interface mesh. This is a well-known topic in planar geometry problems which can be solved by the ray-casting algorithm \cite{haines1994point,o1998computational}. In short, we cast a semi-infinite ray from the node in the positive X direction. Then, we count how many times the ray crosses the surface. Once the ray crosses the surface, the cross number shifts between even and odd. Accordingly, we know that the point is located inside/outside the interface according to the odd/even cross number.

Since the domain has been decomposed as grid cells, we can restrict the calculation to the grid cells where the ray passes through. When the projection in the position $X$-direction is used, for a node $(x_p,y_p,z_p)$ located at grid cell $(i,j,k)$, we only need to consider an array of grid cells $[i,n_x]\times j\times k$. In the grid cell $(i,j,k)$, only the surface elements with at least one vertex located at the positive $X$-direction of the node can be crossed by the ray. For triangle surface elements, this condition can be expressed as $\forall (x_v,y_v,z_v), v=1,2,3, \exists x_v>x_p$. For grid cells $[i+1,n_x]\times j\times k$, all the surface elements need to be considered.

\begin{figure}[h]
	
	\begin{minipage}[b]{0.5\textwidth}
		\centering
		\includegraphics[scale=0.2,trim= 0 0 0 0,clip]{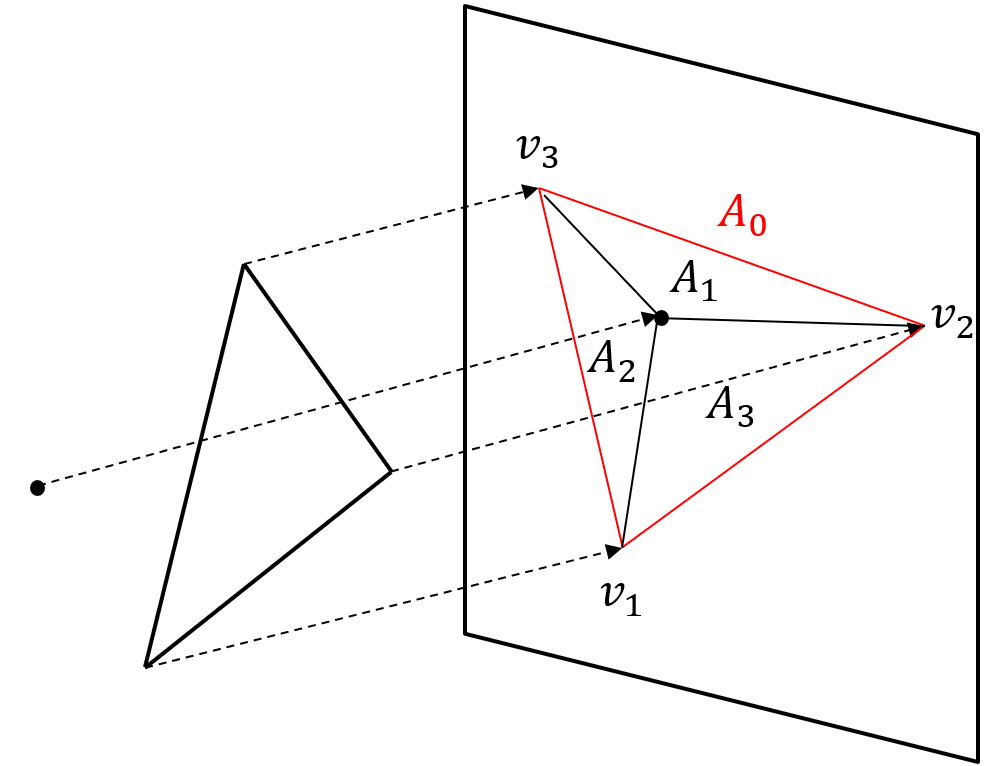}
		\caption*{\hspace{30pt}(a)}
	\end{minipage}
	\begin{minipage}[b]{0.5\textwidth}
		\centering
		\includegraphics[scale=0.2,trim= 0 0 0 0,clip]{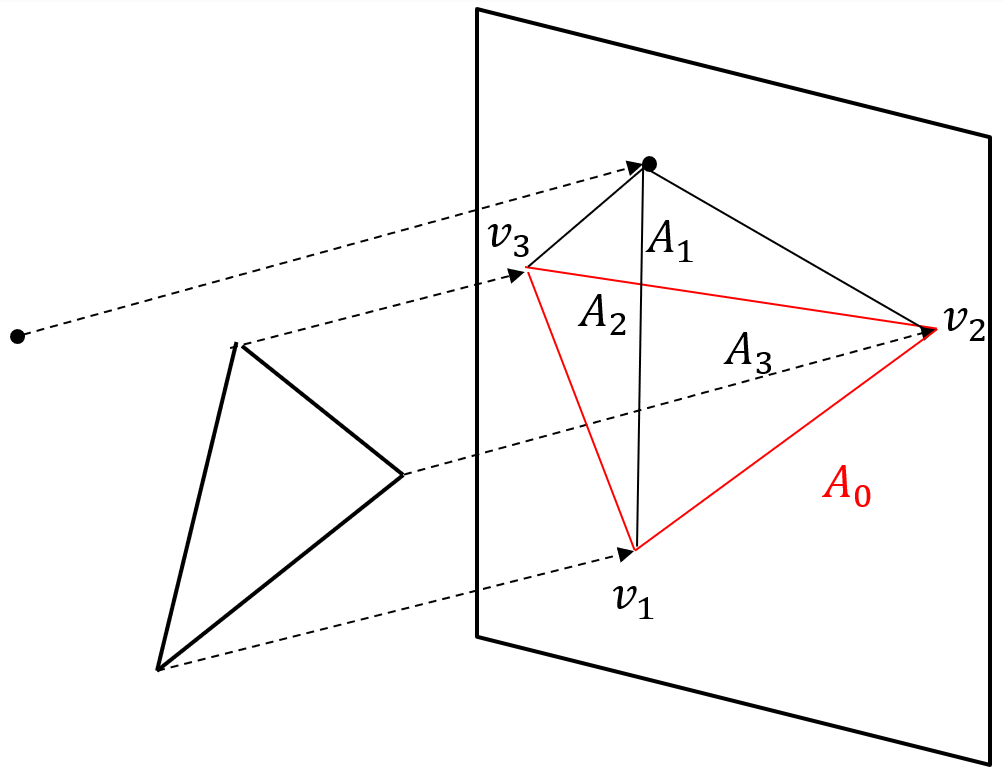}
		\caption*{\hspace{30pt}(b)}
	\end{minipage}
	\caption{Illustration of ray casting algorithm: conditions for ray (a) crossing and (b) missing the surface element. }
	\label{ray}
\end{figure}

After restricting the scope of computation to the node and these triangle elements, we now need to determine whether the ray crosses these triangle elements or not. We first project the node and the triangle elements to the $Y-Z$ plane of $x=x_{\mathrm{max}},x_{\mathrm{max}}=\max\limits_{p=1,2,\cdots,n}\{x_p\}$. As shown in Fig.~\ref{ray}, we denote the area of the projected triangle from the surface element as $A_0$. The areas of the three triangles, which are formed by the projected node and the projected edges of the triangle element, are denoted as $A_1, A_2, A_3$, where $A_i$ denotes the area of the triangle formed by the project node and the opposite edge of vertex $v=i$. If the point lies inside the triangle, we have $A_0=\sum\limits_{i=1}^3A_i$, which indicates that the ray crosses the surface element. For triangle elements that the ray does not cross through, $A_0<\sum\limits_{i=1}^3A_i$ is satisfied. This is illustrated in Fig.~\ref{ray}.

However, the criteria of $A_0=\sum\limits_{i=1}^3A_i$ can be problematic in implementation. The equality may not be strictly satisfied due to the machine truncation error introduced in the area calculations. Therefore, we relax the condition as:
\begin{align}
	A_1+A_2+A_3-A_0\leqslant2\chi,
\end{align}
When the condition is satisfied, we consider that the ray crosses the elements, and the cross number adds one. The relaxation introduces new errors. For example, as illustrated in Fig.~\ref{relax} (a), where the projected node is close to but located outside the projected triangle element. For all positions between $\overrightarrow{v_1v_3}$ and $\overrightarrow{v_1v_2}$ where $A_1\leqslant\chi$ is satisfied, $A_1+A_2+A_3-A_0=2A_1+A_0-A_0\leqslant2\chi$ is true. While the ray does not cross the element, the condition is satisfied and falsely adds one to the cross number. A similar situation happens when the ray is close to the vertices. 

\subsection{Non-duplicated plane partitioning}
\begin{figure}[h]
	\begin{minipage}[b]{0.5\textwidth}
		\centering
		\includegraphics[scale=0.3,trim= 0 0 0 0,clip]{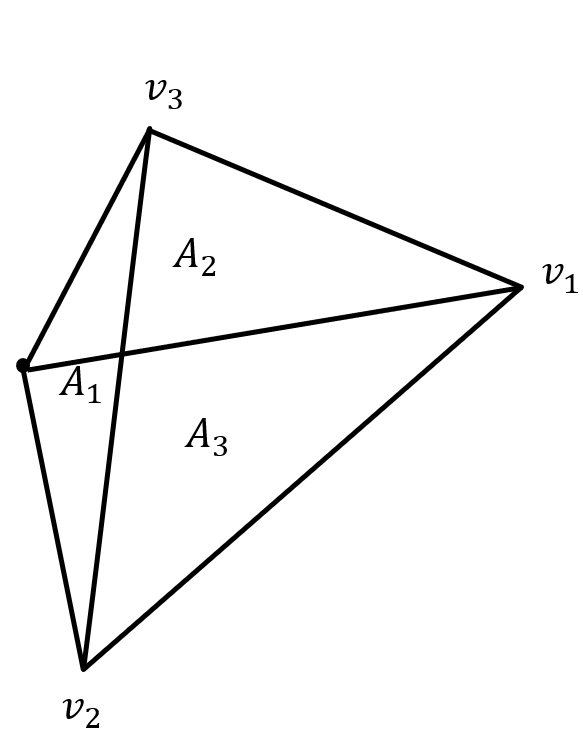}
		\caption*{\hspace{30pt}(a)}
	\end{minipage}
	\begin{minipage}[b]{0.5\textwidth}
		\centering
		\includegraphics[scale=0.3,trim= 0 0 0 0,clip]{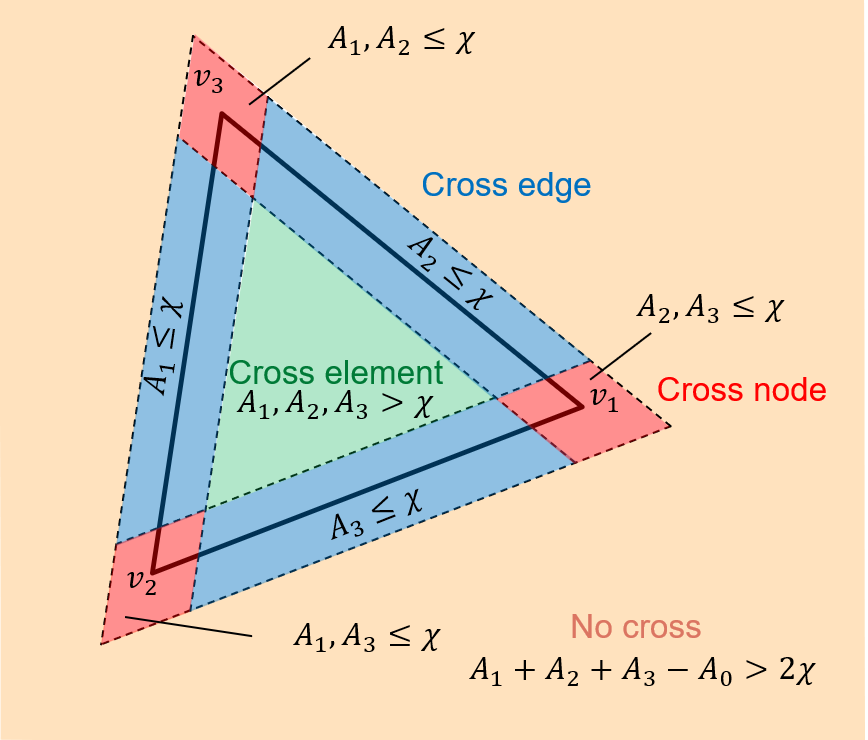}
		\caption*{\hspace{30pt}(b)}
	\end{minipage}
	\caption{Illustration of the non-duplicated plane partitioning: (a) example where the relaxed condition causes duplicate counting on the cross number, (b) non-duplicated plane partitioning and the corresponding conditions. }
	\label{relax}
\end{figure}

To resolve this duplicate counting issue, we classify the cross into three types: element cross, edge cross and node cross. They are defined as follows:
\begin{enumerate}
	\item[a)]If all the areas $A_1, A_2, A_3$ are larger than $\chi$, the projected node is far away from the edges, and the cross is classified as an element cross.
	\item[b)]If there are two areas that are smaller than $\chi$, the cross is classified as a node cross.
	\item[c)]If there exists only one area that is smaller than $\chi$, the cross is classified as an edge cross.
\end{enumerate}
 This classification partitions the plane into eight regions as illustrated in Fig.~\ref{relax} (b). Once the cross-type is identified, we store the relevant vertex indices. Thus three, two, and one indices are recorded for element-cross, edge-cross and node-cross, respectively. The cross number adds one only for non-repeated vortex indices combinations, therefore avoiding the duplicate counting problem. Finally, the total cross number equals the summation of these three types of crosses. 

The proposed plane partitioning reduces the duplicate counts arising from discrete surface mesh to a non-duplicated count, therefore referred to as non-duplicated plane partitioning.
For grid cells that do not contain any surface element, all the nodes inside the grid cell lie on the same side of the interface. In such cases, we only judge the sign for a single node in the grid cell, then assign this sign for all the rest nodes inside the grid cell. The procedures to determine the sign on each Eulerian mesh node are summarized in Algorithm~\ref{sign}.

\begin{algorithm}
	\caption{Determination of signs on Eulerian finite element mesh nodes}
	\label{sign}
	\begin{algorithmic}
\STATE Loop over interface grid cells\\
\INDSTATE Loop over Eulerian mesh nodes in the grid cell\\
\INDSTATE[2] Loop over surface elements in grid cells $([i,i_\mathrm{max}],j,k)$ \\
\INDSTATE[3] Count cross number using ray-casting algorithm and NDP partitioning\\
\INDSTATE[2] Determine the sign according to the cross number\\
Loop over grid cells without surface elements\\
\INDSTATE Pick one Eulerian mesh nodes in the grid cell\\
\INDSTATE Loop over surface elements in grid cells $([i,i_\mathrm{max}],j,k)$ \\
\INDSTATE[2]  Count cross number using ray-casting algorithm and NDP partitioning\\
\INDSTATE  Determine the sign according to the cross number\\
\INDSTATE Assign the sign to all nodes in the grid cell\\
	\end{algorithmic}
\end{algorithm}

\section{Test cases for phase-field interface processing} \label{exp_Init}
In this section, we examine the phase-field interface processing with multiple test cases. We first verify the calculation of the signed distance function in a single grid cell for sphere and cubic interfaces to establish the order of convergence. Then, we test the algorithm for multiple grid cells with the cubic interface case. Finally, we demonstrate the algorithm for the complex geometry of a ship hull.

\subsection{Interface meshes for sphere and cube}   \label{scint}
We start by verifying the calculation of the signed distance function. A domain of $[0,1]\times[0,1]\times[0,1]$ is considered in this problem. A sphere interface with a radius of $r=0.25$ is centered at $(0.5,0.5,0.5)$. The sign is defined as positive inside the sphere interface. A single grid cell is used for the calculation so that the signed distance is calculated throughout the domain. This helps to clarify the convergence of the signed distance function calculation.

\begin{figure}[h]
	
	\begin{minipage}[b]{0.5\textwidth}
		\centering
		\includegraphics[scale=0.15,trim= 0 0 0 0,clip]{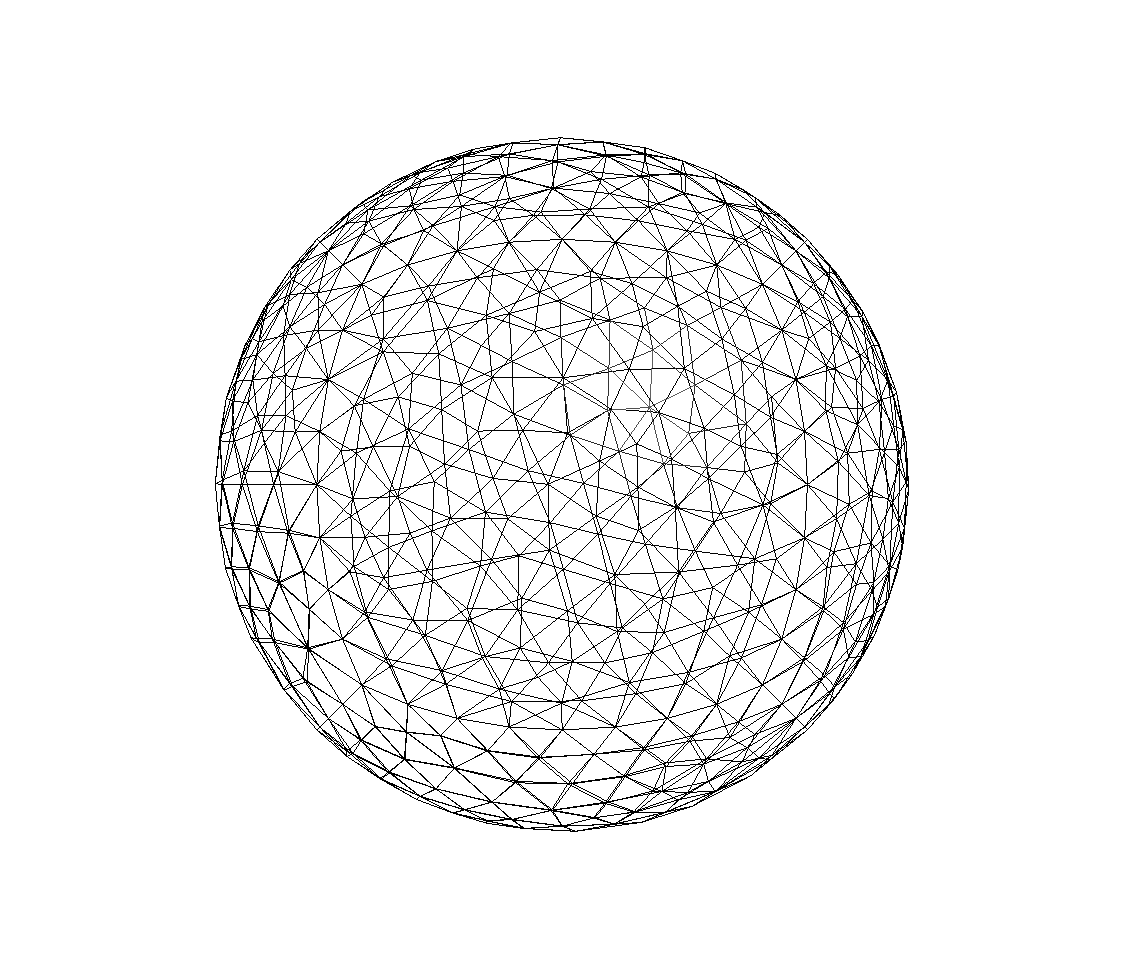}
		\caption*{\hspace{0pt}(a)}
	\end{minipage}
	\hspace{0pt}
	\begin{minipage}[b]{0.5\textwidth}
		\centering
		\includegraphics[scale=0.15,trim= 0 0 0 0,clip]{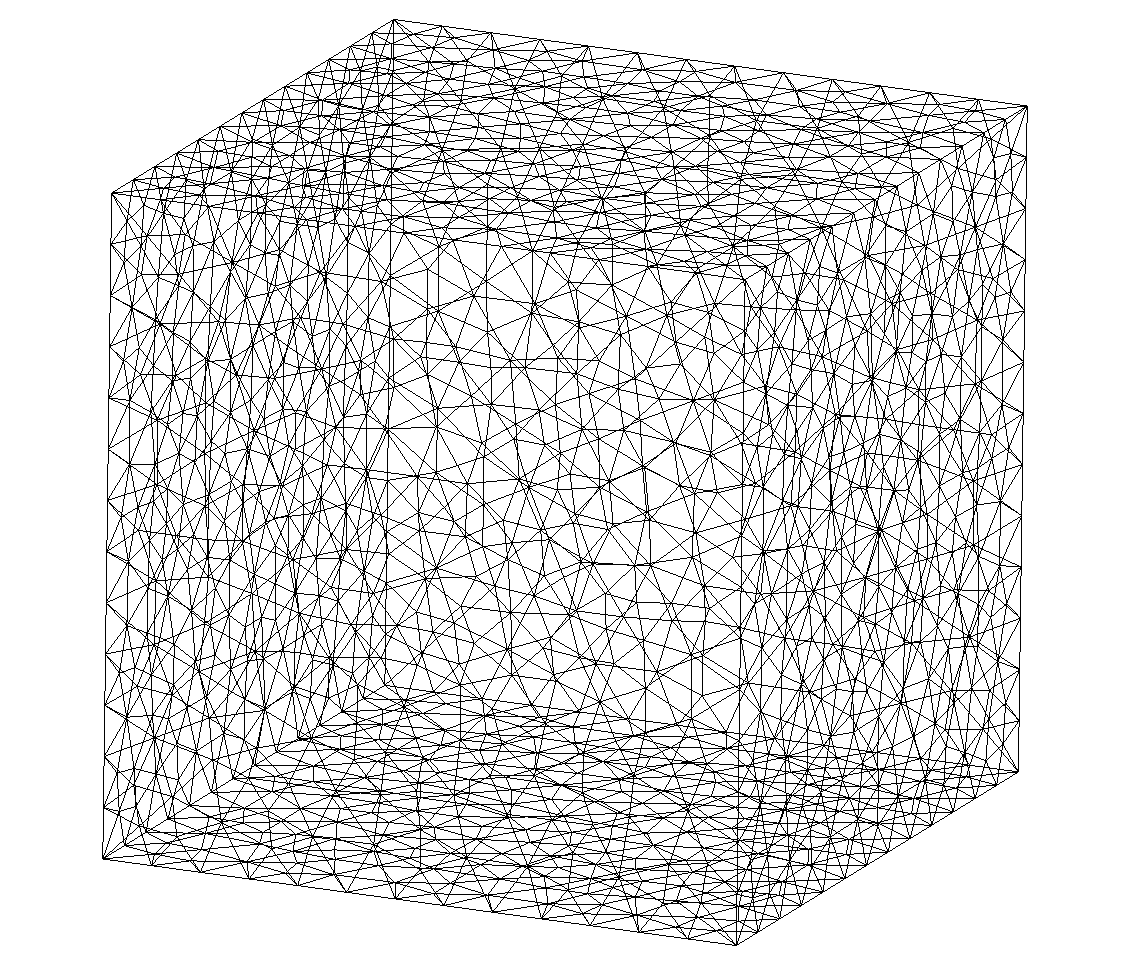}
		\caption*{\hspace{0pt}(b)}
	\end{minipage}
	\caption{Illustration of the interface meshes for two simple geometries: (a) sphere and (b) cube. } 
	\label{int_mesh}
\end{figure}

\begin{figure}[h]
	
	\begin{minipage}[b]{0.5\textwidth}
		\centering
		\includegraphics[scale=0.45,trim= 0 0 0 0,clip]{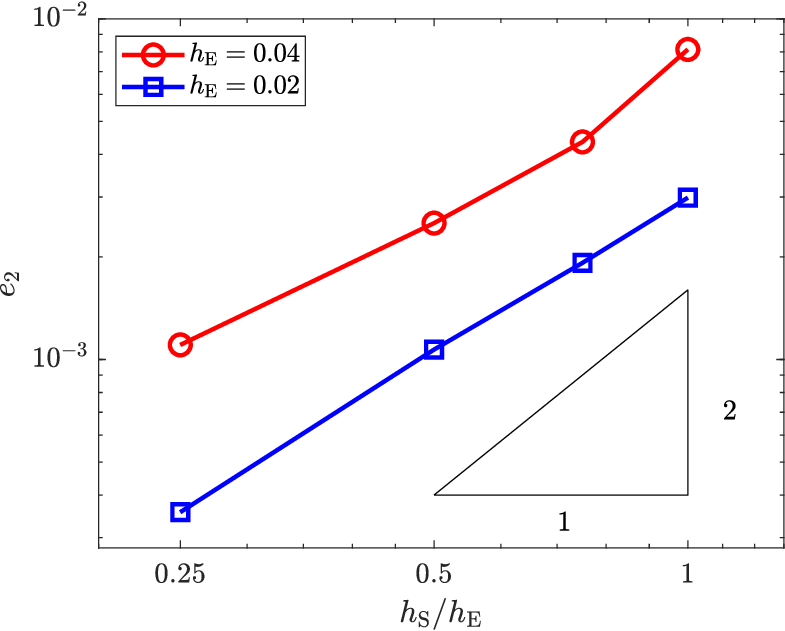}
		\caption*{\hspace{30pt}(a)}
	\end{minipage}
	\hspace{0pt}
	\begin{minipage}[b]{0.5\textwidth}
		\centering
		\includegraphics[scale=0.45,trim= 0 0 0 0,clip]{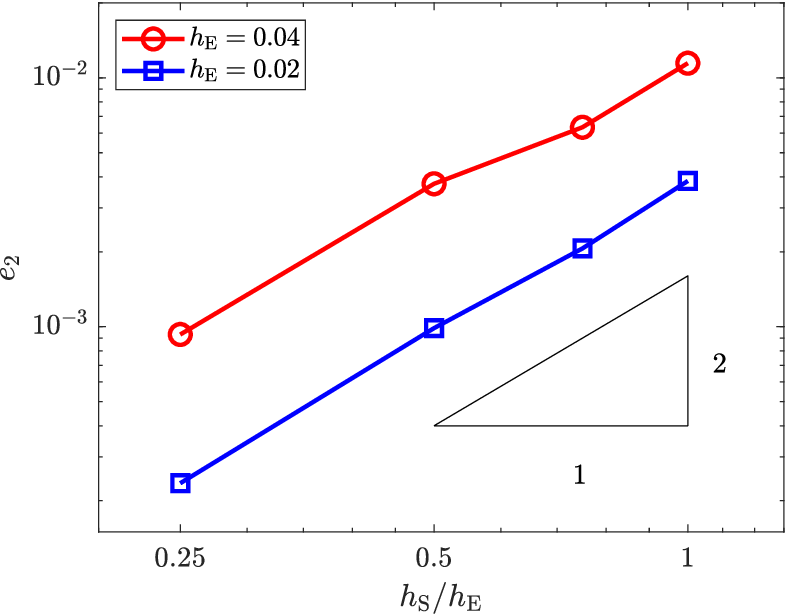}
		\caption*{\hspace{30pt}(b)}
	\end{minipage}
	\caption{ Convergence of the signed distance function for two representative meshes: (a) sphere and (b) cube.} 
	\label{int_conv}
\end{figure}

The computational domain is discretized with uniform structured cubic elements with an edge length of $h_{E}= 0.02$ and $h_{E}=0.01$, where $E$ denotes the Eulerian background mesh. The sphere interface is discretized with unstructured triangular mesh, whose edge length is denoted as $h_S$. The interface mesh is shown in Fig.~\ref{int_mesh} (a). We vary the ratio of the surface mesh size to the Eulerian background mesh size from $h_S/h_E=2$ to $h_S/h_E=0.5$. The error is quantified as $e_2=||\phi-\phi_{ref}||/||\phi_{ref}||$, where $||\cdot||$ denotes the $L_2$ norm. The results are shown in Fig.~\ref{int_conv} (a), where a second-order convergence is observed. We further test the algorithm with a cubic interface mesh, which contains sharp edges and corners. A cubic interface mesh spans through $[0.25,0.75]\times[0.25,0.75]\times[0.25,0.75]$ is used as shown in Fig.~\ref{int_mesh} (b), the rest of the numerical setup is kept the same. The second-order convergence is confirmed in Fig.~\ref{int_conv} (b).

\subsection{Calculation with multiple grid cells}
\begin{figure}[h]
	
	\begin{minipage}[b]{0.5\textwidth}
		\centering
		\includegraphics[scale=0.15,trim= 0 0 0 0,clip]{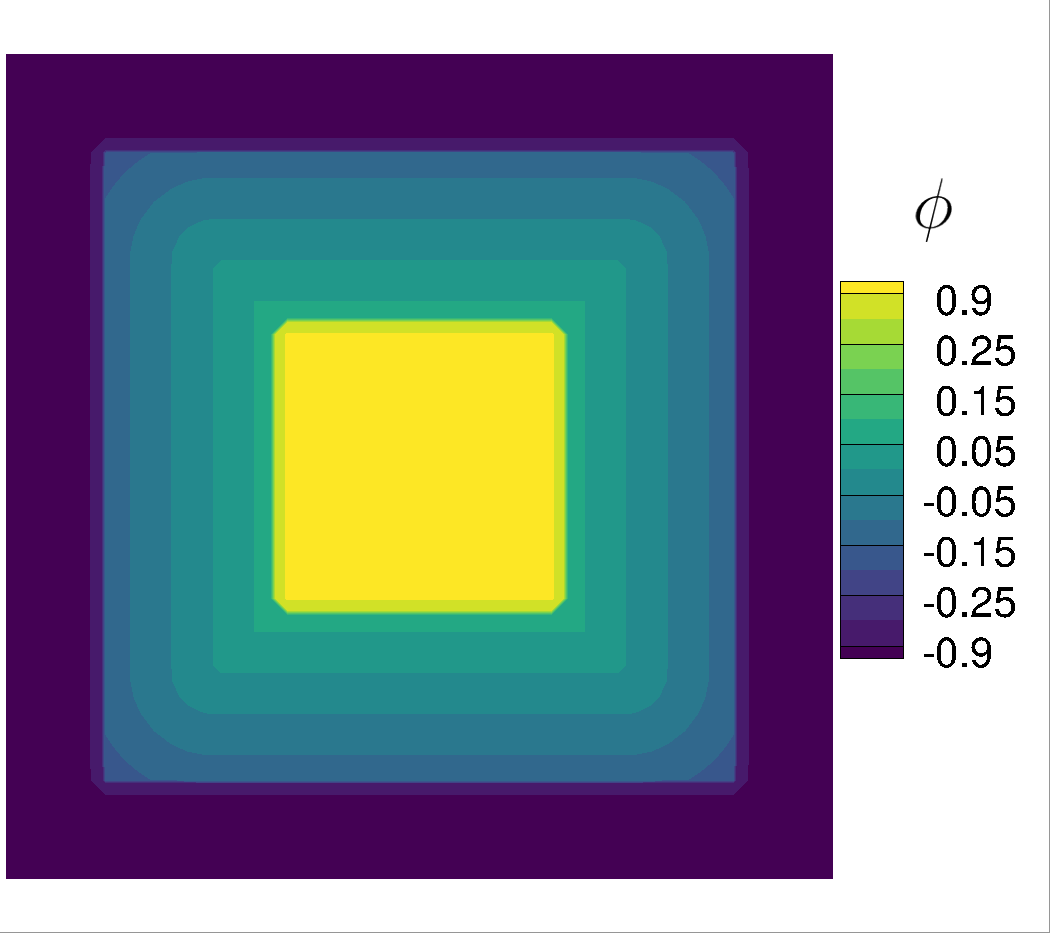}
		\caption*{\hspace{0pt}(a)}
	\end{minipage}
	\hspace{0pt}
	\begin{minipage}[b]{0.5\textwidth}
		\centering
		\includegraphics[scale=0.15,trim= 0 0 0 0,clip]{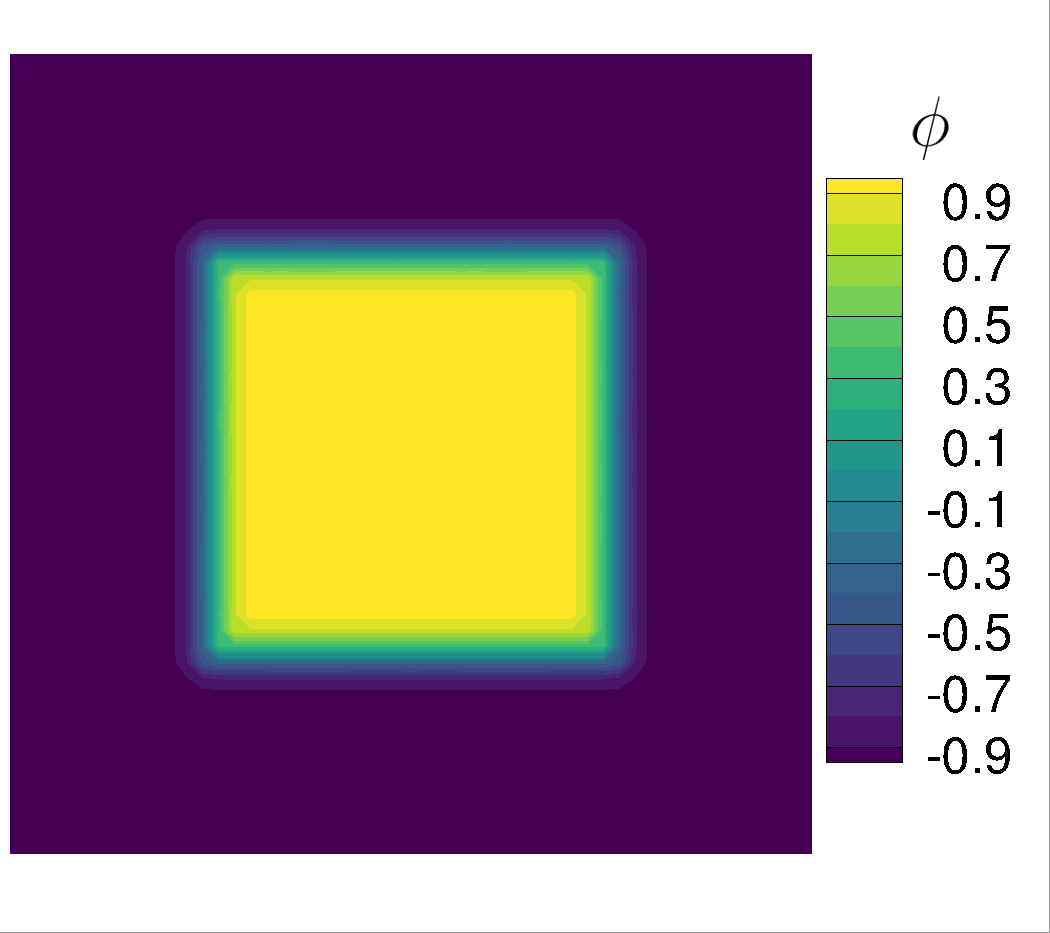}
		\caption*{\hspace{0pt}(b)}
	\end{minipage}
	\caption{Illustration of contours of (a) signed distance function, and (b) order parameter field calculated with multiple grid cells. } 
	\label{sub}
\end{figure}

We then examine the algorithm with multiple grid cells. The cubic interface mesh case discussed in subsection \ref{scint} is used. The domain is discretized with a uniform structured mesh of size $h_E=0.02$, and an unstructured triangle mesh of size $h_S=0.005$ is used for the cubic interface. The domain is decomposed as $9$ segments in the $X, Y$ and $Z$ directions, leading to $729$ grid cells. The slice of $Y$-$Z$ plane at $x=0.5$ is shown in Fig.~\ref{sub} (a). As observed, the calculation of the signed distance function is restricted to grid cells that are close to the interface. In the grid cells that are far from the interface, the distance calculation is omitted, and the values of $1$ and $-1$ are assigned to the order parameter for nodes located inside and outside of the interface, respectively. The last step is to compound the signed distance function with the hyperbolic tangent function. As a result, an order parameter field in which the variation is highly localized to the interface is generated, as shown in Fig.~\ref{sub} (b).

\subsection{Demonstration with a ship hull}
\begin{figure}[h]
	
	\begin{minipage}[b]{0.5\textwidth}
		\centering
		\includegraphics[scale=0.15,trim= 0 0 0 0,clip]{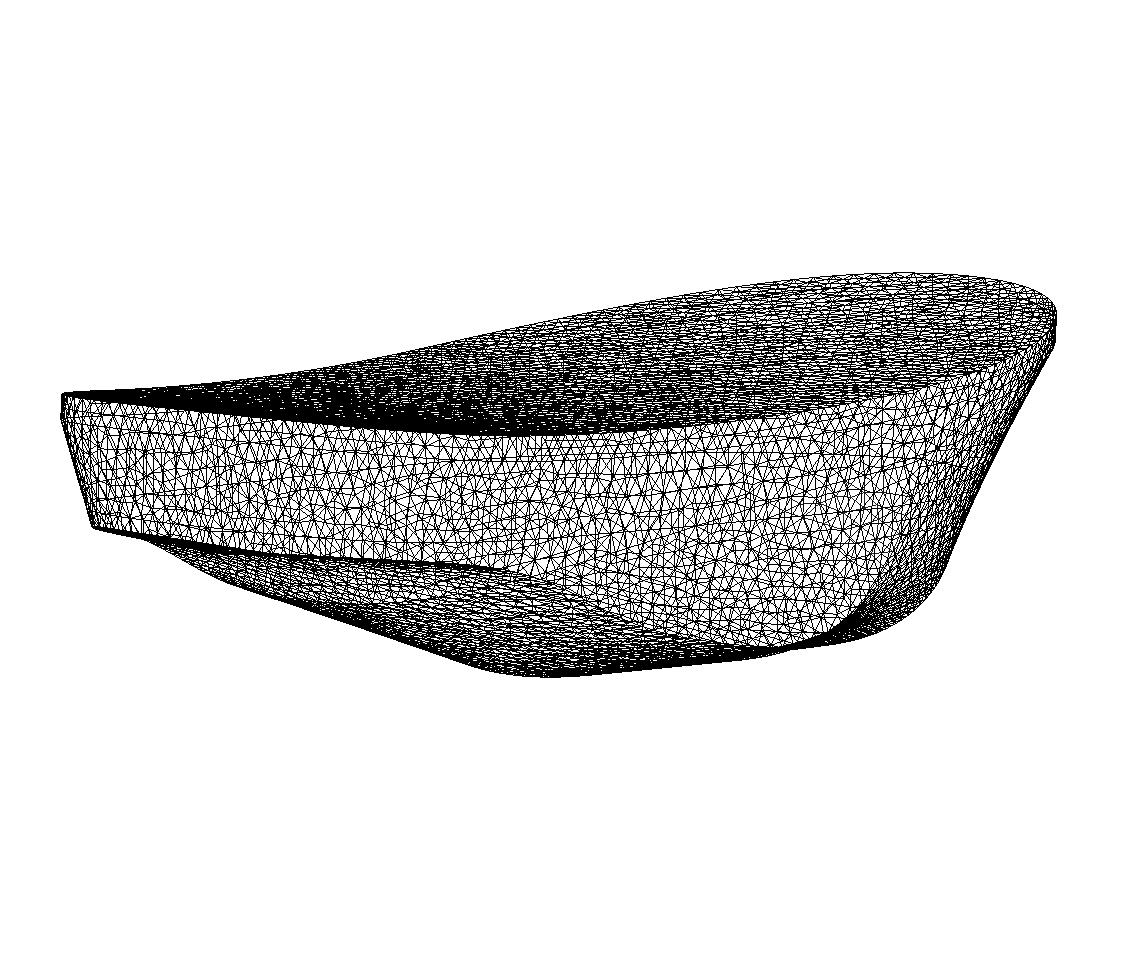}
		\caption*{\hspace{0pt}(a)}
	\end{minipage}
	\hspace{0pt}
	\begin{minipage}[b]{0.5\textwidth}
		\centering
		\includegraphics[scale=0.15,trim= 5 5 5 5,clip]{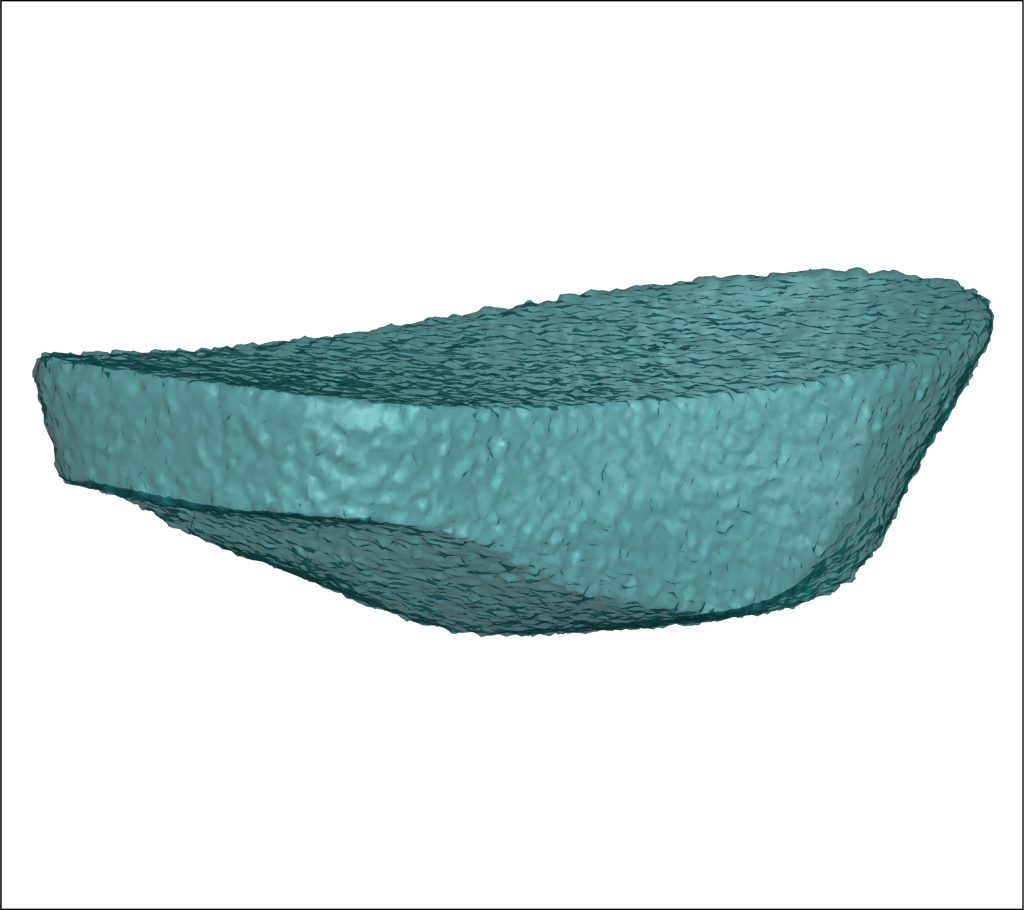}
		\caption*{\hspace{0pt}(b)}
	\end{minipage}
	\caption{An illustration of a tugboat represented by (a) interface mesh and (b) isosurface of $\phi=0$.} 
	\label{int_ship}
\end{figure}

Lastly, we demonstrate the algorithm for interfaces with complex geometry. We employ the vessel STAN TUG 1004 for which the vessel geometry is available online\footnotemark[1]{} . The domain and the interface are discretized with unstructured meshes of size $h_E=h_S=0.02$  The interface mesh is shown in Fig.~\ref{int_ship} (a), while the isosurface of $\phi=0$ of the generated order parameter field is shown in Fig.~\ref{int_ship} (b). As observed, the current algorithm can successfully generate the order parameter field for interfaces with complex geometry. The generated field will be further used as the initial condition for demonstrating the ship-ice interaction problem.

\footnotetext[1]{\url{https://grabcad.com/library/stan-tug-1004-1}}

\section{Test cases for multiphase fluid-solid interaction} \label{test}
In this section, we validate and demonstrate the framework with cases of increasing complexity. We start by verifying the framework with a rotational disk in a lid-driven cavity flow. Then, we showcase the solver with a 3D case of a rotating sphere subject to the lid-driven cavity flow. Finally, the contact between solids is demonstrated with an immersed sphere falling on an elastic block under gravitational force.

\subsection{Rotating disk in lid-driven cavity flow}
\begin{figure}[h]
	\centering
	\trimbox{0 20 0 20}{
		\begin{tikzpicture}[scale=1.4,every node/.style={scale=0.5}]
			
			\begin{axis}[ 
				ymin=-0.2,
				ymax= 1.2,
				xmax=1.5,
				xmin=-0.5,
				xticklabel=\empty,
				yticklabel=\empty,
				ytick style={draw=none},
				xtick style={draw=none},
				axis line style = {draw=none},
				minor tick num=0,
				axis lines = middle,
				label style = {at={(ticklabel cs:1.1)}},
				axis equal
				]
				
				\draw (axis cs: 0,0)--(axis cs: 0,1)--(axis cs: 1,1)--(axis cs: 1,0)--(axis cs: 0,0);
				
				\node at (axis cs: 1.05,0.5) [above right] {\LARGE$v_x=0$};
				\node at (axis cs: 1.05,0.5) [below right] {\LARGE$v_y=0$};
				\node at (axis cs: 0.5,-0.05) [below] {\LARGE$v_x=v_y=0$};
				\node at (axis cs: -0.05,0.5) [above left] {\LARGE$v_x=0$};
				\node at (axis cs: -0.05,0.5) [below left] {\LARGE$v_y=0$};
				\node at (axis cs: 0.5,1.05) [above] {\LARGE$v_x=1,v_y=0$};
				\draw [-stealth] (axis cs: 0,1)--(axis cs: 0.25-0.125,1);
				\draw [-stealth] (axis cs: 0.25-0.125,1)--(axis cs: 0.5-0.125,1);
				\draw [-stealth] (axis cs: 0.5-0.125,1)--(axis cs: 0.75-0.125,1);
				\draw [-stealth] (axis cs: 0.75-0.125,1)--(axis cs: 1-0.125,1);
				
				\node at (0.3,0.75) {\huge$\Omega^{\mathrm{f}}$};
				\node at (0.6,0.5)  {\huge$\Omega^{\mathrm{s}}$};

				\draw [fill=gray,fill opacity=0.5](axis cs: 0.6, 0.5) circle [radius=0.2];
			\end{axis}
	\end{tikzpicture}}
	
	\caption{Schematic diagram of a rotating disk in  lid-driven cavity flow.}
	\label{rd}
\end{figure}
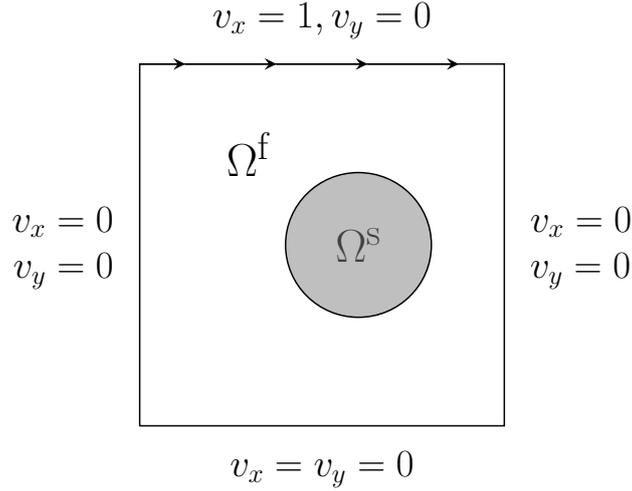

In this section, we examine the fluid-solid interaction aspect of our solver in the case of a rotating disk in lid-driven cavity flow. Consider the computational domain of $[0,1]\times[0,1]$. A circular disk with radius $r=0.2$ is centered at $(0.6,0.5)$. The no-slip boundary condition is applied to the left, right, and bottom boundaries, with a prescribed velocity at the top boundary as $v_x=1$. The order parameter and the left Cauchy-Green tensor on the boundaries are evolved according to the prescribed velocity.  The densities of the solid and fluid are selected as $\rho^\mathrm{s}=\rho^\mathrm{f}=1$. The dynamic viscosity is taken as $\mu^\mathrm{s}=\mu^\mathrm{f}=0.01$. The shear modulus of the solid is chosen as $\mu^{\mathrm{s}}_L=0.1$. The problem setup is illustrated in Fig.~\ref{rd}.

\begin{figure}[h]
	
	\begin{minipage}[b]{0.5\textwidth}
		\centering
		\includegraphics[scale=0.6,trim= 0 0 0 0,clip]{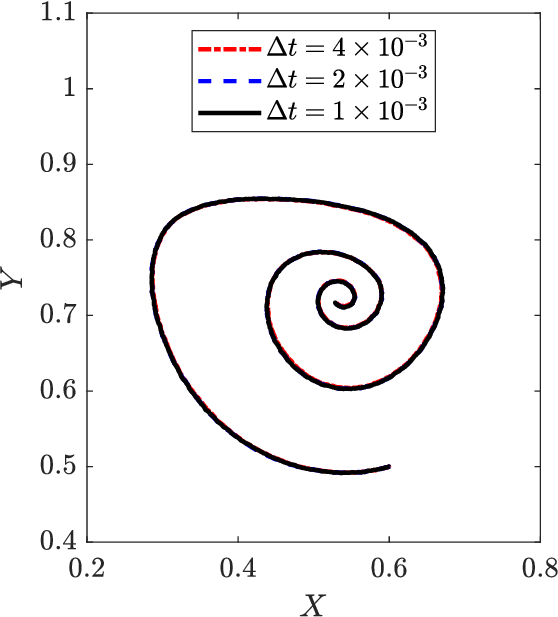}
		\caption*{\hspace{30pt}(a)}
	\end{minipage}
	\hspace{0pt}
	\begin{minipage}[b]{0.5\textwidth}
		\centering
		\includegraphics[scale=0.6,trim= 0 0 0 0,clip]{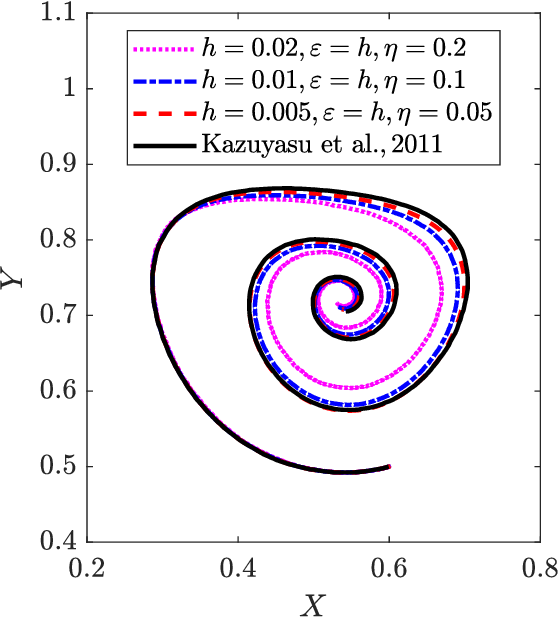}
		\caption*{\hspace{30pt}(b)}
	\end{minipage}
	\caption{Rotating disk in lid-driven cavity flow: (a) temporal and (b) interface convergence study for the center of mass of the disk. } 
	\label{rd_res}
\end{figure}

We first perform a time convergence study. The interface parameters are selected as $\eta=0.2,\varepsilon=0.02$. A uniform structured triangle mesh with edge length $h=0.02$ is used to discretize the computational domain. The time steps are bisected from $\Delta t = 0.004$ to $\Delta t =0.001$, and the case is simulated until $t=20$. The center of mass of the solid is defined as $\boldsymbol{x}_c = \int \boldsymbol{x}d\Omega^s / \int 1 d\Omega^s$, where the solid domain is identified through the corresponding order parameter $d\Omega^s=\alpha^s(\phi^s)d\Omega$. The trajectories of the center of mass with different time steps are plotted in Fig.~\ref{rd_res} (a). As observed, the solution has converged at $\Delta t = 0.001$.

After that,  we perform an interface convergence study\cite{mao2021variational}. With the converged time step $\Delta t = 0.001$, we bisect $\eta$ and $\varepsilon$ simultaneously from $\eta=0.2,\varepsilon=0.02$ to $\eta=0.05,\varepsilon=0.005$. The meshes are refined according to maintain the interface resolution at $h/\varepsilon=1$. The trajectories of the center of mass are plotted in Fig.~\ref{rd_res} (b) and compared with the reference in \cite{sugiyama2011full}. As shown by the figure, the solution matches well with the reference solution.

\subsection{Rotating sphere in lid-driven cavity flow}
\begin{figure}[h]
	\centering
	\trimbox{0 0 0 0}{
	\begin{tikzpicture}[xscale=4.5, yscale=4.5,line width=1]

		\draw [dashed] (0.2,0.2) -- (0.2,1.2);
		\draw [dashed] (0.2,0.2) -- (1.2,0.2);
		\draw [dashed] (0,0) -- (0.2,0.2);
		\draw (0.2,1.2) -- (1.2,1.2);
		\draw (1.2,0.2) -- (1.2,1.2);
		\draw (1,0)--(1.2,0.2);
		\draw (0,1)--(0.2,1.2);
		\draw (1,1)--(1.2,1.2);
		\draw (0,0) rectangle (1,1);
		



		

		
		
		\draw[fill = white,opacity=0.75] (0.7,0.6) circle (0.2);
		\draw [-stealth] (0.7,0.6)--(0.54,0.73);
		\node[above left] at (0.54,0.73) {$R$};
		\draw (0.5,0.6) arc (180:360:0.2 and 0.1);
		\draw[dashed] (0.9,0.6) arc (0:180:0.2 and 0.1);
		\fill[fill=black] (0.7,0.6) circle (0.5pt);
		
		\draw [-stealth,thick] (0.2,0.2) -- (0.1,0.1);
		\node [right] at (0.12,0.1) {Z};
		\draw[-stealth, thick] (0.2,0.2) -- (0.341,0.2);
		\node [below] at (0.341,0.2) {X};
		\draw[-stealth,thick] (0.2,0.2) -- (0.2,0.341);
		\node [left] at (0.2,0.341) {Y};
		\node [right] at (0.72,0.6) {$\Omega_s$};
		\node [right] at (0.3,0.3) {$\Omega_f$};
		
		\draw [-stealth] (0.1,1.1)--(0.1+0.125,1.1);
		\draw [-stealth] (0.1+0.125,1.1)--(0.1+0.375,1.1);
		\draw [-stealth] (0.1+0.375,1.1)--(0.1+0.75-0.125,1.1);
		\draw [-stealth] (0.1+0.75-0.125,1.1)--(1.1-0.125,1.1);
		\draw  (1.1-0.125,1.1)--(1.1,1.1);
		
		\draw [-stealth] (0,1)--(0+0.125,1);
		\draw [-stealth] (0+0.125,1)--(0+0.375,1);
		\draw [-stealth] (0+0.375,1)--(0+0.75-0.125,1.);
		\draw [-stealth] (0+0.75-0.125,1)--(1-0.125,1);

		\draw [-stealth] (0.2,1.2)--(0.2+0.125,1.2);
		\draw [-stealth] (0.2+0.125,1.2)--(0.2+0.375,1.2);
		\draw [-stealth] (0.2+0.375,1.2)--(0.2+0.75-0.125,1.2);
		\draw [-stealth] (0.2+0.75-0.125,1.2)--(1.2-0.125,1.2);
		
	\end{tikzpicture}
}
	\caption{Schematic of a rotating sphere in lid-driven cavity flow.}
	\label{rs}
\end{figure}
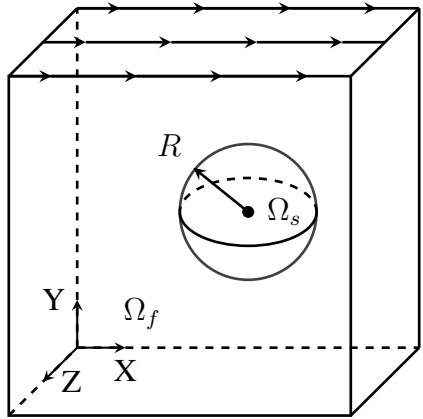

We generalize the rotating disk case as a rotating sphere case to demonstrate the 3D aspect of the solver. A computational domain of $[0,1]\times[0,1]\times[0,1]$ is considered. A solid sphere of radius $r=0.2$ is initially centered at $(0.6,0.5,0.5)$. The no-slip boundary condition is applied to the side and bottom walls, while a constant velocity of $v_x=1$ is applied at the top boundary. The order parameter and the left Cauchy-Green tensor on the boundaries are evolved according to the prescribed velocity.  The densities of the fluid and the solid are chosen as $\rho^\mathrm{s}=\rho^\mathrm{f}=1$. The dynamic viscosity is taken as $\mu^\mathrm{s}=\mu^\mathrm{f}=0.02$. The shear modulus of the solid is selected as $\mu^\mathrm{s}_L = 0.5$. The case setup is illustrated in Fig.~\ref{rs}.

\begin{figure}[h]
	
	\begin{minipage}[b]{0.5\textwidth}
		\centering
		\includegraphics[scale=0.6,trim= 0 0 0 0,clip]{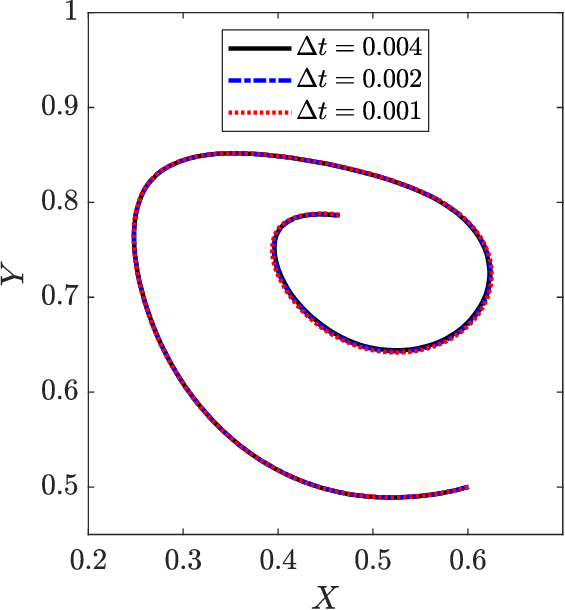}
		\caption*{\hspace{30pt}(a)}
	\end{minipage}
	\hspace{0pt}
	\begin{minipage}[b]{0.5\textwidth}
		\centering
		\includegraphics[scale=0.6,trim= 0 0 0 0,clip]{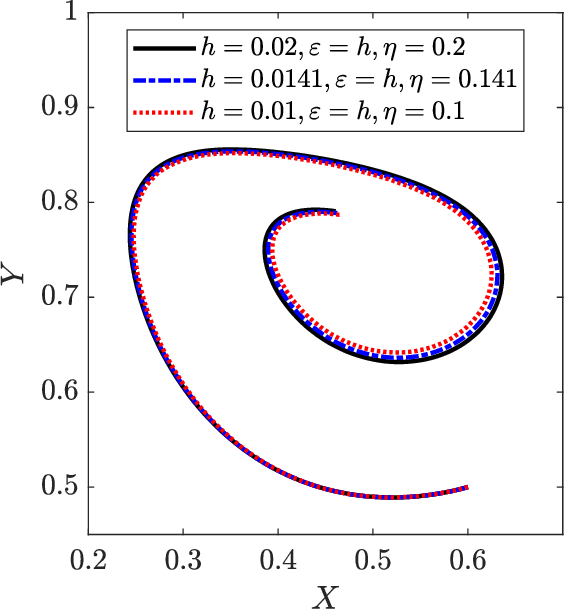}
		\caption*{\hspace{30pt}(b)}
	\end{minipage}
	\caption{Rotating sphere in lid-driven cavity flow: (a) temporal and (b) interface convergence study for the center of mass of the sphere in $X-Z$ plane. } 
	\label{rs_res}
\end{figure}

\begin{figure}[h]
	
	\begin{minipage}[b]{0.3\textwidth}
		\centering
		\includegraphics[scale=0.1,trim= 0 0 0 0,clip]{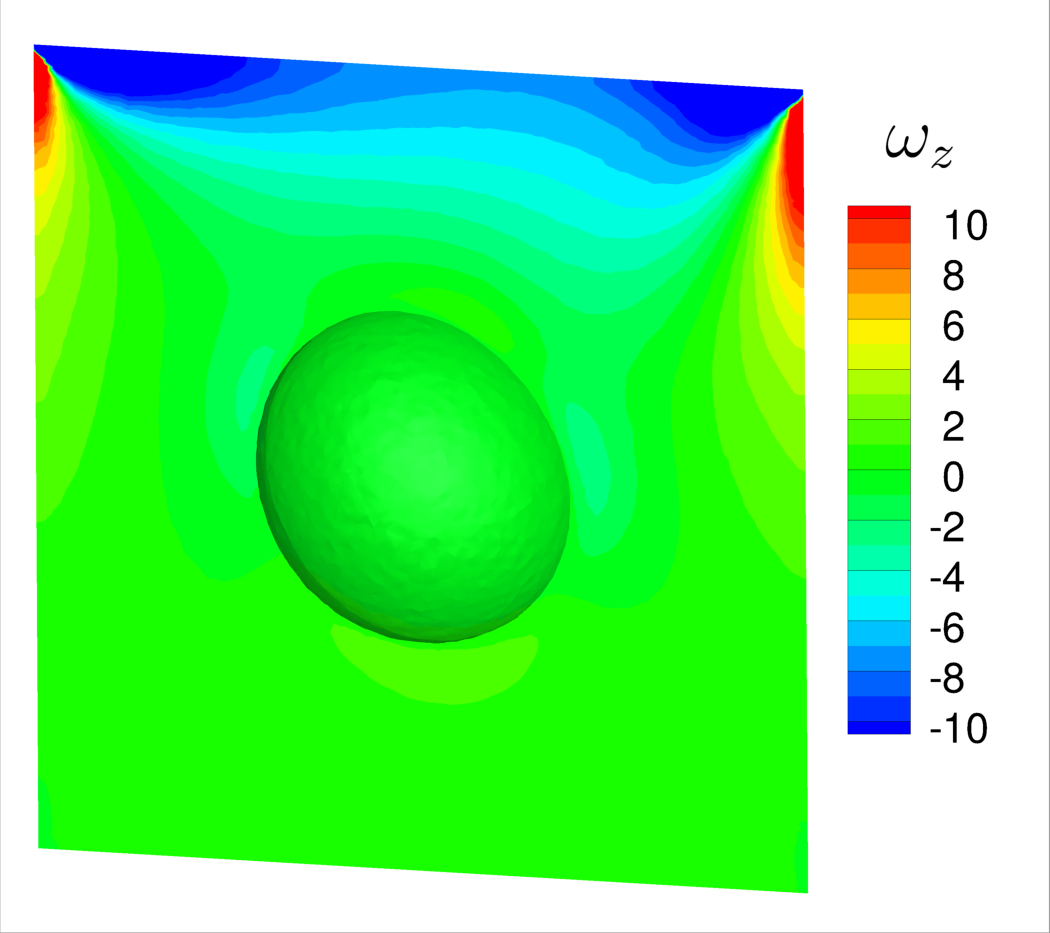}
		\caption*{\hspace{0pt}(a)}
	\end{minipage}
	\begin{minipage}[b]{0.3\textwidth}
		\centering
		\includegraphics[scale=0.1,trim= 0 0 0 0,clip]{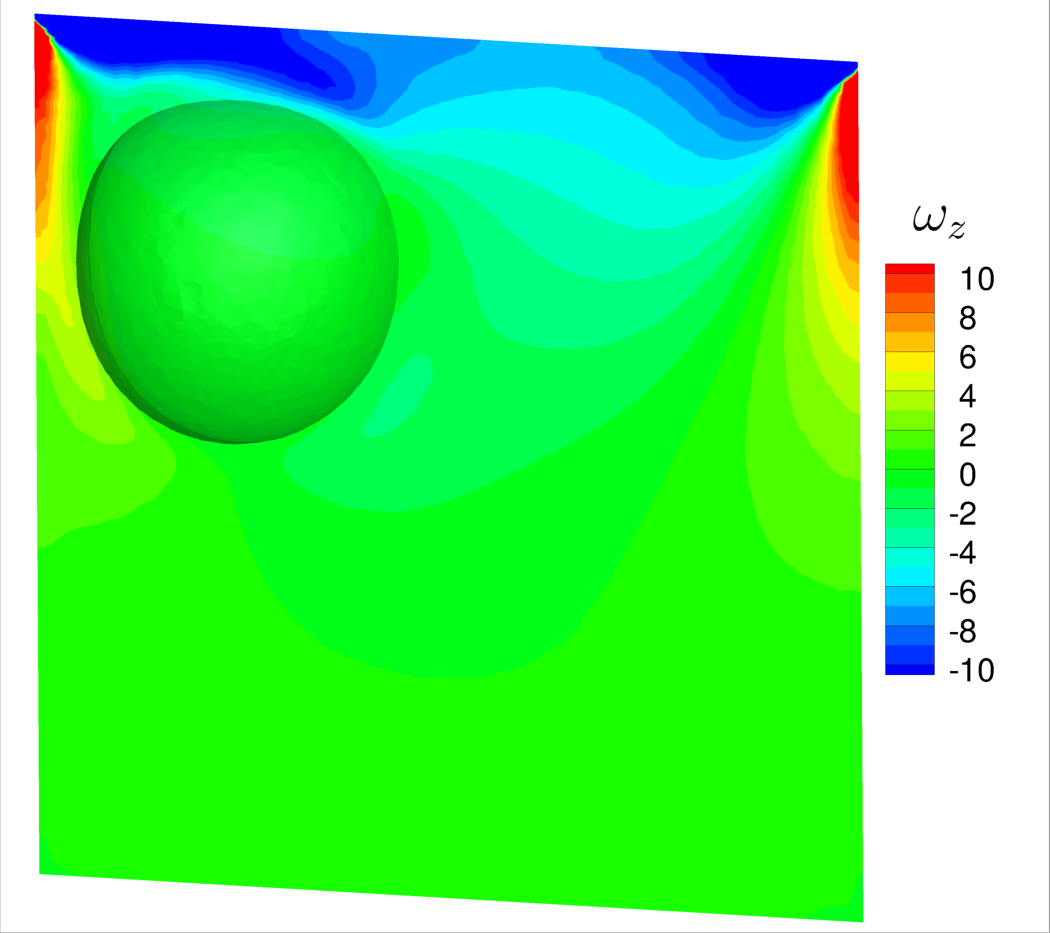}
		\caption*{\hspace{0pt}(b)}
	\end{minipage}
	\begin{minipage}[b]{0.3\textwidth}
		\centering
		\includegraphics[scale=0.1,trim= 0 0 0 0,clip]{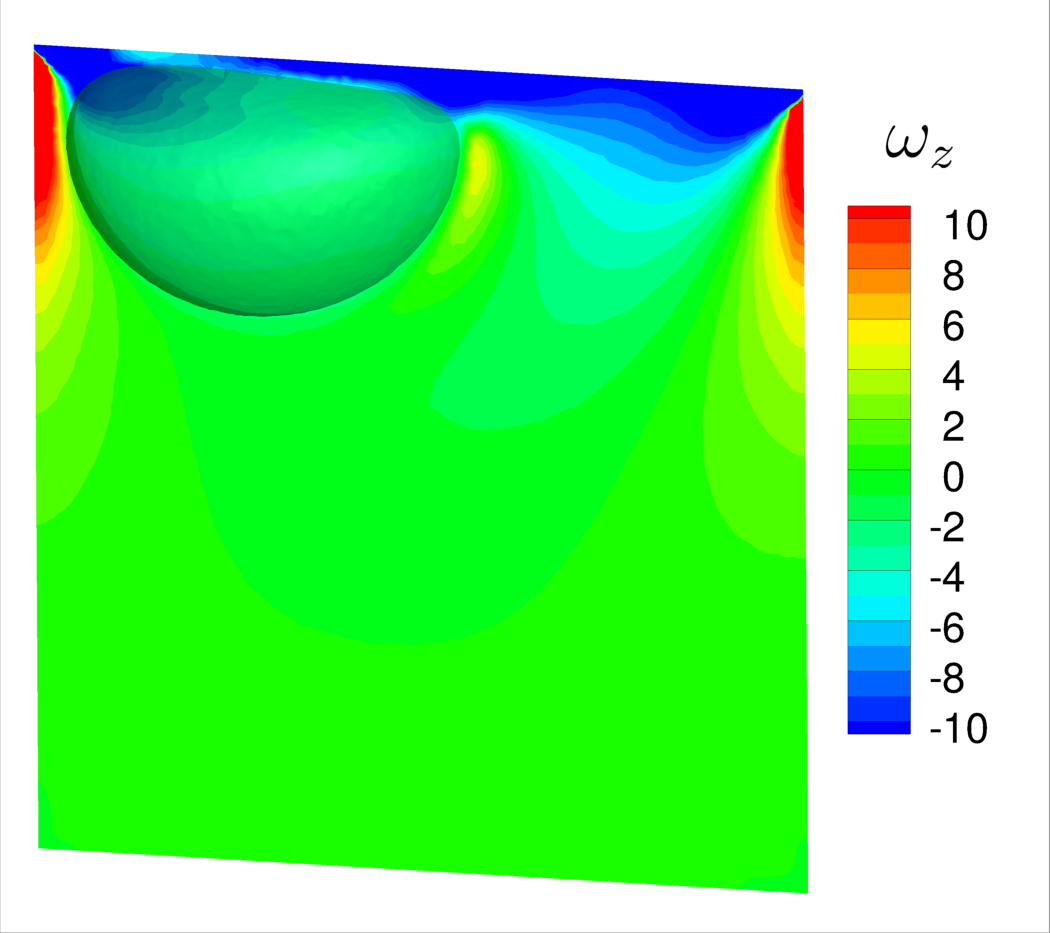}
		\caption*{\hspace{0pt}(c)}
	\end{minipage}
	
	\begin{minipage}[b]{0.3\textwidth}
		\centering
		\includegraphics[scale=0.1,trim= 0 0 0 0,clip]{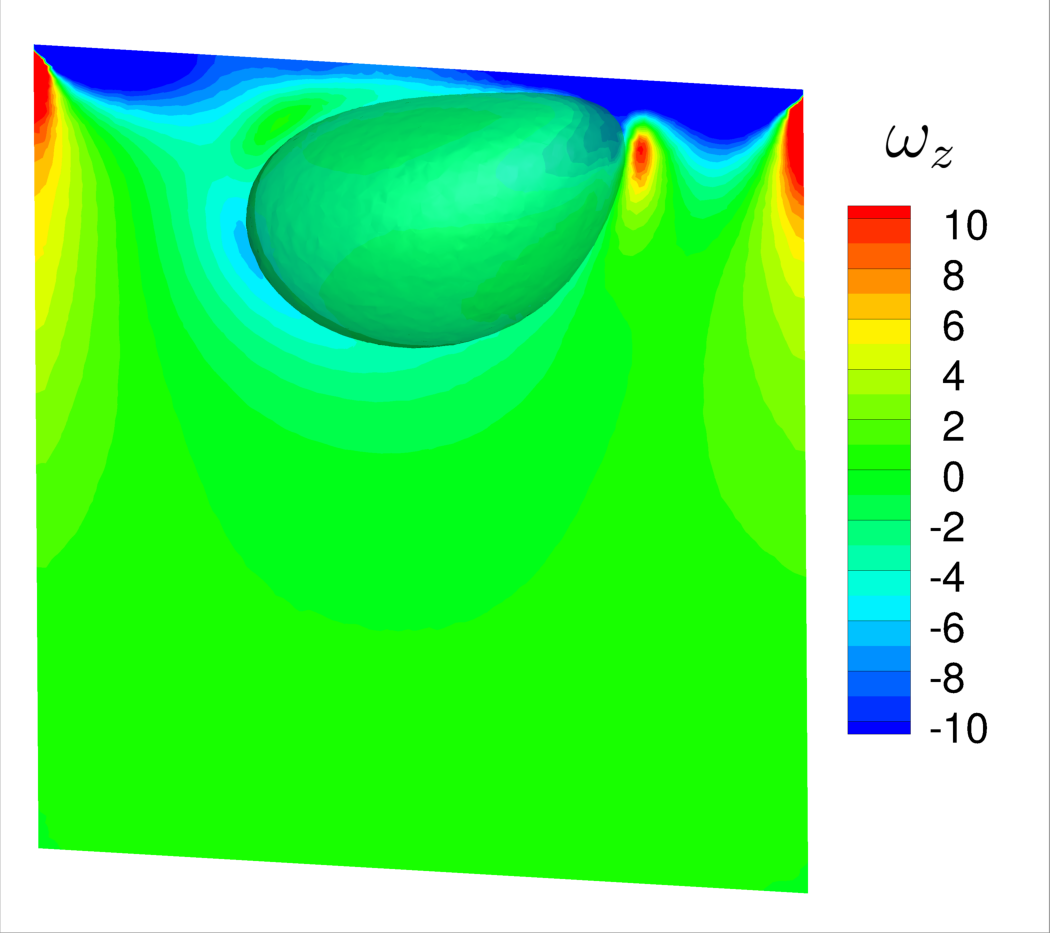}
		\caption*{\hspace{0pt}(d)}
	\end{minipage}
	\begin{minipage}[b]{0.3\textwidth}
		\centering
		\includegraphics[scale=0.1,trim= 0 0 0 0,clip]{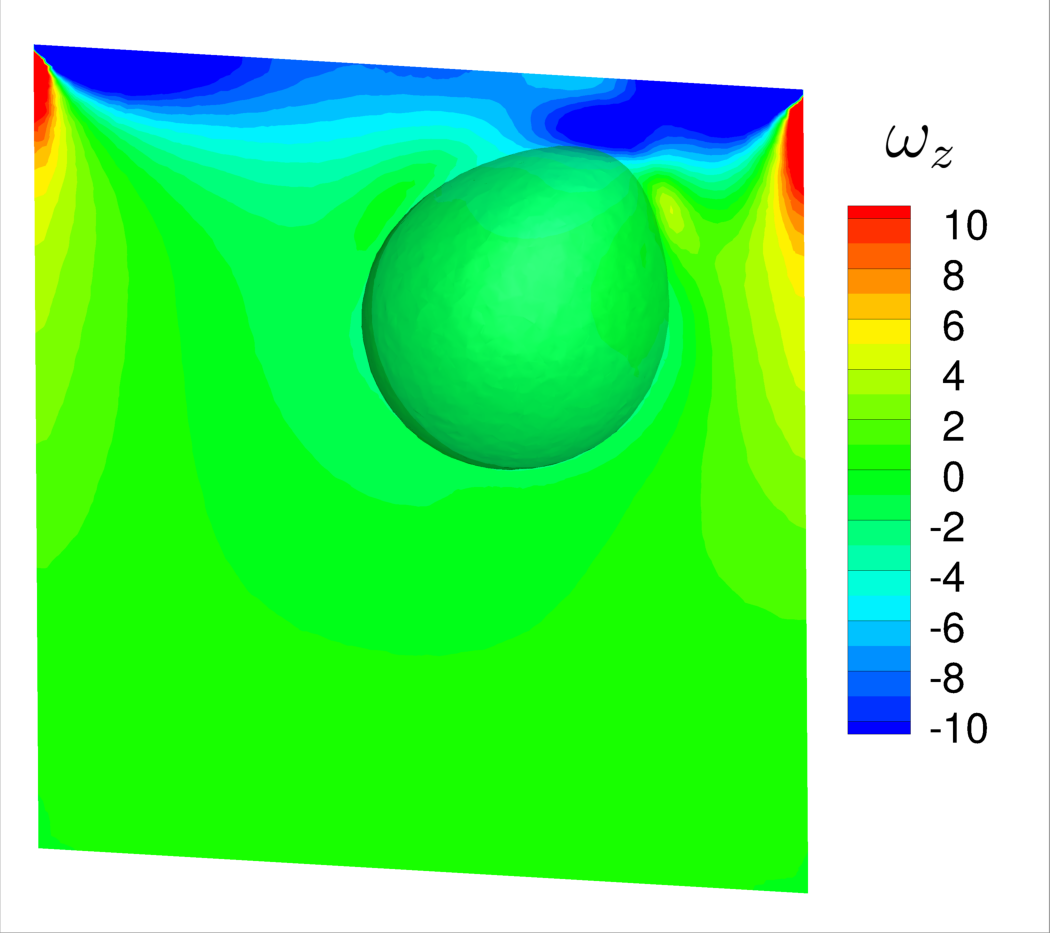}
		\caption*{\hspace{0pt}(e)}
	\end{minipage}
	\begin{minipage}[b]{0.3\textwidth}
		\centering
		\includegraphics[scale=0.1,trim= 0 0 0 0,clip]{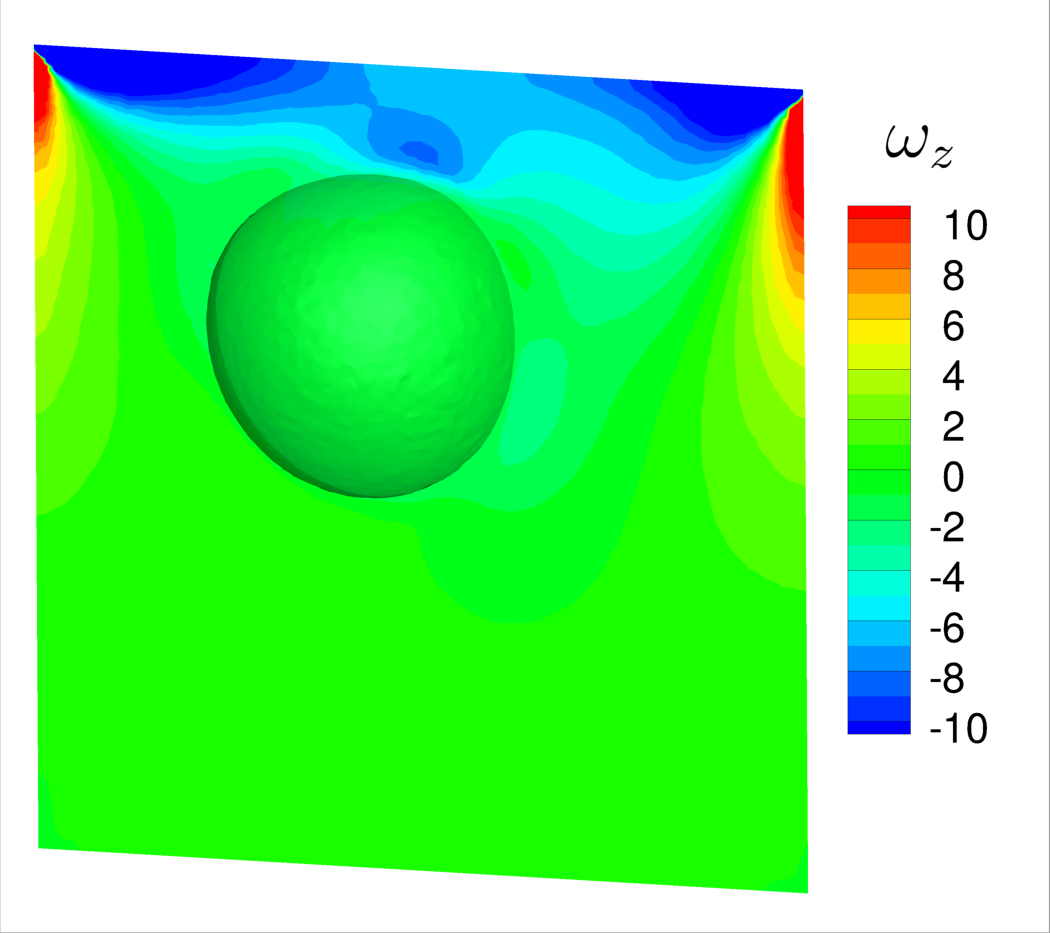}
		\caption*{\hspace{0pt}(f)}
	\end{minipage}
	\caption{Rotating sphere in lid-driven cavity flow: the evolution of the interface $\phi=0$ and the $Z$-vorticity at different time instants $t=$  (a) 1, (b) 3, (c) 4, (d) 5, 
		(e) 6, (f) 8.} 
	\label{rs_ctr}
\end{figure}

Similarly, we perform a temporal and interface convergence study. The interface parameters are selected as $\eta=0.2,\varepsilon=0.02$. The time steps are bisected from $\Delta t = 0.004$ to $\Delta t =0.001$, and the case is simulated until $t=10$. The trajectory of the center of mass on the $X-Y$ plane is plotted in Fig.~\ref{rs_res} (a). As shown by the figure, the solution has converged at $\Delta t = 0.001$ with minor variance. Therefore, we use $\Delta t =0.004$ for the spatial convergence study. With $\Delta t =0.004$, we perform an interface convergence study. The interface parameters $\eta$ and $\varepsilon$ are scaled with $1/\sqrt{2}$ simultaneously from $\eta=0.2,\varepsilon=0.02$ to $\eta=0.1,\varepsilon=0.01$. The meshes are refined according to maintain the interface resolution at $h/\varepsilon=1$. The trajectories of the center of mass are plotted in Fig.~\ref{rs_res} (b). As observed, the solution has converged in the most refined case. To better visualize the field, we show the interface of $\phi = 0$ and $Z$-vorticity at the slice of $z=0.5$ at $t=1,3,4,5,6,8$ in Fig.~\ref{rs_ctr}. As observed, the sphere rotates in the clockwise direction driven by the cavity flow. It touches the wall and experiences large deformation at $t = 4$, and keeps rotating afterward at the upper mid part of the domain, which is reflected by a uniform negative $Z$-vorticity. As demonstrated in the case, the arbitrary rotational motion of an immersed elastic body can be naturally handled in the current framework.

\subsection{Sphere falling on an elastic block}
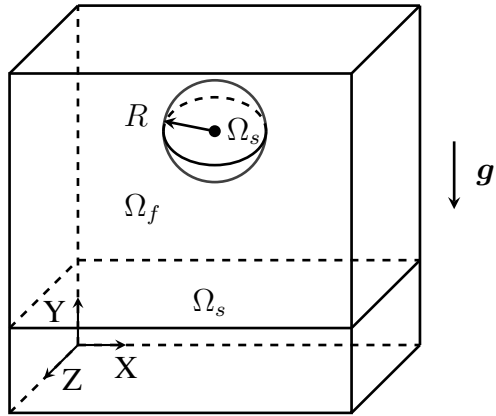
\begin{figure}[h]
	\centering
		\trimbox{0 0 0 0}{
	\begin{tikzpicture}[xscale=4.5, yscale=4.5,line width=1]

		\draw [dashed] (0.2,0.2) -- (0.2,1.2);
		\draw [dashed] (0.2,0.2) -- (1.2,0.2);
		\draw [dashed] (0,0) -- (0.2,0.2);
		\draw (0.2,1.2) -- (1.2,1.2);
		\draw (1.2,0.2) -- (1.2,1.2);
		\draw (1,0)--(1.2,0.2);
		\draw (0,1)--(0.2,1.2);
		\draw (1,1)--(1.2,1.2);
		\draw (0,0) rectangle (1,1);
		



		

		
		
		\draw[fill = white,opacity=0.75] (0.6,0.83) circle (0.15);
		\draw [-stealth] (0.6,0.83)--(0.45,0.86);
		\node[above left] at (0.44,0.81) {$R$};
		\draw (0.45,0.83) arc (180:360:0.15 and 0.1);
		\draw[dashed] (0.75,0.83) arc (0:180:0.15 and 0.1);
		\fill[fill=black] (0.6,0.83) circle (0.5pt);
		\node [right] at (0.6,0.83) {$\Omega_s$};
		
		\draw [-stealth,thick] (0.2,0.2) -- (0.1,0.1);
		\node [right] at (0.12,0.1) {Z};
		\draw[-stealth, thick] (0.2,0.2) -- (0.341,0.2);
		\node [below] at (0.341,0.2) {X};
		\draw[-stealth,thick] (0.2,0.2) -- (0.2,0.341);
		\node [left] at (0.2,0.3) {Y};

		\node [right] at (0.3,0.6) {$\Omega_f$};
		
		\draw [dashed] (0,0.25) -- (0.2,0.45);
		\draw [dashed] (0.2,0.45) -- (1.2,0.45);
		\draw (0,0.25)--(1,0.25);
		\draw (1,0.25)--(1.2,0.45);
		\node [right] at (0.5,0.325) {$\Omega_s$};
		\draw [-stealth] (1.3,0.8)--(1.3,0.6);
		\node [right] at (1.33,0.7){$\boldsymbol{g}$};
	\end{tikzpicture}
}
	\caption{Schematic of an elastic sphere falling on a deformable solid block.}
	\label{fsb}
\end{figure}

After examining the 3D single-phase FSI, we next look into the contact of immersed solids in a fluid environment. A computational domain of $[0,1]\times[0,1]\times [0,1] $ is considered. A solid sphere centered at $(0.5,0.75,0.5)$ at a radius of $r=0.15$ is initially hanging above a solid block filling the computational domain below $y=0.25$ and falls under the gravitational acceleration of $\boldsymbol{g}=(0,-0.98,0)$. The no-slip boundary conditions are applied to the side and bottom walls, while an outflow boundary condition is applied to the top boundary. The order parameter and the left Cauchy-Green tensor on the boundaries are evolved according to the local velocity. The densities of the solid sphere, the solid block and the surrounding fluid are chosen as $\rho^s=5000$, $\rho^b=5000$, $\rho^f=1000$. The viscosities of the solids are set to be zero while the fluid viscosity is selected as $\mu^f=1$. The shear modulus of the solid sphere and the solid block are taken as $\mu^s_L=1000$ and $\mu^b_L=1000$. The problem setup is illustrated in Fig.~\ref{fsb}.

\begin{figure}[h]
	
	\begin{minipage}[b]{0.5\textwidth}
		\centering
		\includegraphics[scale=0.18,trim= 0 0 0 0,clip]{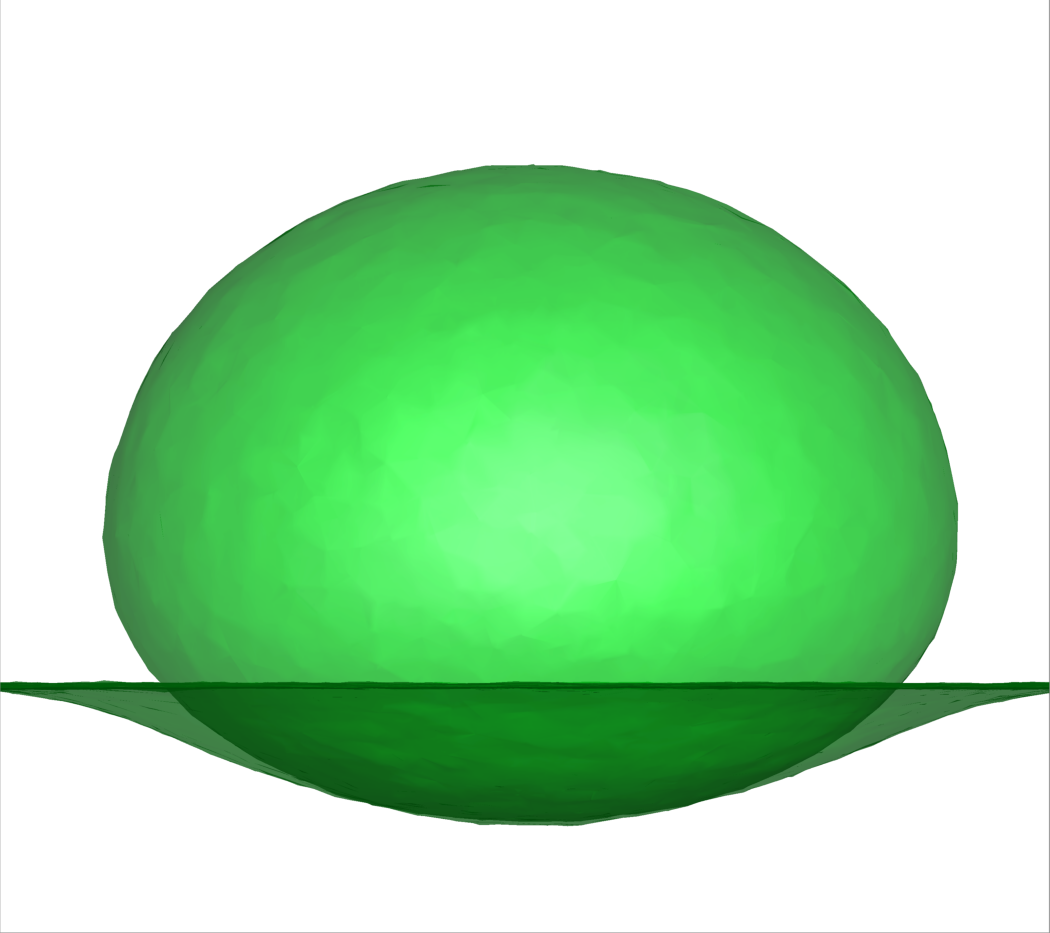}
		\caption*{\hspace{30pt}(a)}
	\end{minipage}
	\hspace{0pt}
	\begin{minipage}[b]{0.5\textwidth}
		\centering
		\includegraphics[scale=0.6,trim= 0 0 0 0,clip]{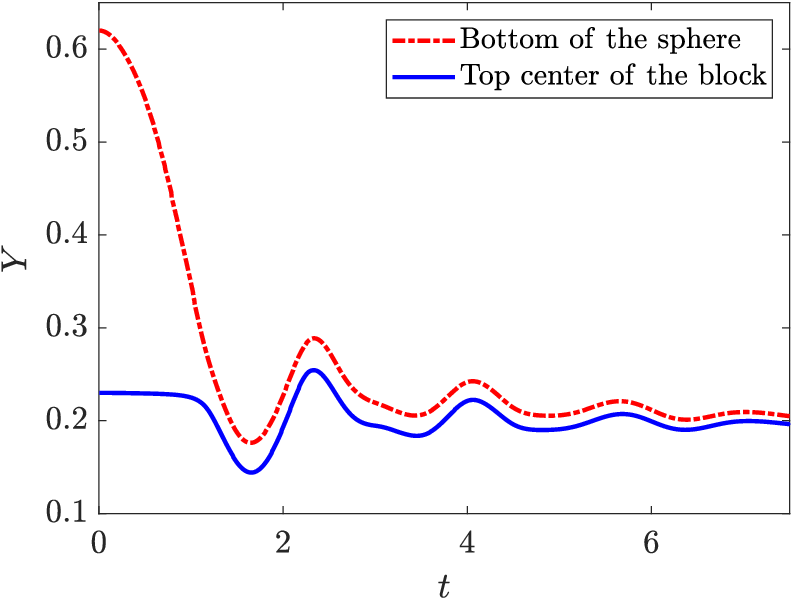}
		\caption*{\hspace{30pt}(b)}
	\end{minipage}
	\caption{Sphere falling on a solid block: (a) Interfaces of solids $\phi=0$ at $t=7.5$, and (b) trajectories of the bottom of the sphere and the top center point of the block } 
	\label{fsb_gap}
\end{figure}

In the computational framework, we set one phase for the falling sphere, one for the solid block, and one for the immersed fluid. The computational domain is discretized with tetrahedron elements of size $h=0.01$. The diffuse interface thickness is chosen as $\varepsilon=h$ and $\eta=0.1$ is used. The time step is selected as $\Delta t =0.005$. The case is simulated until $t=7.5$. The simulation results are visualized at $t=1,1.5,2.5,3.25,4,7.5$ in Fig.~\ref{fsb}. As observed, the sphere starts falling due to gravitational force. It reaches the lowest and highest positions at $t=1,1.5,2.5,3.25,4,7.5$ and gradually settles down as depicted at $t=7.5$. To better monitor the gap during contact between the sphere and the block, we track the trajectory of the bottom of the sphere and the top center of the solid block. To avoid the influence from the diffuse interface modeling, we take a $2\varepsilon$ shift towards the internal of the solids. As a result, the monitor points are initially located at $(0.5,0.6+2\varepsilon,0.5)$ and $(0.5,0.25-2\varepsilon,0.5)$. The points are evolved by a parallelized particle tracking algorithm, which is presented in \ref{PT}. The zoomed-in view of the contact interface at $t=7.5$ and the time history of $Y$ coordinates of the monitor points are plotted in Fig.~\ref{fsb_gap} (a) and (b), respectively. As observed, the no-penetration condition in the process of contact is naturally satisfied. As demonstrated in this case, large translational motion and contact between solids can be naturally handled by the current framework.

\begin{figure}[h]
	
	\begin{minipage}[b]{0.3\textwidth}
		\centering
		\includegraphics[scale=0.13,trim= 0 0 0 100,clip]{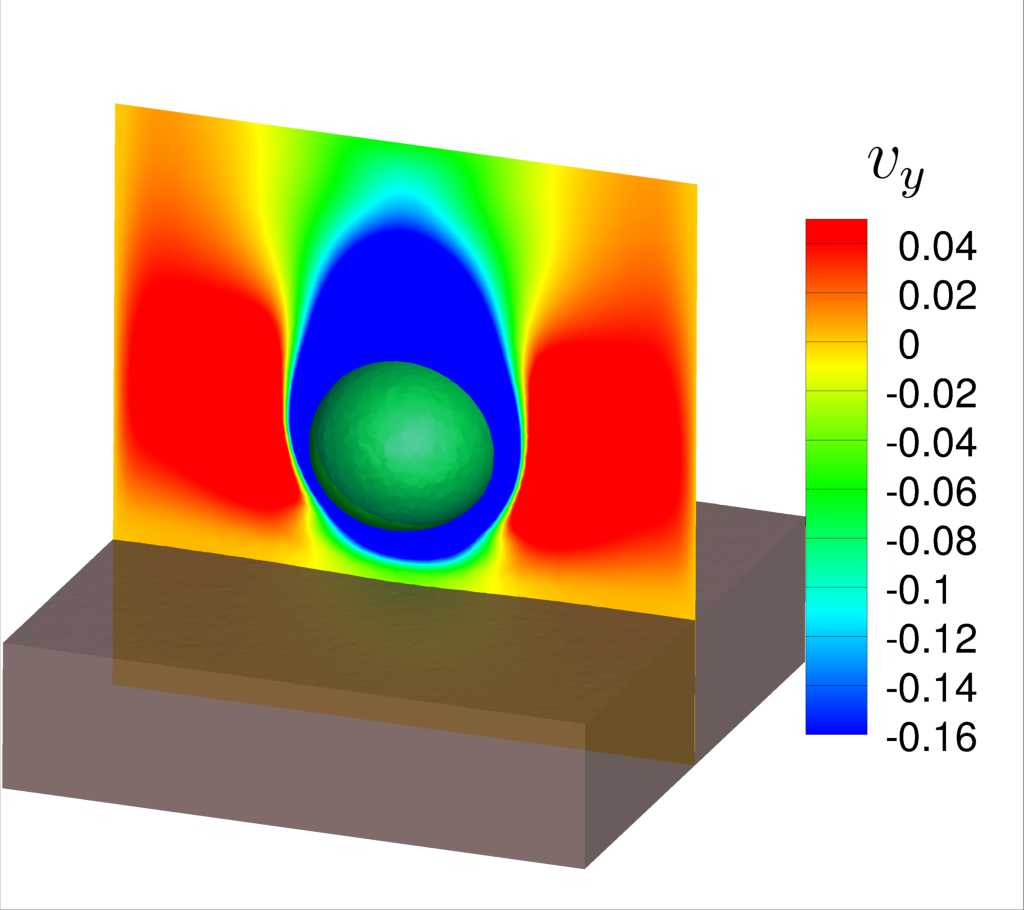}
		\caption*{\hspace{0pt}(a)}
	\end{minipage}
	\begin{minipage}[b]{0.3\textwidth}
		\centering
		\includegraphics[scale=0.13,trim= 0 0 0 100,clip]{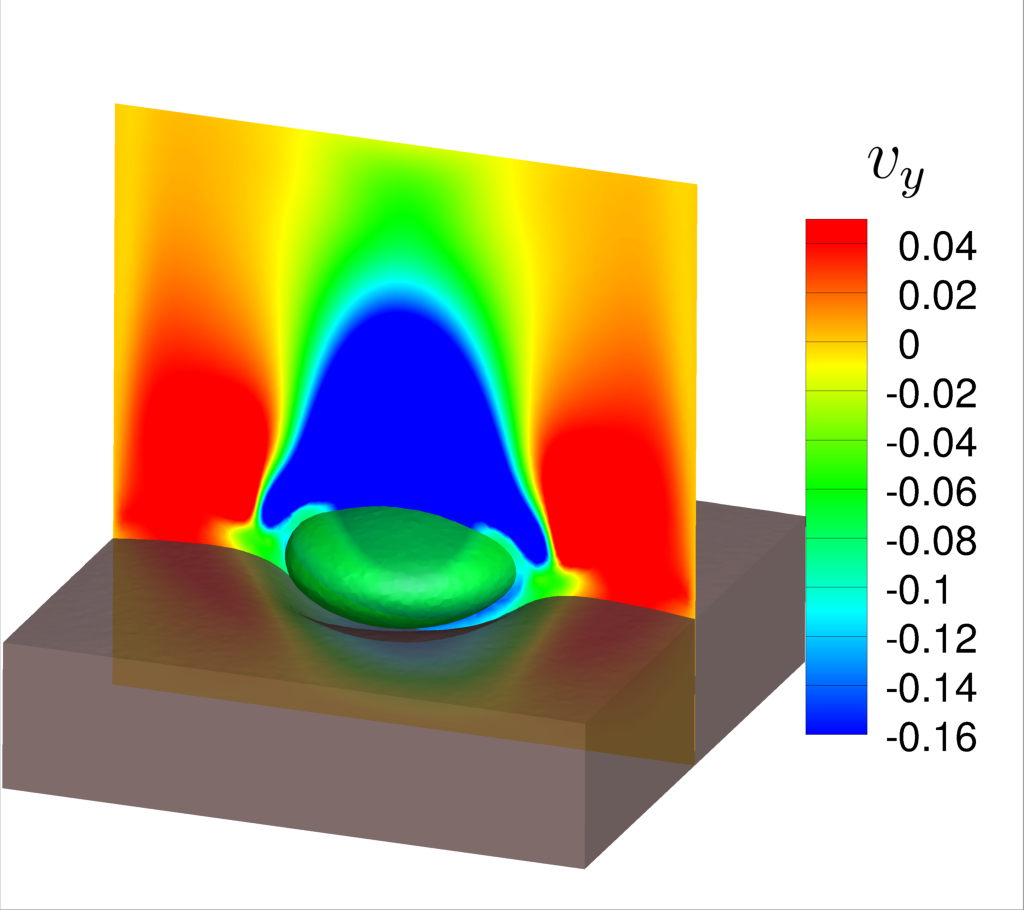}
		\caption*{\hspace{0pt}(b)}
	\end{minipage}
	\begin{minipage}[b]{0.3\textwidth}
		\centering
		\includegraphics[scale=0.13,trim= 0 0 0 100,clip]{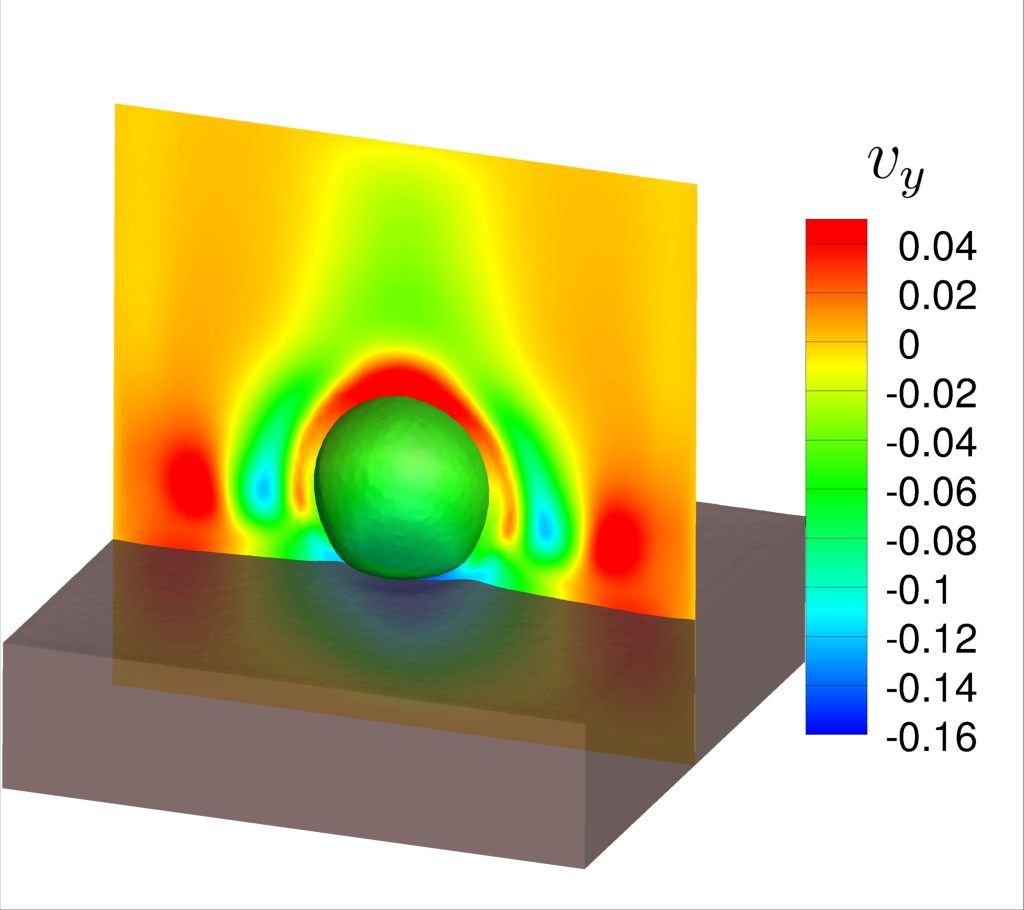}
		\caption*{\hspace{0pt}(c)}
	\end{minipage}
	
	\begin{minipage}[b]{0.3\textwidth}
		\centering
		\includegraphics[scale=0.13,trim= 0 0 0 0,clip]{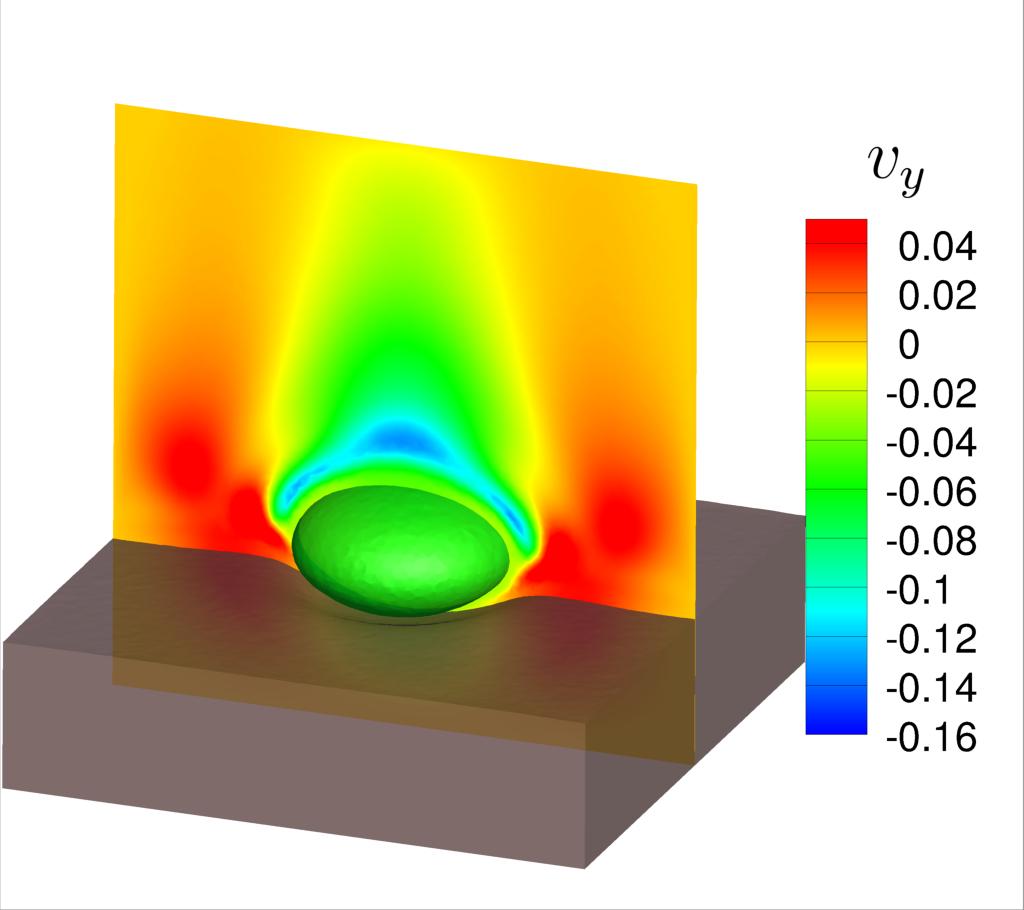}
		\caption*{\hspace{0pt}(d)}
	\end{minipage}
	\begin{minipage}[b]{0.3\textwidth}
		\centering
		\includegraphics[scale=0.13,trim= 0 0 0 0,clip]{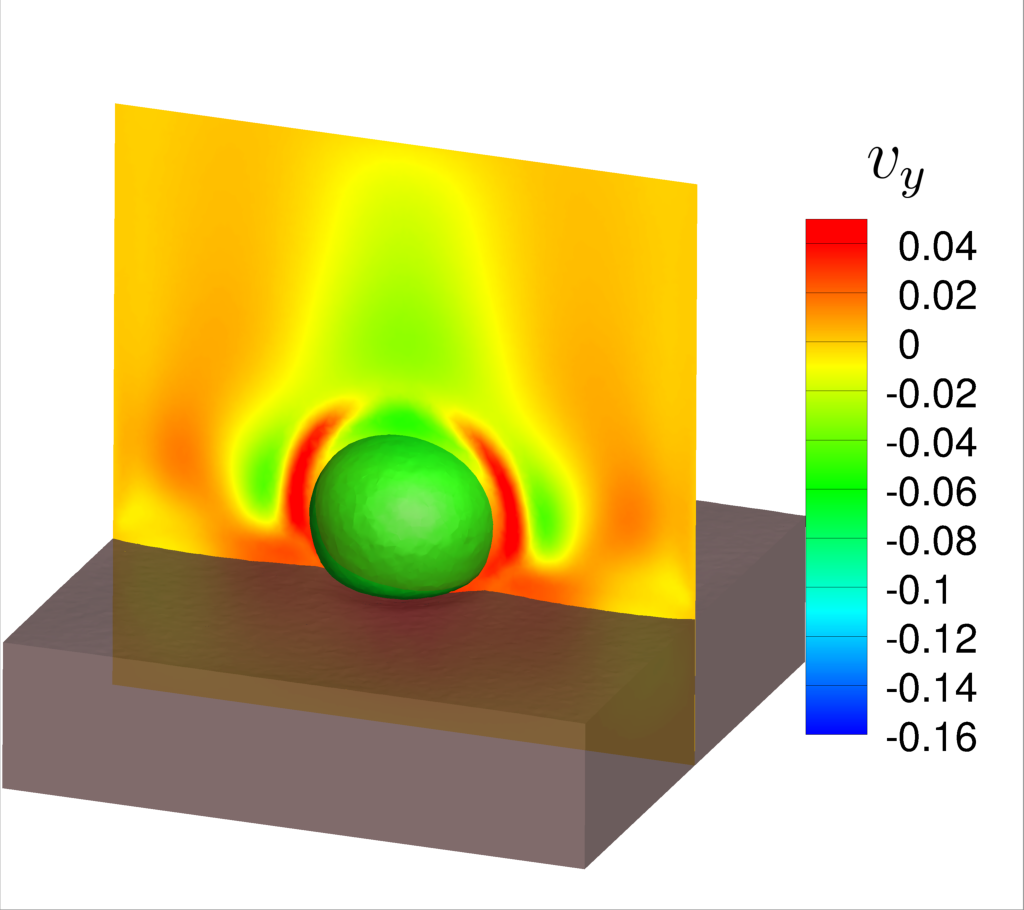}
		\caption*{\hspace{0pt}(e)}
	\end{minipage}
	\begin{minipage}[b]{0.3\textwidth}
		\centering
		\includegraphics[scale=0.13,trim= 0 0 0 0,clip]{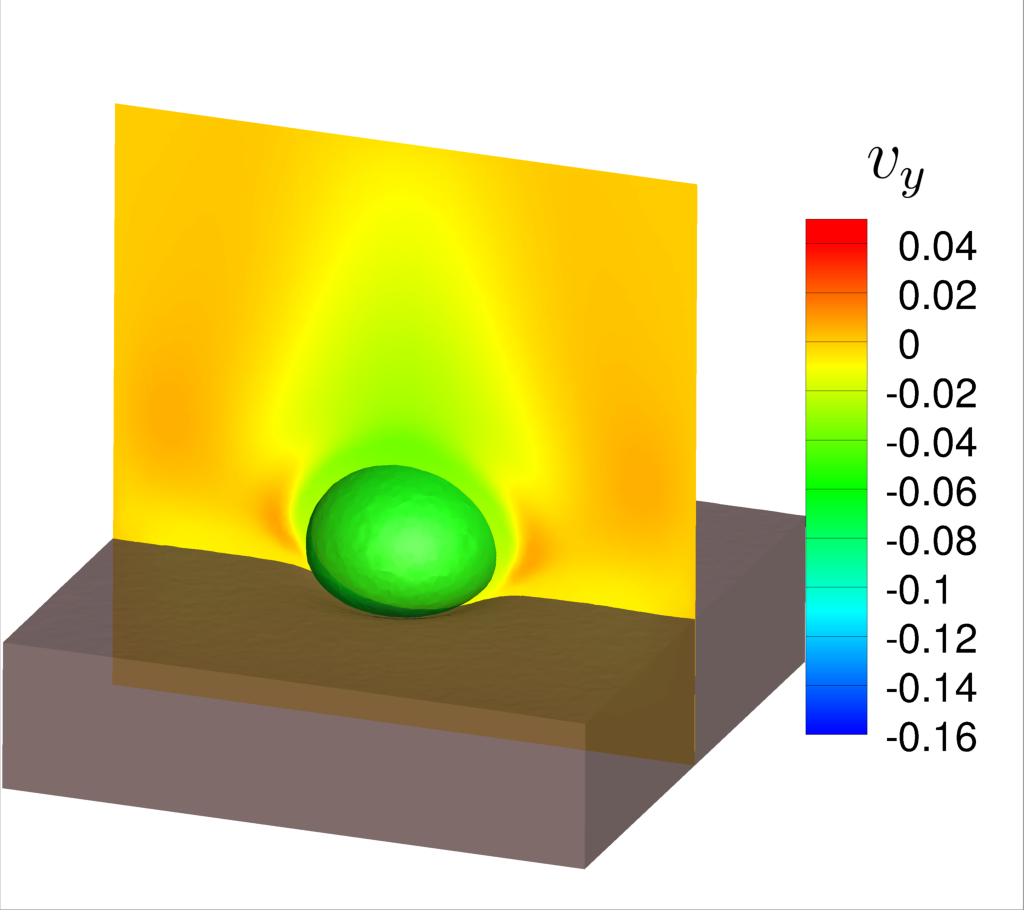}
		\caption*{\hspace{0pt}(f)}
	\end{minipage}
	\caption{An elastic sphere falling on a deformable solid block: the evolution of the interfaces $\phi=0$ and $v_y$ contour at $Z=0.5$ at different time instants $t=$  (a) 1, (b) 1.5, (c) 2.5, (d) 3.25, (e) 4, (f) 7.5.} 
	\label{fso_ctr}
\end{figure}

\section{Application to ship-ice interaction}\label{Sec:ship}
Of particular interest is the ship-ice interaction in the Arctic region. Located at the north pole and surrounded by lands, the Arctic region encompassing various terrains, enormous biodiversity and rich resources, is now the home to indigenous people of over 40 communities and a total population of four million \cite{stadtlander2020book,schoolmeester2019global,larsen2015arctic}. Under the effect of global warming, the Arctic Ocean may be seasonally ice-free by 2050 \cite{timmermans2020understanding}. This creates opportunities for extracting natural resources and finding new routes for transcontinental transportation \cite{lavissiere2020transportation}. With the growing strategic importance of Arctic transportation, it is required to design and manufacture marine vessels that are reliable and safe when sailing across ocean environments with ice sheets and floes. High-fidelity simulation is of great importance to understanding the physical process involved and aiding the design and optimization processes.

\begin{figure}
	\begin{minipage}{0.5\textwidth}
		\centering
		\includegraphics[scale=0.35,trim=0 0 0 0,clip]{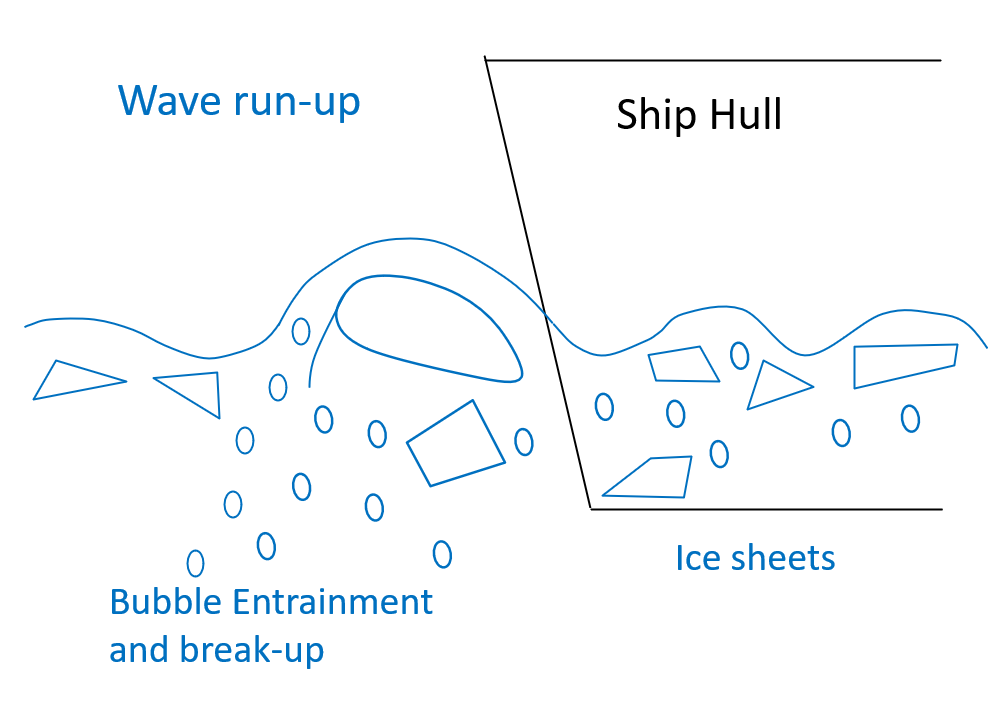}
		\caption*{(a)}
	\end{minipage}
	\begin{minipage}{0.5\textwidth}
	\centering
	\includegraphics[scale=0.35,trim=0 0 0 0,clip]{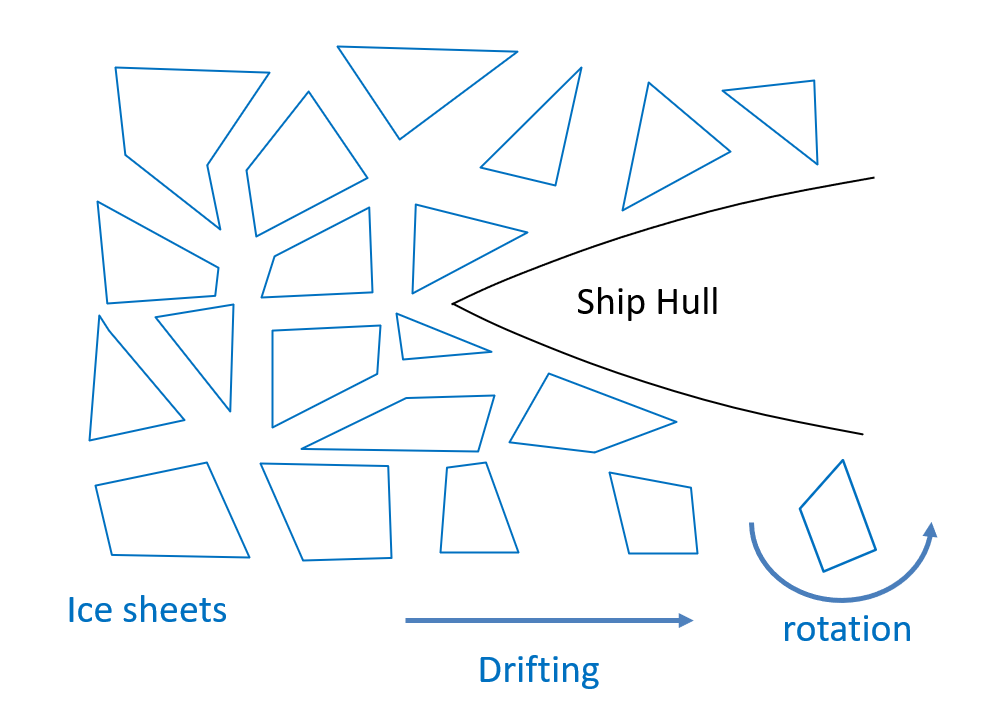}
	\caption*{(b)}
	\end{minipage}
	\caption{Illustration of ship-ice interaction showing combined effects of hydrodynamics and ice floes interacting with a ship hull: (b) side view and (c) top view. }
	\label{iceship}
\end{figure}

The problem of ship-ice interaction involves interactions between ice mechanics, ship structure, hydrodynamics, and wind flow. The ice floating on the sea level can vary in size and geometry: from small-scale ice rubble to large-scale ice ridges and floes \cite{li2022review}. During the sailing of ships, they are in contact with the ship hull, accelerated, pushed aside, or submerged, which creates loads and deformation on both sides. Then, as a source of disturbance, the moving ice can interact with surrounding static ice and cause drifting and rotating motions. The ice and ship hull interaction happens near the water-air interface under the gravitational force. As the ship moves forward, the wave runs up and entrains air bubbles inside. The bubbles then break apart and become sea foam. The sea foam propagates along with waves surrounding the ship and interacts with floating ice and ship hull \cite{huang2020ship}, as illustrated in Fig.~\ref{iceship}. They are all coupled and interact through interfaces of different types, forming a highly complex nonlinear dynamic system of multiphase and multi-physics nature, resulting in both the global resistance and local load on the ship. They play important roles in different ways: the global resistance affects the operation of a ship, such as maneuverability and fuel consumption, while the local load is critical for the safety of the ship structure.


 While research efforts have been devoted to understanding the ice-ship interaction, the current state-of-the-art methods for calculating coupled hydrodynamics-structure-ice interactions
	are semi-empirical and highly phenomenological\cite{xue2020review}. Using our multiphase FSI framework, for the first time, we will demonstrate ship-ice interaction with nonlinear hydrodynamics effects. With first-principle-based continuum mechanics, the simulation captures relevant physics completely, including the interactions between floating solids and surrounding free surface, and contact between the structures. This is crucial to understand the dynamics involved in the process, and can be utilized to provide better resistance and extreme load prediction.

 In the demonstration case, we employ a vessel STAN TUG 1004 for which the vessel geometry is available online\footnotemark[2]{}. The domain of $[-1,1]\times[-2,2]\times[-7.5,2]$ is considered. The free water level is the $X-Y$ plane of $x=0$. The leading point of the cross-plane of the ship and the free water level is placed at $(0,0,-1.1157)$. Six pieces of ice floes in total are posed in front of the boat. They are formed by extruding the cross section from $x=0$ to $x=-0.3$. The geometries of cross sections of the ice floes are listed in Table~\ref{ice_geo}. 
 
 \footnotetext[2]{\url{https://grabcad.com/library/stan-tug-1004-1}}
 
\begin{table}[h]
	\caption{Geometry of the ice floes}
	\label{ice_geo}
	\begin{tabular}{|c| c| c| }
		\hline	Index & shape of cross section & coordinate of cross sections in $(z,y)$ \\
		\hline 	1     & triangle               &  $(-1.1,0.6),(-1.7,0.2),(-1.7,0.9)$\\ 
		\hline 	2     & quadrilateral          &  $(-1.6,-0.2),(-2.1,-0.6),(-1.8,-1),(-1,-0.6)$\\
		\hline 	3     & quadrilateral          &  $(-3,1.3),(-3.7,1),(-3.2,0.3),(-2.7,0.7)$ \\
		\hline  4     & quadrilateral          &  $(-4,-0.4),(-4.4,-0.9),(-3.8,-1.7),(-3.2,-0.7)$ \\
		\hline 	5     & quadrilateral          &  $(-4.8,1),(-5.4,0.6),(-4.7,0.4),(-4.1,0.5)$ \\
		\hline 	6     & quadrilateral          &  $(-5.3,-0.4),(-6,-0.7),(-5.9,-1.2),(-5.1,-1)$\\
		\hline	
	\end{tabular}
	
\end{table}

\begin{figure}[h]
	\begin{minipage}[b]{0.3\textwidth}
		\centering
		\includegraphics[scale=0.4,trim= 0 0 0 0,clip]{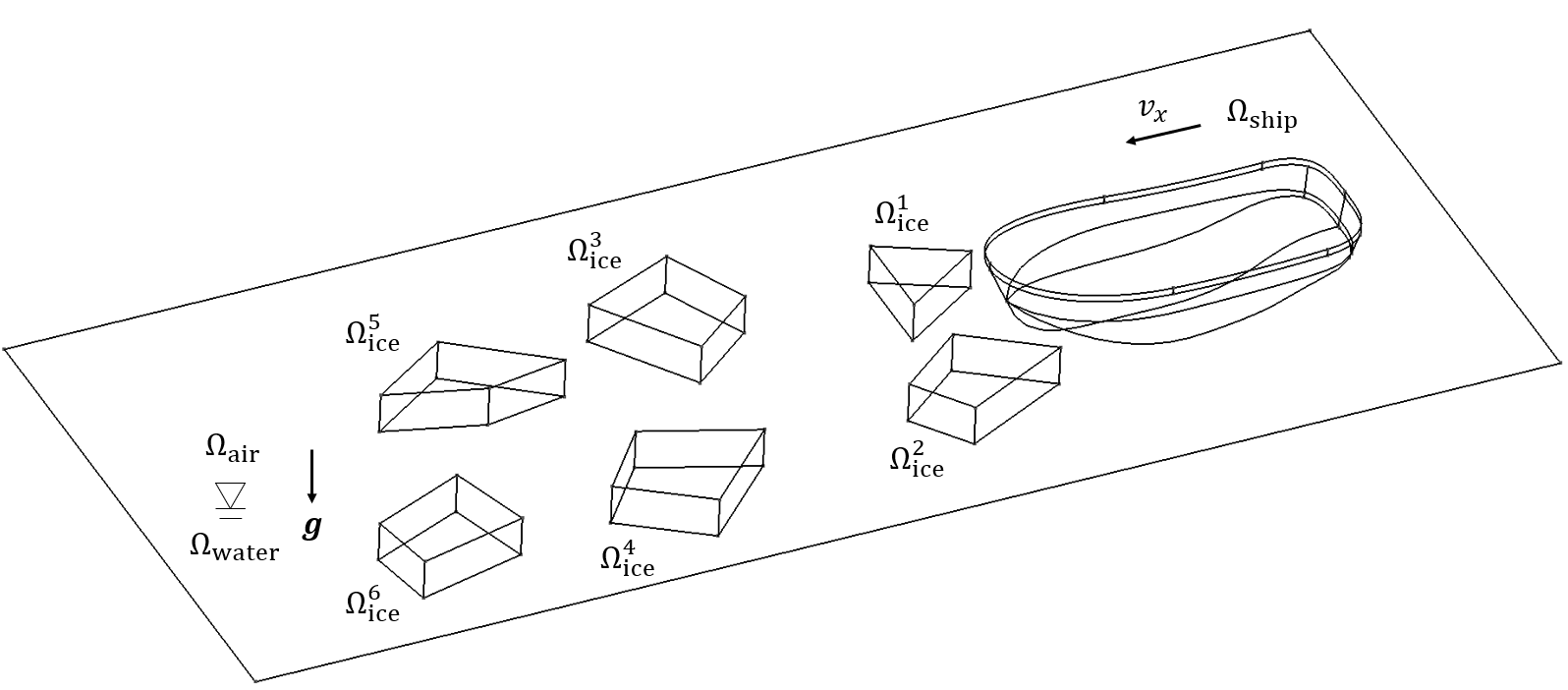}
	\end{minipage}
	
	\caption{Schematic diagram of an ice-going ship with free surface and ice floes.} 
	\label{ship}
\end{figure}

The boat accelerate from stationary to $v_z=-1$ sinusoidally: $v_z=-\sin(2/5\pi t), t<1.25.
v_z=-1, t\geq 1.25.$
This is realized by finding the nodes inside the boat and specifying the velocity on these nodes. The nodes are identified by $-0.2-d(t)\leq z\leq0.2-d(t),|y|\leq0.2,|x|\leq0.2$, where $d(t)$ is the moving distance of the ship integrated from the given velocity $v_z$. The slip boundary condition is applied to side boundaries. The no-slip boundary condition is applied on the bottom boundary, and the outlet boundary condition is used for the top boundary. The zero Neumann boundary condition is applied for all the order parameters and left Cauchy-Green tensors on all the boundaries. The physical properties are chosen such that the overall behavior is captured while the computation is eased for demonstration purposes. Densities of the ship, ice, water, and air phases are selected as $\rho^s=600, \rho^i=800, \rho^w=1000, \rho^a=100$. The viscosities of the phases are selected as $\mu^s=\mu^i=\mu^w=\mu^a=1$ to stabilize the numerical simulation. The shear modulus of the ship and the ice floes are selected as $\mu^s_L=\mu^i_L=1\times 10^6$ to mimic the nearly rigid contact. The gravitational acceleration is chosen as $\boldsymbol{g}=(-0.98, 0, 0 )$ to amplify the free surface motion. The case is simulated until $t=4$. The case setup is illustrated in Fig.~\ref{ship}.

\begin{figure}[h]
	
	\begin{minipage}[b]{0.5\textwidth}
		\centering
		\includegraphics[scale=0.1,trim= 0 2 2 0,clip]{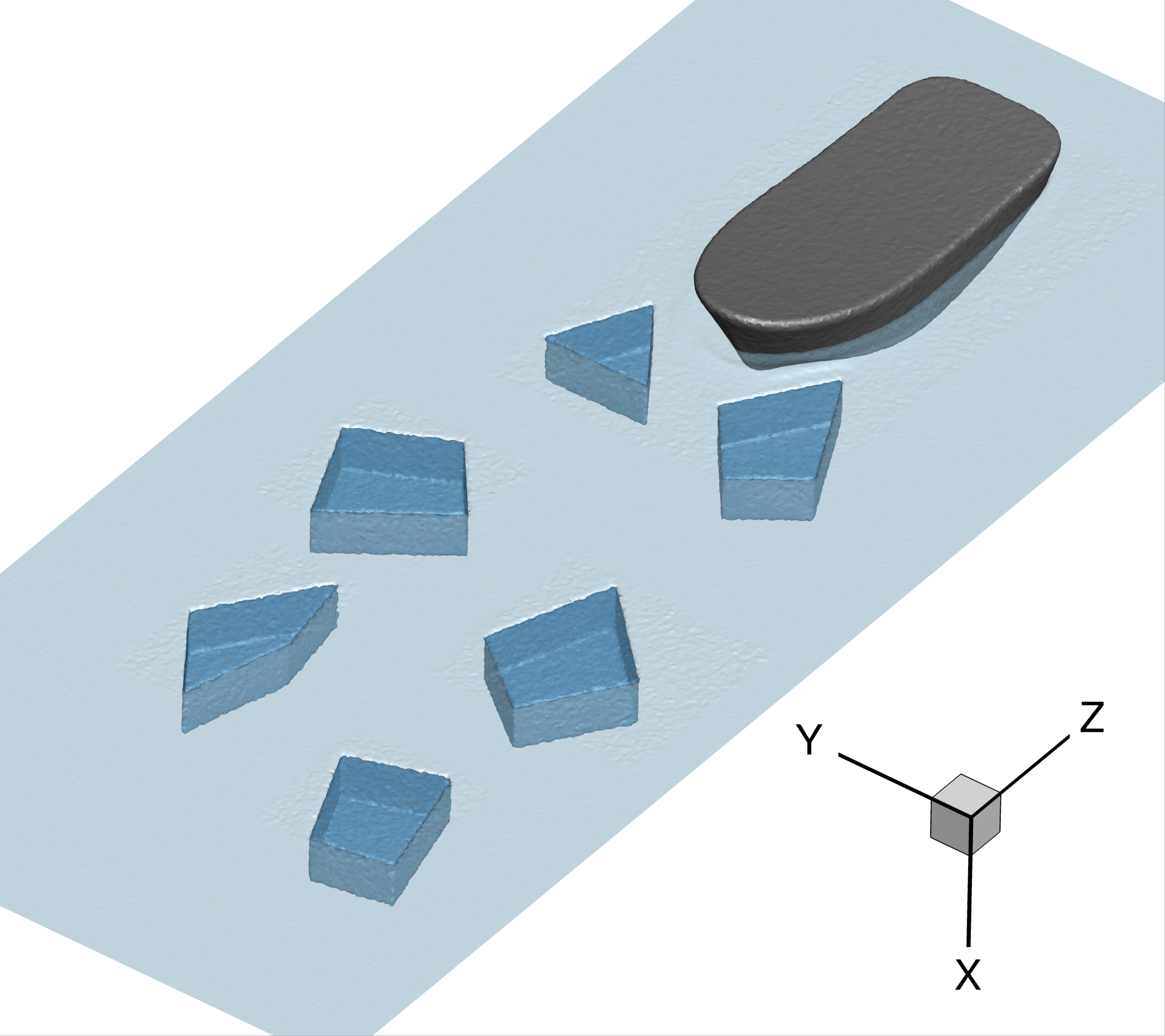}
		\caption*{\hspace{30pt}(a)}
	\end{minipage}
	\begin{minipage}[b]{0.5\textwidth}
		\centering
		\includegraphics[scale=0.1,trim= 0 3 3 0,clip]{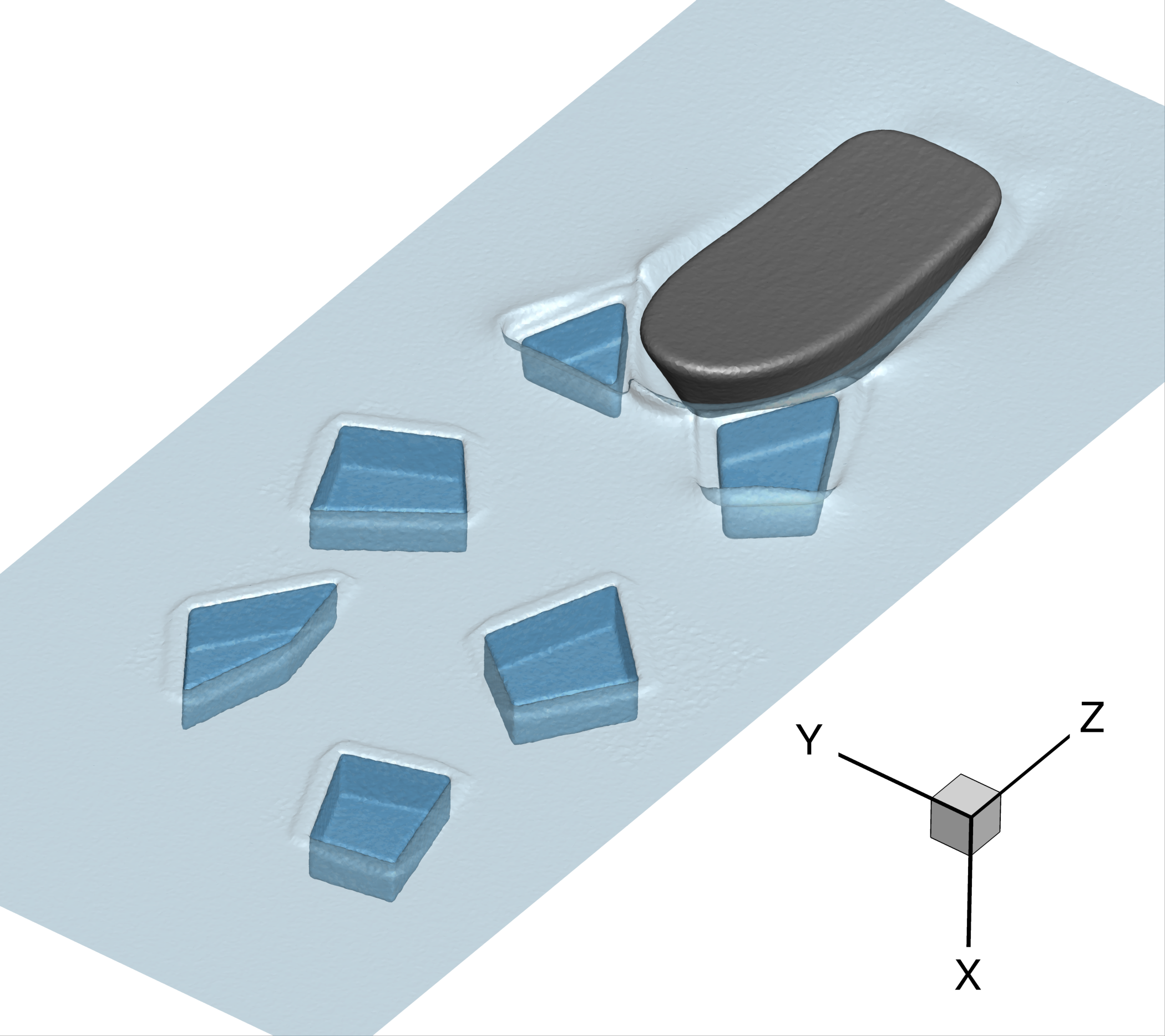}
		\caption*{\hspace{30pt}(b)}
	\end{minipage}

	\begin{minipage}[b]{0.5\textwidth}
		\centering
		\includegraphics[scale=0.1,trim= 0 3 3 0,clip]{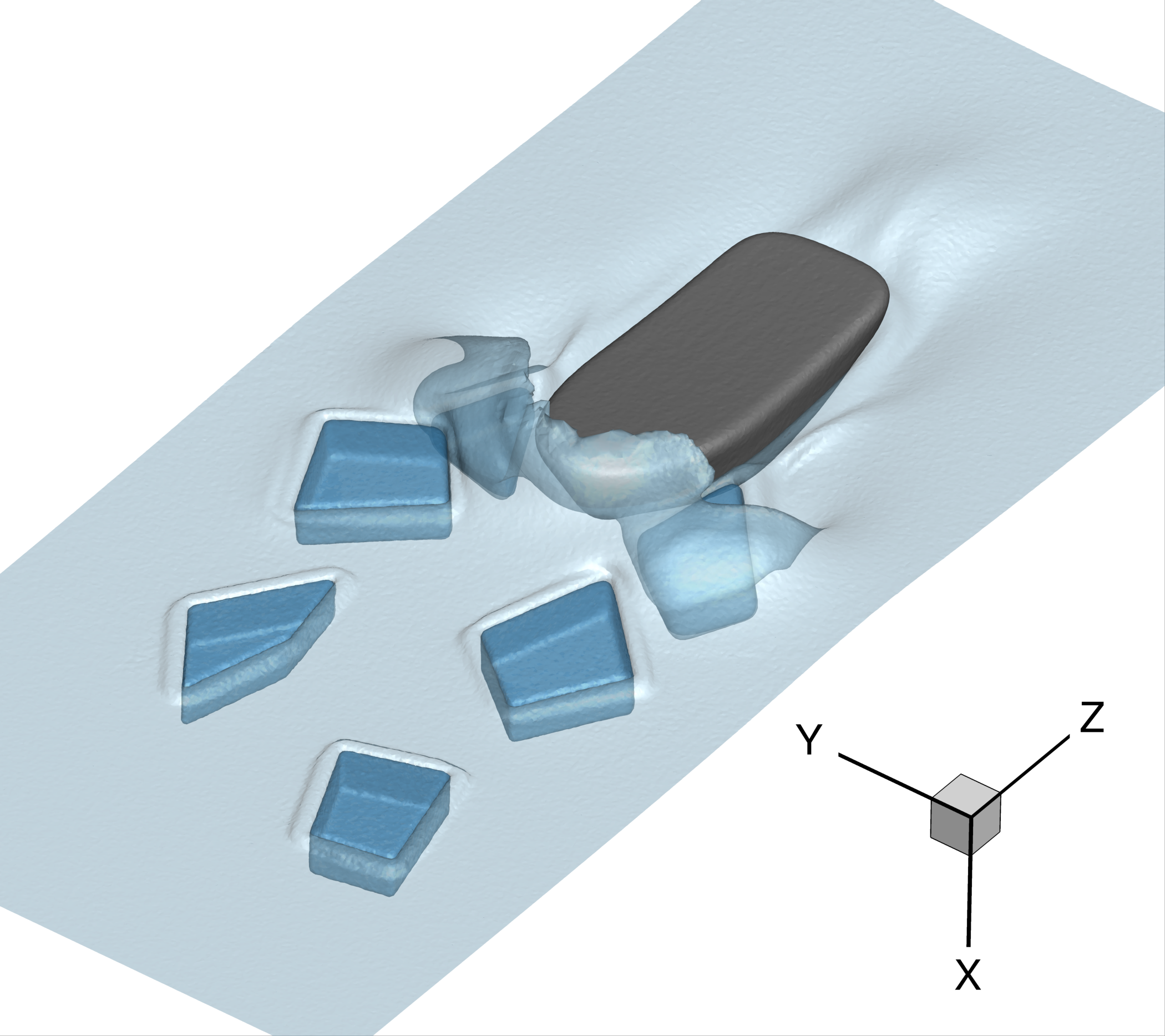}
		\caption*{\hspace{30pt}(c)}
	\end{minipage}
	\begin{minipage}[b]{0.5\textwidth}
		\centering
		\includegraphics[scale=0.1,trim= 0 3 3 0,clip]{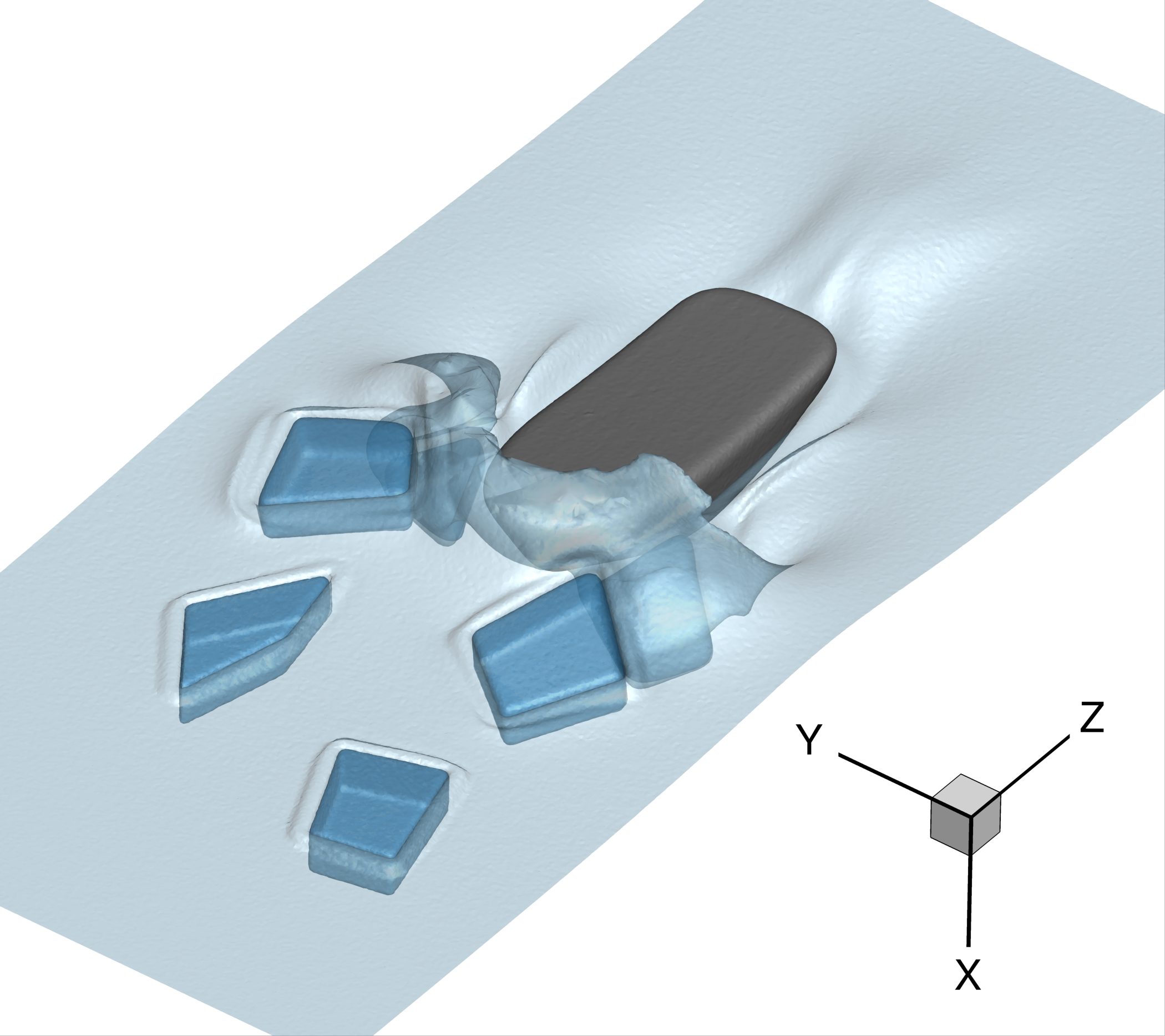}
		\caption*{\hspace{30pt}(d)}
	\end{minipage}

	\begin{minipage}[b]{0.5\textwidth}
		\centering
		\includegraphics[scale=0.1,trim= 0 3 3 0,clip]{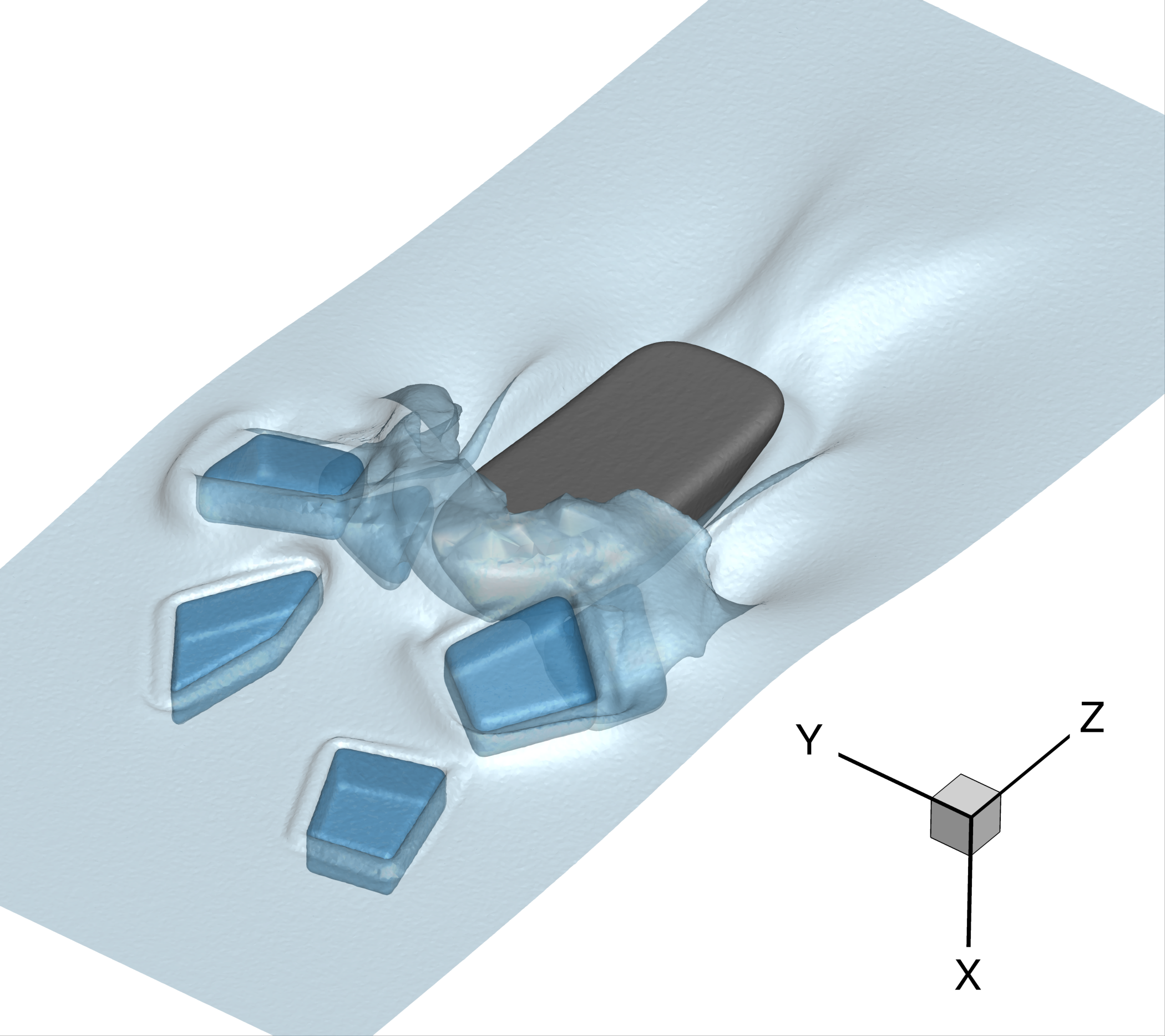}
		\caption*{\hspace{30pt}(e)}
	\end{minipage}
	\begin{minipage}[b]{0.5\textwidth}
		\centering
		\includegraphics[scale=0.1,trim= 0 3 3 0,clip]{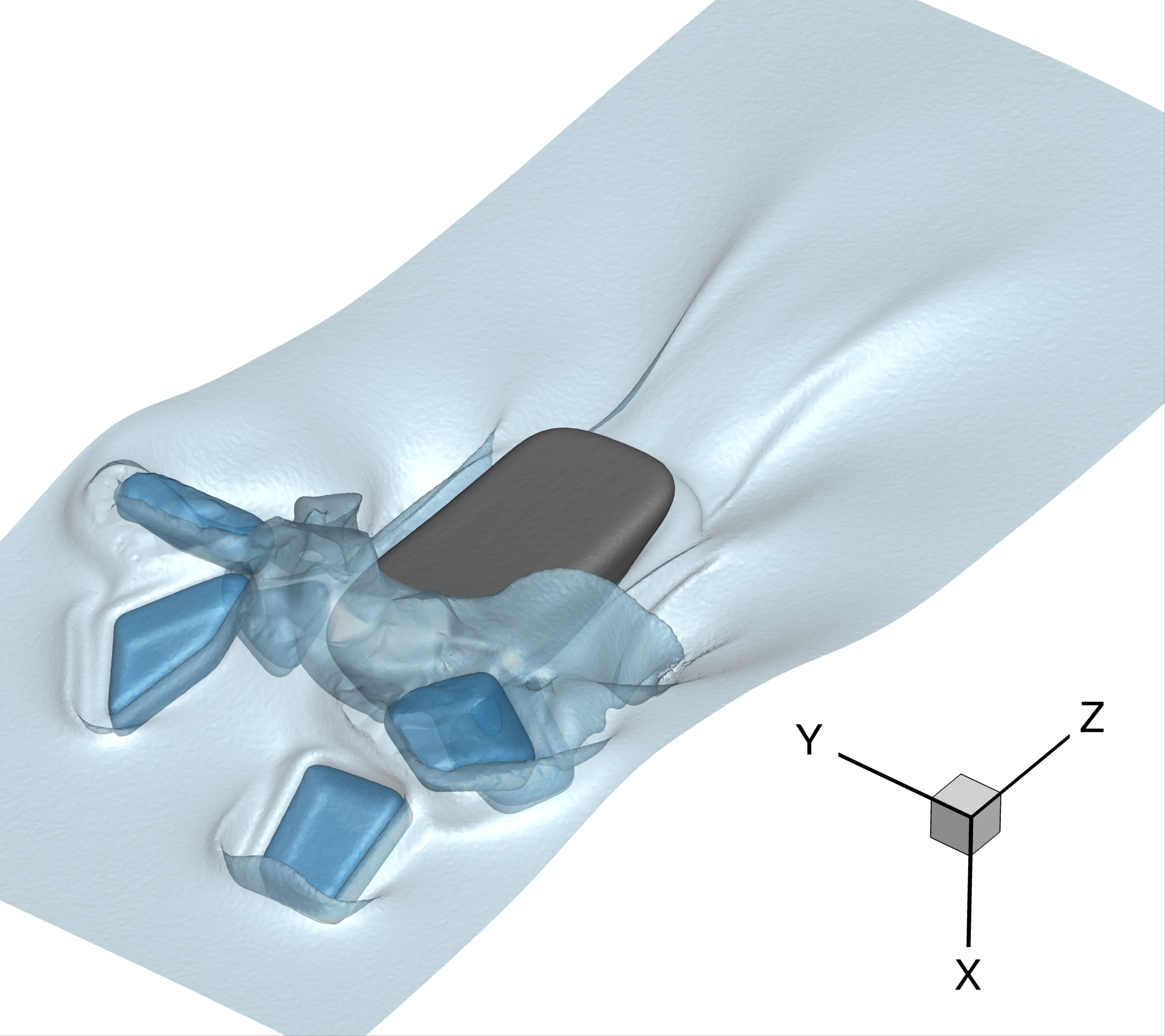}
		\caption*{\hspace{30pt}(f)}
	\end{minipage}
	\caption{Representative results of ship-ice interaction: superposition of the interfaces $\phi^i=0$ at different time instants $t$ = (a) 0.25, (b) 1, (c) 2, (d) 2.5, (e) 3, (f) 4.} 
	\label{ship_iso}
\end{figure}

In the computational framework, we set one fluid phase for water, one fluid phase for air, one solid phase for the ship, and one solid phase for all the ice floes. The computational domain is discretized with tetrahedron elements of size $h=0.02$ around the interacting phases. The diffuse interface thickness is chosen as $\varepsilon = h$. With the focus on the geometry of the solids, we set $\eta = 1000$. The time step is selected as $\Delta t =0.005$. The simulation results are visualized at $t=0.25, 1, 2, 2.5, 3, 4$ in Fig.~\ref{ship_iso}. As observed, the rich physics of interactions between the ship and free surface, the ship-ice interaction, and the coupling of ice floes and free surface are captured. An interesting dynamics of ship-ice interaction at the bow area can be observed qualitatively. It is worth mentioning that
the ice mechanics model is very simplified without considering detailed aspects of constitutive modeling and fracture mechanics. Further parametric investigations and validation should be carried out for more realistic modeling of ice mechanics for the ship-ice interaction with nonlinear hydrodynamics effects.

\clearpage
\section{Conclusion} \label{conclusion}
In this paper, we presented a novel unified Eulerian variational framework for multiphase FSI problems based on the phase-field method. 
By employing the phase-field method on a fixed mesh,  the proposed fully Eulerian framework offers numerous attractive characteristics 
to simulate multiphase fluid-structure interactions with solid-solid contact.  
 Some of the key features of the proposed computational framework are:
\begin{itemize}
	\item Phase-field interface processing based on grid cell method, ray-casting algorithm and non-duplicated plane partitioning for order parameter initialization of complex interface geometries;
	\item Unified momentum and mass conservation equations for the dynamics of general multiphase FSI problems;
	\item Improved accuracy of interface representation with interface-preserving phase-field model;
	\item High robustness in handling interface motion, including large translation/rotational motion, interface merging/breaking-up and solid-solid contact interactions;
	\item 3D parallelized implementation for geometric preprocessing of phase-field method and fully-Eulerian finite element simulation.
\end{itemize} 

We demonstrated the versatility and validity of the proposed framework for the increasing complexity of problems. To begin, we verified the framework against reference results with a 2D rotational disk case and generalized the case to the 3D rotational sphere. The cases showed that rotational motion and large deformation can be handled easily in the current framework. Next, we simulated an immersed sphere falling on an elastic block to demonstrate the capability to deal with translational motion and contact between solids. As observed, no-penetration contact can be naturally handled by the framework. Finally, we solved the ship-ice interaction problem involving free surface between air and water and contact between ship and ice floes. The hydraulic effects and contact between solids are faithfully captured in the case.
This application demonstrates the capability of the proposed framework to simulate practical multiphysics/multiphase systems with complex structures.
The framework allows the incorporation of additional physical fields and any number of solid objects thus enabling a valuable tool for analyzing a broad range of industrial applications.

\section*{Acknowledgement}
The authors would like to acknowledge the Natural Sciences and Engineering Research Council of Canada (NSERC) and Seaspan Shipyards for the funding. This research was supported in part through computational resources and services provided by Advanced Research Computing at the University of British Columbia.

\appendix
\setcounter{equation}{0} 
\setcounter{figure}{0}

\section{Parallel implementation for $L^2$-projection}\label{backward projection}
In the implementation of a time-dependent mobility model, we need to calculate the first-order derivatives of velocities and order parameters at nodes. A shape function based weighted average is used to achieve this, which is referred to as the $L^2$-projection. It projects the first-order derivatives from quadrature points to nodes. For node point $p$ and the shape function on this point $N_p$, the formulation is given as:
\begin{align}\label{bp}
(\nabla \varphi)_p = \frac{\int N_p \nabla\varphi d\Omega}{\int N_p d\Omega}
\end{align}
This step takes the information from all directions for nodes inside the domain. However, the information can only come from a single side for nodes located at boundaries. This will cause inconsistency in the order of accuracy.

For example, consider the domain of $x\in[0,2]$ with three nodes $x=0,1,2$. Denote the domain of $x\in[0,1]$ as $\Omega_L$ and $x\in[1,2]$ as $\Omega_R$. Piece-wise linear shape functions are used for Galerkin projection. We test the $L^2$-projection for the function of $\varphi = x^2$, the analytical solution of the first order derivative is $\frac{\partial\varphi}{\partial x}=2x$, which is a linear function. When multiplied with linear shape function at the numerator of Eq.~\ref{bp}, the integrated function is a second-order polynomial, which can be calculated exactly with first-order quadrature. As a result, the first-order derivative values calculated from the $L^2$-projection are:
\begin{align}
\frac{\partial\varphi}{\partial x}\Big|_{x=0}= \frac{\int(1-x)2xd\Omega_L}{\int(1-x)d\Omega_L}\approx&\frac{\sum\limits_{\mathrm{qp}=1}^2 w_{qp}(1-x_{\mathrm{qp}})2x_{qp}}{\sum\limits_{\mathrm{qp}=1}^2 w_{qp}(1-x_{\mathrm{qp}})}\approx0.667
\end{align}
\begin{align}
\frac{\partial\varphi}{\partial x}\Big|_{x=1}= \frac{\int(x)2xd\Omega_L+\int(2-x)2xd\Omega_R}{\int(x)d\Omega_L+\int(2-x)d\Omega_R}\approx&\frac{\sum\limits_{\mathrm{qp}=1}^2 w_{qp}(x_{\mathrm{qp}})2x_{qp}+\sum\limits_{\mathrm{qp}=1}^2 w_{qp}(2-x_{\mathrm{qp}})2x_{qp}}{\sum\limits_{\mathrm{qp}=1}^2 w_{qp}(x_{\mathrm{qp}})+\sum\limits_{\mathrm{qp}=1}^2 w_{qp}(2-x_{\mathrm{qp}})}\approx2
\end{align}
\begin{align}
\frac{\partial\varphi}{\partial x}\Big|_{x=2}= \frac{\int(x-1)2xd\Omega_R}{\int(x-1)d\Omega_R}\approx&\frac{\sum\limits_{\mathrm{qp}=1}^2 w_{qp}(x_{\mathrm{qp}}-1)2x_{qp}}{\sum\limits_{\mathrm{qp}=1}^2 w_{qp}(x_{\mathrm{qp}}-1)}\approx3.333
\end{align}
As observed, while the result at the interior node $x=1$ is identical with the exact value, errors are introduced at the boundary points of $x=0$ and $x=2$. This can be further clarified by expanding the domain to $x\in[-1,3]$. Following the same procedures, the first order derivatives can be calculated as: $\frac{\partial \varphi}{\partial x}\Big|_{x=-1}=-1.333,\frac{\partial \varphi}{\partial x}\Big|_{x=0}=0,\frac{\partial \varphi}{\partial x}\Big|_{x=1}=2,\frac{\partial \varphi}{\partial x}\Big|_{x=2}=4,\frac{\partial \varphi}{\partial x}\Big|_{x=3}=5.333$. As we can see, the error only occurs at the boundaries of the domain.

When the $L^2$-projection is implemented in parallel, the computational grid will be decomposed as multiple mesh blocks. At the boundaries of the mesh blocks, the error mentioned in previous paragraphs is introduced. To fix this issue, we first calculate the numerator $\int N_p \nabla\varphi d\Omega$ and the denominator $\int N_p d\Omega$ of Eq.~\ref{bp} at each mesh block. Then, we add the values at the mesh block boundary nodes together to take information from all directions. After this information exchange, we perform a node-wise division to complete the $L^2$-projection.

\section{Jacobian terms of solid stress in Eulerian frame}\label{solid stress}
In this appendix, we present a 2D component form of the Jacobian terms of the solid stress in the momentum conservation equation with a focus on the shear components. We start from the convection of the left Cauchy-Green tensor:

\begin{align}
	\frac{\partial B_{11}}{\partial t}=\left(-v_x\frac{\partial B_{11}}{\partial x}-v_y\frac{\partial B_{11}}{\partial y}+\left( \frac{\partial v_x}{\partial  x}B_{11}+ \frac{\partial v_x}{\partial y} B_{21}\right) +\left(B_{11} \frac{\partial v_x}{\partial x} +B_{12}\frac{\partial v_x}{\partial y} \right)\right),\nonumber\\
	\frac{\partial B_{12}}{\partial t}=\left(-v_x\frac{\partial B_{12}}{\partial x}-v_y\frac{\partial B_{12}}{\partial y}+\left( \frac{\partial v_x}{\partial  x}B_{12}+ \frac{\partial v_x}{\partial y} B_{22}\right) +\left(B_{11} \frac{\partial v_y}{\partial x} +B_{12}\frac{\partial v_y}{\partial y} \right)\right),\nonumber\\
	\frac{\partial B_{21}}{\partial t}=\left(-v_x\frac{\partial B_{21}}{\partial x}-v_y\frac{\partial B_{21}}{\partial y}+\left( \frac{\partial v_y}{\partial  x}B_{11}+ \frac{\partial v_y}{\partial y} B_{21}\right) +\left(B_{21} \frac{\partial v_x}{\partial x} +B_{22}\frac{\partial v_x}{\partial y} \right)\right),\nonumber\\
	\frac{\partial B_{22}}{\partial t}=\left(-v_x\frac{\partial B_{22}}{\partial x}-v_y\frac{\partial B_{22}}{\partial y}+\left( \frac{\partial v_y}{\partial  x}B_{12}+ \frac{\partial v_y}{\partial y} B_{22}\right) +\left(B_{21} \frac{\partial v_y}{\partial x} +B_{22}\frac{\partial v_y}{\partial y} \right)\right).\nonumber
\end{align}

To keep the symmetry of the left Cauchy-Green tensor, we let $B_{21}=B_{12}$ in the practical implementation. Using the generalized-$\alpha$ method, $\boldsymbol{B}^{n+\alpha}$ can be written as a function of $\boldsymbol{v}^{n+\alpha}$ as follows:
\begin{align}
	B_{11}^{n+\alpha}&=B_{11}^n+\alpha\Delta t \left(1-\frac{\gamma}{\alpha_m}\right)\partial_t B_{11}^n\nonumber\\
	&+\frac{\Delta t\alpha\gamma}{\alpha_m}\left(-v_x^{n+\alpha}\frac{\partial B_{11}^{n+\alpha}}{\partial x}-v_y^{n+\alpha}\frac{\partial B_{11}^{n+\alpha}}{\partial y}+2\frac{\partial v_x^{n+\alpha}}{\partial x}B_{11}^{n+\alpha}+2\frac{\partial v_x^{n+\alpha}}{\partial y}B_{12}^{n+\alpha}\right),\nonumber\\
	B_{12}^{n+\alpha}&=B_{12}^n+\alpha\Delta t \left(1-\frac{\gamma}{\alpha_m}\right)\partial_t B_{12}^n\nonumber\\
	&+\frac{\Delta t\alpha\gamma}{\alpha_m}\left(-v_x^{n+\alpha}\frac{\partial B_{12}^{n+\alpha}}{\partial x}-v_y^{n+\alpha}\frac{\partial B_{12}^{n+\alpha}}{\partial y}\right)\nonumber\\
	&+\frac{\Delta t\alpha\gamma}{\alpha_m}\left(\frac{\partial v_x^{n+\alpha}}{\partial x}B_{12}^{n+\alpha}+\frac{\partial v_x^{n+\alpha}}{\partial y}B_{22}^{n+\alpha}+\frac{\partial v_y^{n+\alpha}}{\partial x}B_{11}^{n+\alpha}+\frac{\partial v_y^{n+\alpha}}{\partial y}B_{12}^{n+\alpha}\right),\nonumber\\
	B_{21}^{n+\alpha}&=B_{12}^{n+\alpha},\nonumber\\
	B_{22}^{n+\alpha}&=B_{22}^n+\alpha\Delta t \left(1-\frac{\gamma}{\alpha_m}\right)\partial_t B_{22}^n\nonumber\\
	&+\frac{\Delta t\alpha\gamma}{\alpha_m}\left(-v_x^{n+\alpha}\frac{\partial B_{22}^{n+\alpha}}{\partial x}-v_y^{n+\alpha}\frac{\partial B_{22}^{n+\alpha}}{\partial y}+2\frac{\partial v_y^{n+\alpha}}{\partial y}B_{22}^{n+\alpha}+2\frac{\partial v_y^{n+\alpha}}{\partial x}B_{12}^{n+\alpha}\right).\nonumber
\end{align}

Finally, the Jacobian terms of the shear components in the solid stress are given by:
\begin{align}
	\mu^{\mathrm{s}}_L\frac{\delta(N_{,x}B^{n+\alpha}_{11}+N_{,y}B^{n+\alpha}_{12})}{\delta v_x^{n+\alpha}}
	&=\mu^{\mathrm{s}}_LN_{,x}\frac{\Delta t\alpha\gamma}{\alpha_m}\left(-N\frac{\partial B_{11}^{n+\alpha}}{\partial x}+2N_{,x}B_{11}^{n+\alpha}+2N_{,y}B_{12}^{n+\alpha}\right)\nonumber\\
	&+\mu^{\mathrm{s}}_LN_{,y}\frac{\Delta t\alpha\gamma}{\alpha_m}\left(-N\frac{\partial B_{12}^{n+\alpha}}{\partial x}+N_{,x}B_{12}^{n+\alpha}+N_{,y}B_{22}^{n+\alpha}\right),\nonumber
\end{align}
\begin{align}
	\mu^{\mathrm{s}}_L\frac{\delta(N_{,x}B^{n+\alpha}_{11}+N_{,y}B^{n+\alpha}_{12})}{\delta v_y^{n+\alpha}}&=\mu^{\mathrm{s}}_LN_{,x}\frac{\Delta t\alpha\gamma}{\alpha_m}\left(-N\frac{\partial B_{11}^{n+\alpha}}{\partial y}\right)\nonumber\\
	&+\mu^{\mathrm{s}}_LN_{,y}\frac{\Delta t\alpha\gamma}{\alpha_m}\left(-N\frac{\partial B_{12}^{n+\alpha}}{\partial y}+N_{,x}B_{11}^{n+\alpha}+N_{,y}B_{12}^{n+\alpha}\right),\nonumber
\end{align}
\begin{align}
	\mu^{\mathrm{s}}_L\frac{\delta(N_{,x}B^{n+\alpha}_{12}+N_{,y}B^{n+\alpha}_{22})}{\delta v_x^{n+\alpha}}&=\mu^{\mathrm{s}}_LN_{,x}\frac{\Delta t\alpha\gamma}{\alpha_m}\left(-N\frac{\partial B_{12}^{n+\alpha}}{\partial x}+N_{,x}B_{12}^{n+\alpha}+N_{,y}B_{22}^{n+\alpha}\right)\nonumber\\
	&+\mu^{\mathrm{s}}_LN_{,y}\frac{\Delta t\alpha\gamma}{\alpha_m}\left(-N\frac{\partial B_{22}^{n+\alpha}}{\partial x}\right),\nonumber
\end{align}
\begin{align}
	\mu^{\mathrm{s}}_L\frac{\delta(N_{,x}B^{n+\alpha}_{12}+N_{,y}B^{n+\alpha}_{22})}{\delta v_y^{n+\alpha}}&=\mu^{\mathrm{s}}_LN_{,x}\frac{\Delta t\alpha\gamma}{\alpha_m}\left(-N\frac{\partial B_{12}^{n+\alpha}}{\partial y}+N_{,x} B_{11}^{n+\alpha}+N_{,y}B_{12}^{n+\alpha}\right)\nonumber\\
	&+\mu^{\mathrm{s}}_LN_{,y}\frac{\Delta t\alpha\gamma}{\alpha_m}\left(-N\frac{\partial B_{22}^{n+\alpha}}{\partial y}+2N_{,y}\partial B_{22}^{n+\alpha}+2N_{,x}B_{12}^{n+\alpha}\right).\nonumber
\end{align}
\section{Parallel implementation of particle tracking method} \label{PT}
In an Eulerian frame of reference, tracking a solid particle may be required to know the exact motion of the solid. To achieve this, we need to search for the solid particle in the entire domain according to its coordinates, interpolate the velocity at the particle position, and evolve the particle accordingly. This procedure is repeated at each time step, which can become a large overhead without efficient implementation.
To avoid this issue, we adopt a parallelized implementation. We first broadcast the particle position to all the processors. In each processor, we split the elements into tetrahedrons. Then, the particle is searched in all the tetrahedrons.
Inside each tetrahedron, assuming linear basis function, we have four basis functions corresponding to the four vertices $N_i(x,y,z)=a_ix+b_iy+c_iz+d_i, i=1,2,3,4$. According to the definition of the shape function, $N_i(x,y,z)$ takes the value of $1$ at the node $i$, and takes the value of $0$ at the rest nodes. Due to the linearity, for all the points $(x,y,z)$ inside the tetrahedron, $N_i(x,y,z)\in[0,1],i = 1,2,3,4$.

With this in mind, we solve for the coefficients $a_i,b_i,c_i,d_i, i=1,2,3,4$. Considering the first shape function, which takes the value of $1$ at the first node, and becomes zero at the rest nodes:
\begin{align}
	\begin{bmatrix}
		x_1&y_1&z_1&1\\
		x_2&y_2&z_2&1\\
		x_3&y_3&z_3&1\\
		x_4&y_4&z_4&1
	\end{bmatrix}	\begin{bmatrix}
		a_1\\
		b_1\\
		c_1\\
		d_1
	\end{bmatrix}=	\begin{bmatrix}
		1\\
		0\\
		0\\
		0
	\end{bmatrix}.
\end{align}
Considering all the four nodes, we have:
\begin{align}
	\begin{bmatrix}
		x_1&y_1&z_1&1\\
		x_2&y_2&z_2&1\\
		x_3&y_3&z_3&1\\
		x_4&y_4&z_4&1
	\end{bmatrix}	\begin{bmatrix}
		a_1&a_2&a_3&a_4\\
		b_1&b_2&b_3&b_4\\
		c_1&c_2&c_3&c_4\\
		d_1&d_2&d_3&d_4
	\end{bmatrix}=	\begin{bmatrix}
		1&0&0&0\\
		0&1&0&0\\
		0&0&1&0\\
		0&0&0&1
	\end{bmatrix}.
\end{align}
The coefficients can be solved as:
\begin{align}
	\begin{bmatrix}
		a_1&a_2&a_3&a_4\\
		b_1&b_2&b_3&b_4\\
		c_1&c_2&c_3&c_4\\
		d_1&d_2&d_3&d_4
	\end{bmatrix} = 	\begin{bmatrix}
		x_1&y_1&z_1&1\\
		x_2&y_2&z_2&1\\
		x_3&y_3&z_3&1\\
		x_4&y_4&z_4&1
	\end{bmatrix}^{-1}.
\end{align}
After getting the coefficients, the value of the shape function on the particle position $(x,y,z)$ can be calculated as:
\begin{align}
	\begin{bmatrix}
		N_1(x,y,z)\\
		N_2(x,y,z)\\
		N_3(x,y,z)\\
		N_4(x,y,z)\\
	\end{bmatrix}=	\begin{bmatrix}
		a_1&a_2&a_3&a_4\\
		b_1&b_2&b_3&b_4\\
		c_1&c_2&c_3&c_4\\
		d_1&d_2&d_3&d_4
	\end{bmatrix}^{T}\begin{bmatrix}
		x\\
		y\\
		z\\
		1
	\end{bmatrix}.
\end{align}
if $\forall i=1,2,3,4,\ N_i(x,y,z)\in[0,1]$, the particle is inside the tetrahedron. Once the particle is found, its velocity can be interpolated as $\boldsymbol{v}=\sum\limits_{i=1}^4N_i(x,y,z)\boldsymbol{v}_i$. The solid particle is then evolved by the velocity and passed back to the root processor.

\bibliography{bibfile}
\end{document}